\documentclass[longauth]{aa}

\usepackage{graphicx}
\usepackage{natbib}
\usepackage{txfonts}
\usepackage{bm}
\usepackage{amsmath,amssymb,amsfonts}
\usepackage{graphicx}
\usepackage{txfonts}
\usepackage{enumerate}
\usepackage{colordvi}
\usepackage{acronym}
\usepackage{epsfig}
\usepackage[pdfencoding=auto,psdextra]{hyperref}
\hypersetup{
    colorlinks=true,
    linkcolor=blue,
    filecolor=magenta,      
    urlcolor=blue,
    citecolor=blue
}
\makeatletter
\renewcommand*\aa@pageof{, page \thepage{} of \pageref{LastPage}}
\makeatother

\usepackage[utf8]{inputenc}

\usepackage[dvipsnames]{xcolor}
\usepackage{color, colortbl}
\usepackage[T1]{fontenc}
\usepackage{braket}
\usepackage{placeins}
\usepackage{tabularx}
\usepackage{comment}
\usepackage{euclid}
\usepackage[switch, mathlines]{lineno}
\usepackage{textcomp}

\definecolor{amaranth}{rgb}{0.9, 0.17, 0.31}
\definecolor{forestgreen(web)}{rgb}{0.13, 0.55, 0.13}
\definecolor{lavender(web)}{rgb}{0.9, 0.9, 0.98}
\definecolor{cosmiclatte}{rgb}{1.0, 0.97, 0.91}
\definecolor{jonquil}{rgb}{0.98, 0.85, 0.37}
\definecolor{khaki(x11)(lightkhaki)}{rgb}{0.94, 0.9, 0.55}
\definecolor{thistle}{rgb}{0.85, 0.75, 0.85}

\newcommand{\bk}{\bm{k}}

\newcommand{\fnl}{f_{\rm NL}}
\newcommand{\fnll}{f_{\mathrm{NL}}^{\mathrm{loc}}}
\newcommand{\fnle}{f_{\mathrm{NL}}^{\mathrm{equil}}}
\newcommand{\fnloC}{f_{\mathrm{NL}}^{\mathrm{ortho-CMB}}}
\newcommand{\fnloL}{f_{\mathrm{NL}}^{\mathrm{ortho-LSS}}}

\newcommand*{\veps} {\varepsilon}
\newcommand{\GCsp}{{\text{GC}\ensuremath{_\mathrm{sp}}}}
\newcommand{\GCph}{{\text{GC}\ensuremath{_\mathrm{ph}}}}

\newcommand{\lcdm}{\ensuremath{\Lambda\mathrm{CDM}}}

\newcommand{\de}{\mathrm{d}}
\newcommand{\bpd}{{b_{\phi , \, \delta}}}

\newcommand{\borg}{\texttt{BORG}\xspace}
\newcommand{\Mpch}{\ensuremath{h^{-1}\,\text{Mpc}}}
\newcommand{\dd}{\mathrm{d}}

\begin{document} 

\title{\vspace*{-0.3cm}Euclid preparation. Expected constraints on initial conditions}

\newcommand{\orcid}[1]{}

\author{Euclid Collaboration: F.~Finelli\orcid{0000-0002-6694-3269}\thanks{\email{fabio.finelli@inaf.it}}\inst{\ref{aff1},\ref{aff2}}
\and Y.~Akrami\orcid{0000-0002-2407-7956}\inst{\ref{aff3},\ref{aff4}}
\and A.~Andrews\inst{\ref{aff1},\ref{aff2}}
\and M.~Ballardini\orcid{0000-0003-4481-3559}\inst{\ref{aff5},\ref{aff6},\ref{aff1}}
\and S.~Casas\orcid{0000-0002-4751-5138}\inst{\ref{aff7}}
\and D.~Karagiannis\orcid{0000-0002-4927-0816}\inst{\ref{aff5},\ref{aff8}}
\and Z.~Sakr\orcid{0000-0002-4823-3757}\inst{\ref{aff9},\ref{aff10},\ref{aff11}}
\and J.~Valiviita\orcid{0000-0001-6225-3693}\inst{\ref{aff12},\ref{aff13}}
\and G.~Alestas\orcid{0000-0003-1790-4914}\inst{\ref{aff3}}
\and N.~Bartolo\inst{\ref{aff14},\ref{aff15},\ref{aff16}}
\and J.~R.~Bermejo-Climent\inst{\ref{aff17},\ref{aff18}}
\and S.~Nesseris\orcid{0000-0002-0567-0324}\inst{\ref{aff3}}
\and D.~Paoletti\orcid{0000-0003-4761-6147}\inst{\ref{aff1},\ref{aff2}}
\and D.~Sapone\orcid{0000-0001-7089-4503}\inst{\ref{aff19}}
\and I.~Tutusaus\orcid{0000-0002-3199-0399}\inst{\ref{aff10}}
\and A.~Ach\'ucarro\inst{\ref{aff20},\ref{aff21}}
\and G.~Ca\~nas-Herrera\orcid{0000-0003-2796-2149}\inst{\ref{aff22},\ref{aff20},\ref{aff23}}
\and J.~Jasche\orcid{0000-0002-4677-5843}\inst{\ref{aff24},\ref{aff25}}
\and G.~Lavaux\orcid{0000-0003-0143-8891}\inst{\ref{aff26}}
\and N.~Aghanim\orcid{0000-0002-6688-8992}\inst{\ref{aff27}}
\and B.~Altieri\orcid{0000-0003-3936-0284}\inst{\ref{aff28}}
\and A.~Amara\inst{\ref{aff29}}
\and L.~Amendola\orcid{0000-0002-0835-233X}\inst{\ref{aff9}}
\and S.~Andreon\orcid{0000-0002-2041-8784}\inst{\ref{aff30}}
\and N.~Auricchio\orcid{0000-0003-4444-8651}\inst{\ref{aff1}}
\and C.~Baccigalupi\orcid{0000-0002-8211-1630}\inst{\ref{aff31},\ref{aff32},\ref{aff33},\ref{aff34}}
\and D.~Bagot\inst{\ref{aff35}}
\and M.~Baldi\orcid{0000-0003-4145-1943}\inst{\ref{aff36},\ref{aff1},\ref{aff37}}
\and S.~Bardelli\orcid{0000-0002-8900-0298}\inst{\ref{aff1}}
\and P.~Battaglia\orcid{0000-0002-7337-5909}\inst{\ref{aff1}}
\and A.~Biviano\orcid{0000-0002-0857-0732}\inst{\ref{aff32},\ref{aff31}}
\and E.~Branchini\orcid{0000-0002-0808-6908}\inst{\ref{aff38},\ref{aff39},\ref{aff30}}
\and M.~Brescia\orcid{0000-0001-9506-5680}\inst{\ref{aff40},\ref{aff41}}
\and S.~Camera\orcid{0000-0003-3399-3574}\inst{\ref{aff42},\ref{aff43},\ref{aff44}}
\and V.~Capobianco\orcid{0000-0002-3309-7692}\inst{\ref{aff44}}
\and C.~Carbone\orcid{0000-0003-0125-3563}\inst{\ref{aff45}}
\and J.~Carretero\orcid{0000-0002-3130-0204}\inst{\ref{aff46},\ref{aff47}}
\and M.~Castellano\orcid{0000-0001-9875-8263}\inst{\ref{aff48}}
\and G.~Castignani\orcid{0000-0001-6831-0687}\inst{\ref{aff1}}
\and S.~Cavuoti\orcid{0000-0002-3787-4196}\inst{\ref{aff41},\ref{aff49}}
\and K.~C.~Chambers\orcid{0000-0001-6965-7789}\inst{\ref{aff50}}
\and A.~Cimatti\inst{\ref{aff51}}
\and C.~Colodro-Conde\inst{\ref{aff17}}
\and G.~Congedo\orcid{0000-0003-2508-0046}\inst{\ref{aff52}}
\and C.~J.~Conselice\orcid{0000-0003-1949-7638}\inst{\ref{aff53}}
\and L.~Conversi\orcid{0000-0002-6710-8476}\inst{\ref{aff54},\ref{aff28}}
\and Y.~Copin\orcid{0000-0002-5317-7518}\inst{\ref{aff55}}
\and F.~Courbin\orcid{0000-0003-0758-6510}\inst{\ref{aff56},\ref{aff57}}
\and H.~M.~Courtois\orcid{0000-0003-0509-1776}\inst{\ref{aff58}}
\and M.~Cropper\orcid{0000-0003-4571-9468}\inst{\ref{aff59}}
\and A.~Da~Silva\orcid{0000-0002-6385-1609}\inst{\ref{aff60},\ref{aff61}}
\and H.~Degaudenzi\orcid{0000-0002-5887-6799}\inst{\ref{aff62}}
\and S.~de~la~Torre\inst{\ref{aff63}}
\and G.~De~Lucia\orcid{0000-0002-6220-9104}\inst{\ref{aff32}}
\and A.~M.~Di~Giorgio\orcid{0000-0002-4767-2360}\inst{\ref{aff64}}
\and H.~Dole\orcid{0000-0002-9767-3839}\inst{\ref{aff27}}
\and M.~Douspis\orcid{0000-0003-4203-3954}\inst{\ref{aff27}}
\and F.~Dubath\orcid{0000-0002-6533-2810}\inst{\ref{aff62}}
\and C.~A.~J.~Duncan\orcid{0009-0003-3573-0791}\inst{\ref{aff52},\ref{aff53}}
\and X.~Dupac\inst{\ref{aff28}}
\and S.~Dusini\orcid{0000-0002-1128-0664}\inst{\ref{aff15}}
\and S.~Escoffier\orcid{0000-0002-2847-7498}\inst{\ref{aff65}}
\and M.~Farina\orcid{0000-0002-3089-7846}\inst{\ref{aff64}}
\and R.~Farinelli\inst{\ref{aff1}}
\and F.~Faustini\orcid{0000-0001-6274-5145}\inst{\ref{aff48},\ref{aff66}}
\and S.~Ferriol\inst{\ref{aff55}}
\and P.~Fosalba\orcid{0000-0002-1510-5214}\inst{\ref{aff67},\ref{aff68}}
\and M.~Frailis\orcid{0000-0002-7400-2135}\inst{\ref{aff32}}
\and E.~Franceschi\orcid{0000-0002-0585-6591}\inst{\ref{aff1}}
\and M.~Fumana\orcid{0000-0001-6787-5950}\inst{\ref{aff45}}
\and S.~Galeotta\orcid{0000-0002-3748-5115}\inst{\ref{aff32}}
\and K.~George\orcid{0000-0002-1734-8455}\inst{\ref{aff69}}
\and B.~Gillis\orcid{0000-0002-4478-1270}\inst{\ref{aff52}}
\and C.~Giocoli\orcid{0000-0002-9590-7961}\inst{\ref{aff1},\ref{aff37}}
\and J.~Gracia-Carpio\inst{\ref{aff70}}
\and A.~Grazian\orcid{0000-0002-5688-0663}\inst{\ref{aff16}}
\and F.~Grupp\inst{\ref{aff70},\ref{aff69}}
\and S.~V.~H.~Haugan\orcid{0000-0001-9648-7260}\inst{\ref{aff71}}
\and W.~Holmes\inst{\ref{aff72}}
\and I.~M.~Hook\orcid{0000-0002-2960-978X}\inst{\ref{aff73}}
\and F.~Hormuth\inst{\ref{aff74}}
\and A.~Hornstrup\orcid{0000-0002-3363-0936}\inst{\ref{aff75},\ref{aff76}}
\and K.~Jahnke\orcid{0000-0003-3804-2137}\inst{\ref{aff77}}
\and M.~Jhabvala\inst{\ref{aff78}}
\and B.~Joachimi\orcid{0000-0001-7494-1303}\inst{\ref{aff79}}
\and E.~Keih\"anen\orcid{0000-0003-1804-7715}\inst{\ref{aff80}}
\and S.~Kermiche\orcid{0000-0002-0302-5735}\inst{\ref{aff65}}
\and A.~Kiessling\orcid{0000-0002-2590-1273}\inst{\ref{aff72}}
\and B.~Kubik\orcid{0009-0006-5823-4880}\inst{\ref{aff55}}
\and M.~K\"ummel\orcid{0000-0003-2791-2117}\inst{\ref{aff69}}
\and M.~Kunz\orcid{0000-0002-3052-7394}\inst{\ref{aff81}}
\and H.~Kurki-Suonio\orcid{0000-0002-4618-3063}\inst{\ref{aff12},\ref{aff13}}
\and A.~M.~C.~Le~Brun\orcid{0000-0002-0936-4594}\inst{\ref{aff82}}
\and S.~Ligori\orcid{0000-0003-4172-4606}\inst{\ref{aff44}}
\and P.~B.~Lilje\orcid{0000-0003-4324-7794}\inst{\ref{aff71}}
\and V.~Lindholm\orcid{0000-0003-2317-5471}\inst{\ref{aff12},\ref{aff13}}
\and I.~Lloro\orcid{0000-0001-5966-1434}\inst{\ref{aff83}}
\and G.~Mainetti\orcid{0000-0003-2384-2377}\inst{\ref{aff84}}
\and D.~Maino\inst{\ref{aff85},\ref{aff45},\ref{aff86}}
\and E.~Maiorano\orcid{0000-0003-2593-4355}\inst{\ref{aff1}}
\and O.~Mansutti\orcid{0000-0001-5758-4658}\inst{\ref{aff32}}
\and S.~Marcin\inst{\ref{aff87}}
\and O.~Marggraf\orcid{0000-0001-7242-3852}\inst{\ref{aff88}}
\and M.~Martinelli\orcid{0000-0002-6943-7732}\inst{\ref{aff48},\ref{aff89}}
\and N.~Martinet\orcid{0000-0003-2786-7790}\inst{\ref{aff63}}
\and F.~Marulli\orcid{0000-0002-8850-0303}\inst{\ref{aff90},\ref{aff1},\ref{aff37}}
\and R.~J.~Massey\orcid{0000-0002-6085-3780}\inst{\ref{aff91}}
\and E.~Medinaceli\orcid{0000-0002-4040-7783}\inst{\ref{aff1}}
\and S.~Mei\orcid{0000-0002-2849-559X}\inst{\ref{aff92},\ref{aff93}}
\and Y.~Mellier\inst{\ref{aff24},\ref{aff26}}
\and M.~Meneghetti\orcid{0000-0003-1225-7084}\inst{\ref{aff1},\ref{aff37}}
\and E.~Merlin\orcid{0000-0001-6870-8900}\inst{\ref{aff48}}
\and G.~Meylan\inst{\ref{aff94}}
\and A.~Mora\orcid{0000-0002-1922-8529}\inst{\ref{aff95}}
\and M.~Moresco\orcid{0000-0002-7616-7136}\inst{\ref{aff90},\ref{aff1}}
\and L.~Moscardini\orcid{0000-0002-3473-6716}\inst{\ref{aff90},\ref{aff1},\ref{aff37}}
\and C.~Neissner\orcid{0000-0001-8524-4968}\inst{\ref{aff96},\ref{aff47}}
\and S.-M.~Niemi\orcid{0009-0005-0247-0086}\inst{\ref{aff22}}
\and C.~Padilla\orcid{0000-0001-7951-0166}\inst{\ref{aff96}}
\and S.~Paltani\orcid{0000-0002-8108-9179}\inst{\ref{aff62}}
\and F.~Pasian\orcid{0000-0002-4869-3227}\inst{\ref{aff32}}
\and K.~Pedersen\inst{\ref{aff97}}
\and W.~J.~Percival\orcid{0000-0002-0644-5727}\inst{\ref{aff98},\ref{aff99},\ref{aff100}}
\and V.~Pettorino\inst{\ref{aff22}}
\and S.~Pires\orcid{0000-0002-0249-2104}\inst{\ref{aff101}}
\and G.~Polenta\orcid{0000-0003-4067-9196}\inst{\ref{aff66}}
\and M.~Poncet\inst{\ref{aff35}}
\and L.~A.~Popa\inst{\ref{aff102}}
\and L.~Pozzetti\orcid{0000-0001-7085-0412}\inst{\ref{aff1}}
\and F.~Raison\orcid{0000-0002-7819-6918}\inst{\ref{aff70}}
\and R.~Rebolo\orcid{0000-0003-3767-7085}\inst{\ref{aff17},\ref{aff103},\ref{aff18}}
\and A.~Renzi\orcid{0000-0001-9856-1970}\inst{\ref{aff14},\ref{aff15}}
\and J.~Rhodes\orcid{0000-0002-4485-8549}\inst{\ref{aff72}}
\and G.~Riccio\inst{\ref{aff41}}
\and E.~Romelli\orcid{0000-0003-3069-9222}\inst{\ref{aff32}}
\and M.~Roncarelli\orcid{0000-0001-9587-7822}\inst{\ref{aff1}}
\and C.~Rosset\orcid{0000-0003-0286-2192}\inst{\ref{aff92}}
\and R.~Saglia\orcid{0000-0003-0378-7032}\inst{\ref{aff69},\ref{aff70}}
\and B.~Sartoris\orcid{0000-0003-1337-5269}\inst{\ref{aff69},\ref{aff32}}
\and M.~Schirmer\orcid{0000-0003-2568-9994}\inst{\ref{aff77}}
\and T.~Schrabback\orcid{0000-0002-6987-7834}\inst{\ref{aff104}}
\and A.~Secroun\orcid{0000-0003-0505-3710}\inst{\ref{aff65}}
\and E.~Sefusatti\orcid{0000-0003-0473-1567}\inst{\ref{aff32},\ref{aff31},\ref{aff33}}
\and G.~Seidel\orcid{0000-0003-2907-353X}\inst{\ref{aff77}}
\and M.~Seiffert\orcid{0000-0002-7536-9393}\inst{\ref{aff72}}
\and S.~Serrano\orcid{0000-0002-0211-2861}\inst{\ref{aff67},\ref{aff105},\ref{aff68}}
\and P.~Simon\inst{\ref{aff88}}
\and C.~Sirignano\orcid{0000-0002-0995-7146}\inst{\ref{aff14},\ref{aff15}}
\and G.~Sirri\orcid{0000-0003-2626-2853}\inst{\ref{aff37}}
\and A.~Spurio~Mancini\orcid{0000-0001-5698-0990}\inst{\ref{aff106}}
\and L.~Stanco\orcid{0000-0002-9706-5104}\inst{\ref{aff15}}
\and J.~Steinwagner\orcid{0000-0001-7443-1047}\inst{\ref{aff70}}
\and P.~Tallada-Cresp\'{i}\orcid{0000-0002-1336-8328}\inst{\ref{aff46},\ref{aff47}}
\and D.~Tavagnacco\orcid{0000-0001-7475-9894}\inst{\ref{aff32}}
\and A.~N.~Taylor\inst{\ref{aff52}}
\and I.~Tereno\orcid{0000-0002-4537-6218}\inst{\ref{aff60},\ref{aff107}}
\and N.~Tessore\orcid{0000-0002-9696-7931}\inst{\ref{aff79}}
\and S.~Toft\orcid{0000-0003-3631-7176}\inst{\ref{aff108},\ref{aff109}}
\and R.~Toledo-Moreo\orcid{0000-0002-2997-4859}\inst{\ref{aff110}}
\and F.~Torradeflot\orcid{0000-0003-1160-1517}\inst{\ref{aff47},\ref{aff46}}
\and L.~Valenziano\orcid{0000-0002-1170-0104}\inst{\ref{aff1},\ref{aff2}}
\and T.~Vassallo\orcid{0000-0001-6512-6358}\inst{\ref{aff69},\ref{aff32}}
\and G.~Verdoes~Kleijn\orcid{0000-0001-5803-2580}\inst{\ref{aff111}}
\and A.~Veropalumbo\orcid{0000-0003-2387-1194}\inst{\ref{aff30},\ref{aff39},\ref{aff38}}
\and Y.~Wang\orcid{0000-0002-4749-2984}\inst{\ref{aff112}}
\and J.~Weller\orcid{0000-0002-8282-2010}\inst{\ref{aff69},\ref{aff70}}
\and A.~Zacchei\orcid{0000-0003-0396-1192}\inst{\ref{aff32},\ref{aff31}}
\and G.~Zamorani\orcid{0000-0002-2318-301X}\inst{\ref{aff1}}
\and F.~M.~Zerbi\inst{\ref{aff30}}
\and E.~Zucca\orcid{0000-0002-5845-8132}\inst{\ref{aff1}}
\and V.~Allevato\orcid{0000-0001-7232-5152}\inst{\ref{aff41}}
\and E.~Bozzo\orcid{0000-0002-8201-1525}\inst{\ref{aff62}}
\and C.~Burigana\orcid{0000-0002-3005-5796}\inst{\ref{aff113},\ref{aff2}}
\and R.~Cabanac\orcid{0000-0001-6679-2600}\inst{\ref{aff10}}
\and M.~Calabrese\orcid{0000-0002-2637-2422}\inst{\ref{aff114},\ref{aff45}}
\and A.~Cappi\inst{\ref{aff1},\ref{aff115}}
\and D.~Di~Ferdinando\inst{\ref{aff37}}
\and J.~A.~Escartin~Vigo\inst{\ref{aff70}}
\and L.~Gabarra\orcid{0000-0002-8486-8856}\inst{\ref{aff116}}
\and J.~Mart\'{i}n-Fleitas\orcid{0000-0002-8594-569X}\inst{\ref{aff117}}
\and S.~Matthew\orcid{0000-0001-8448-1697}\inst{\ref{aff52}}
\and N.~Mauri\orcid{0000-0001-8196-1548}\inst{\ref{aff51},\ref{aff37}}
\and R.~B.~Metcalf\orcid{0000-0003-3167-2574}\inst{\ref{aff90},\ref{aff1}}
\and A.~A.~Nucita\inst{\ref{aff118},\ref{aff119},\ref{aff120}}
\and A.~Pezzotta\orcid{0000-0003-0726-2268}\inst{\ref{aff121},\ref{aff70}}
\and M.~P\"ontinen\orcid{0000-0001-5442-2530}\inst{\ref{aff12}}
\and C.~Porciani\orcid{0000-0002-7797-2508}\inst{\ref{aff88}}
\and I.~Risso\orcid{0000-0003-2525-7761}\inst{\ref{aff122}}
\and V.~Scottez\orcid{0009-0008-3864-940X}\inst{\ref{aff24},\ref{aff123}}
\and M.~Sereno\orcid{0000-0003-0302-0325}\inst{\ref{aff1},\ref{aff37}}
\and M.~Tenti\orcid{0000-0002-4254-5901}\inst{\ref{aff37}}
\and M.~Viel\orcid{0000-0002-2642-5707}\inst{\ref{aff31},\ref{aff32},\ref{aff34},\ref{aff33},\ref{aff124}}
\and M.~Wiesmann\orcid{0009-0000-8199-5860}\inst{\ref{aff71}}
\and I.~T.~Andika\orcid{0000-0001-6102-9526}\inst{\ref{aff125},\ref{aff126}}
\and M.~Archidiacono\orcid{0000-0003-4952-9012}\inst{\ref{aff85},\ref{aff86}}
\and F.~Atrio-Barandela\orcid{0000-0002-2130-2513}\inst{\ref{aff127}}
\and S.~Avila\orcid{0000-0001-5043-3662}\inst{\ref{aff46}}
\and A.~Balaguera-Antolinez\orcid{0000-0001-5028-3035}\inst{\ref{aff17},\ref{aff128}}
\and D.~Bertacca\orcid{0000-0002-2490-7139}\inst{\ref{aff14},\ref{aff16},\ref{aff15}}
\and M.~Bethermin\orcid{0000-0002-3915-2015}\inst{\ref{aff129}}
\and A.~Blanchard\orcid{0000-0001-8555-9003}\inst{\ref{aff10}}
\and L.~Blot\orcid{0000-0002-9622-7167}\inst{\ref{aff130},\ref{aff82}}
\and H.~B\"ohringer\orcid{0000-0001-8241-4204}\inst{\ref{aff70},\ref{aff131},\ref{aff132}}
\and S.~Borgani\orcid{0000-0001-6151-6439}\inst{\ref{aff133},\ref{aff31},\ref{aff32},\ref{aff33},\ref{aff124}}
\and M.~L.~Brown\orcid{0000-0002-0370-8077}\inst{\ref{aff53}}
\and S.~Bruton\orcid{0000-0002-6503-5218}\inst{\ref{aff134}}
\and A.~Calabro\orcid{0000-0003-2536-1614}\inst{\ref{aff48}}
\and B.~Camacho~Quevedo\orcid{0000-0002-8789-4232}\inst{\ref{aff31},\ref{aff34},\ref{aff32},\ref{aff67},\ref{aff68}}
\and F.~Caro\inst{\ref{aff48}}
\and C.~S.~Carvalho\inst{\ref{aff107}}
\and T.~Castro\orcid{0000-0002-6292-3228}\inst{\ref{aff32},\ref{aff33},\ref{aff31},\ref{aff124}}
\and F.~Cogato\orcid{0000-0003-4632-6113}\inst{\ref{aff90},\ref{aff1}}
\and S.~Conseil\orcid{0000-0002-3657-4191}\inst{\ref{aff55}}
\and A.~R.~Cooray\orcid{0000-0002-3892-0190}\inst{\ref{aff135}}
\and S.~Davini\orcid{0000-0003-3269-1718}\inst{\ref{aff39}}
\and F.~De~Paolis\orcid{0000-0001-6460-7563}\inst{\ref{aff118},\ref{aff119},\ref{aff120}}
\and G.~Desprez\orcid{0000-0001-8325-1742}\inst{\ref{aff111}}
\and A.~D\'iaz-S\'anchez\orcid{0000-0003-0748-4768}\inst{\ref{aff136}}
\and J.~J.~Diaz\orcid{0000-0003-2101-1078}\inst{\ref{aff17}}
\and S.~Di~Domizio\orcid{0000-0003-2863-5895}\inst{\ref{aff38},\ref{aff39}}
\and J.~M.~Diego\orcid{0000-0001-9065-3926}\inst{\ref{aff137}}
\and P.~Dimauro\orcid{0000-0001-7399-2854}\inst{\ref{aff138},\ref{aff48}}
\and A.~Enia\orcid{0000-0002-0200-2857}\inst{\ref{aff36},\ref{aff1}}
\and Y.~Fang\inst{\ref{aff69}}
\and A.~G.~Ferrari\orcid{0009-0005-5266-4110}\inst{\ref{aff37}}
\and A.~Finoguenov\orcid{0000-0002-4606-5403}\inst{\ref{aff12}}
\and A.~Fontana\orcid{0000-0003-3820-2823}\inst{\ref{aff48}}
\and A.~Franco\orcid{0000-0002-4761-366X}\inst{\ref{aff119},\ref{aff118},\ref{aff120}}
\and K.~Ganga\orcid{0000-0001-8159-8208}\inst{\ref{aff92}}
\and J.~Garc\'ia-Bellido\orcid{0000-0002-9370-8360}\inst{\ref{aff3}}
\and T.~Gasparetto\orcid{0000-0002-7913-4866}\inst{\ref{aff32}}
\and V.~Gautard\inst{\ref{aff139}}
\and E.~Gaztanaga\orcid{0000-0001-9632-0815}\inst{\ref{aff68},\ref{aff67},\ref{aff140}}
\and F.~Giacomini\orcid{0000-0002-3129-2814}\inst{\ref{aff37}}
\and F.~Gianotti\orcid{0000-0003-4666-119X}\inst{\ref{aff1}}
\and G.~Gozaliasl\orcid{0000-0002-0236-919X}\inst{\ref{aff141},\ref{aff12}}
\and A.~Gruppuso\orcid{0000-0001-9272-5292}\inst{\ref{aff1},\ref{aff37}}
\and M.~Guidi\orcid{0000-0001-9408-1101}\inst{\ref{aff36},\ref{aff1}}
\and C.~M.~Gutierrez\orcid{0000-0001-7854-783X}\inst{\ref{aff142}}
\and S.~Hemmati\orcid{0000-0003-2226-5395}\inst{\ref{aff143}}
\and C.~Hern\'andez-Monteagudo\orcid{0000-0001-5471-9166}\inst{\ref{aff18},\ref{aff17}}
\and H.~Hildebrandt\orcid{0000-0002-9814-3338}\inst{\ref{aff144}}
\and J.~Hjorth\orcid{0000-0002-4571-2306}\inst{\ref{aff97}}
\and S.~Joudaki\orcid{0000-0001-8820-673X}\inst{\ref{aff46}}
\and J.~J.~E.~Kajava\orcid{0000-0002-3010-8333}\inst{\ref{aff145},\ref{aff146}}
\and Y.~Kang\orcid{0009-0000-8588-7250}\inst{\ref{aff62}}
\and V.~Kansal\orcid{0000-0002-4008-6078}\inst{\ref{aff147},\ref{aff148}}
\and K.~Kiiveri\inst{\ref{aff80}}
\and C.~C.~Kirkpatrick\inst{\ref{aff80}}
\and S.~Kruk\orcid{0000-0001-8010-8879}\inst{\ref{aff28}}
\and M.~Lattanzi\orcid{0000-0003-1059-2532}\inst{\ref{aff6}}
\and V.~Le~Brun\orcid{0000-0002-5027-1939}\inst{\ref{aff63}}
\and J.~Le~Graet\orcid{0000-0001-6523-7971}\inst{\ref{aff65}}
\and L.~Legrand\orcid{0000-0003-0610-5252}\inst{\ref{aff149},\ref{aff150}}
\and M.~Lembo\orcid{0000-0002-5271-5070}\inst{\ref{aff26}}
\and F.~Lepori\orcid{0009-0000-5061-7138}\inst{\ref{aff151}}
\and G.~Leroy\orcid{0009-0004-2523-4425}\inst{\ref{aff152},\ref{aff91}}
\and G.~F.~Lesci\orcid{0000-0002-4607-2830}\inst{\ref{aff90},\ref{aff1}}
\and J.~Lesgourgues\orcid{0000-0001-7627-353X}\inst{\ref{aff7}}
\and L.~Leuzzi\orcid{0009-0006-4479-7017}\inst{\ref{aff1}}
\and T.~I.~Liaudat\orcid{0000-0002-9104-314X}\inst{\ref{aff153}}
\and J.~Macias-Perez\orcid{0000-0002-5385-2763}\inst{\ref{aff154}}
\and G.~Maggio\orcid{0000-0003-4020-4836}\inst{\ref{aff32}}
\and M.~Magliocchetti\orcid{0000-0001-9158-4838}\inst{\ref{aff64}}
\and F.~Mannucci\orcid{0000-0002-4803-2381}\inst{\ref{aff155}}
\and R.~Maoli\orcid{0000-0002-6065-3025}\inst{\ref{aff156},\ref{aff48}}
\and C.~J.~A.~P.~Martins\orcid{0000-0002-4886-9261}\inst{\ref{aff157},\ref{aff158}}
\and L.~Maurin\orcid{0000-0002-8406-0857}\inst{\ref{aff27}}
\and M.~Migliaccio\inst{\ref{aff159},\ref{aff160}}
\and M.~Miluzio\inst{\ref{aff28},\ref{aff161}}
\and P.~Monaco\orcid{0000-0003-2083-7564}\inst{\ref{aff133},\ref{aff32},\ref{aff33},\ref{aff31}}
\and C.~Moretti\orcid{0000-0003-3314-8936}\inst{\ref{aff34},\ref{aff124},\ref{aff32},\ref{aff31},\ref{aff33}}
\and G.~Morgante\inst{\ref{aff1}}
\and S.~Nadathur\orcid{0000-0001-9070-3102}\inst{\ref{aff140}}
\and K.~Naidoo\orcid{0000-0002-9182-1802}\inst{\ref{aff140}}
\and A.~Navarro-Alsina\orcid{0000-0002-3173-2592}\inst{\ref{aff88}}
\and L.~Pagano\orcid{0000-0003-1820-5998}\inst{\ref{aff5},\ref{aff6}}
\and F.~Passalacqua\orcid{0000-0002-8606-4093}\inst{\ref{aff14},\ref{aff15}}
\and K.~Paterson\orcid{0000-0001-8340-3486}\inst{\ref{aff77}}
\and L.~Patrizii\inst{\ref{aff37}}
\and A.~Pisani\orcid{0000-0002-6146-4437}\inst{\ref{aff65}}
\and D.~Potter\orcid{0000-0002-0757-5195}\inst{\ref{aff151}}
\and S.~Quai\orcid{0000-0002-0449-8163}\inst{\ref{aff90},\ref{aff1}}
\and M.~Radovich\orcid{0000-0002-3585-866X}\inst{\ref{aff16}}
\and P.~Reimberg\orcid{0000-0003-3410-0280}\inst{\ref{aff24}}
\and P.-F.~Rocci\inst{\ref{aff27}}
\and G.~Rodighiero\orcid{0000-0002-9415-2296}\inst{\ref{aff14},\ref{aff16}}
\and S.~Sacquegna\orcid{0000-0002-8433-6630}\inst{\ref{aff118},\ref{aff119},\ref{aff120}}
\and M.~Sahl\'en\orcid{0000-0003-0973-4804}\inst{\ref{aff162}}
\and D.~B.~Sanders\orcid{0000-0002-1233-9998}\inst{\ref{aff50}}
\and E.~Sarpa\orcid{0000-0002-1256-655X}\inst{\ref{aff34},\ref{aff124},\ref{aff33}}
\and A.~Schneider\orcid{0000-0001-7055-8104}\inst{\ref{aff151}}
\and D.~Sciotti\orcid{0009-0008-4519-2620}\inst{\ref{aff48},\ref{aff89}}
\and E.~Sellentin\inst{\ref{aff163},\ref{aff23}}
\and L.~C.~Smith\orcid{0000-0002-3259-2771}\inst{\ref{aff164}}
\and K.~Tanidis\orcid{0000-0001-9843-5130}\inst{\ref{aff116}}
\and C.~Tao\orcid{0000-0001-7961-8177}\inst{\ref{aff65}}
\and G.~Testera\inst{\ref{aff39}}
\and R.~Teyssier\orcid{0000-0001-7689-0933}\inst{\ref{aff165}}
\and S.~Tosi\orcid{0000-0002-7275-9193}\inst{\ref{aff38},\ref{aff39},\ref{aff30}}
\and A.~Troja\orcid{0000-0003-0239-4595}\inst{\ref{aff14},\ref{aff15}}
\and M.~Tucci\inst{\ref{aff62}}
\and C.~Valieri\inst{\ref{aff37}}
\and A.~Venhola\orcid{0000-0001-6071-4564}\inst{\ref{aff166}}
\and D.~Vergani\orcid{0000-0003-0898-2216}\inst{\ref{aff1}}
\and F.~Vernizzi\orcid{0000-0003-3426-2802}\inst{\ref{aff167}}
\and G.~Verza\orcid{0000-0002-1886-8348}\inst{\ref{aff168}}
\and P.~Vielzeuf\orcid{0000-0003-2035-9339}\inst{\ref{aff65}}
\and N.~A.~Walton\orcid{0000-0003-3983-8778}\inst{\ref{aff164}}}
										   
\institute{INAF-Osservatorio di Astrofisica e Scienza dello Spazio di Bologna, Via Piero Gobetti 93/3, 40129 Bologna, Italy\label{aff1}
\and
INFN-Bologna, Via Irnerio 46, 40126 Bologna, Italy\label{aff2}
\and
Instituto de F\'isica Te\'orica UAM-CSIC, Campus de Cantoblanco, 28049 Madrid, Spain\label{aff3}
\and
CERCA/ISO, Department of Physics, Case Western Reserve University, 10900 Euclid Avenue, Cleveland, OH 44106, USA\label{aff4}
\and
Dipartimento di Fisica e Scienze della Terra, Universit\`a degli Studi di Ferrara, Via Giuseppe Saragat 1, 44122 Ferrara, Italy\label{aff5}
\and
Istituto Nazionale di Fisica Nucleare, Sezione di Ferrara, Via Giuseppe Saragat 1, 44122 Ferrara, Italy\label{aff6}
\and
Institute for Theoretical Particle Physics and Cosmology (TTK), RWTH Aachen University, 52056 Aachen, Germany\label{aff7}
\and
Department of Physics and Astronomy, University of the Western Cape, Bellville, Cape Town, 7535, South Africa\label{aff8}
\and
Institut f\"ur Theoretische Physik, University of Heidelberg, Philosophenweg 16, 69120 Heidelberg, Germany\label{aff9}
\and
Institut de Recherche en Astrophysique et Plan\'etologie (IRAP), Universit\'e de Toulouse, CNRS, UPS, CNES, 14 Av. Edouard Belin, 31400 Toulouse, France\label{aff10}
\and
Universit\'e St Joseph; Faculty of Sciences, Beirut, Lebanon\label{aff11}
\and
Department of Physics, P.O. Box 64, 00014 University of Helsinki, Finland\label{aff12}
\and
Helsinki Institute of Physics, Gustaf H{\"a}llstr{\"o}min katu 2, University of Helsinki, Helsinki, Finland\label{aff13}
\and
Dipartimento di Fisica e Astronomia "G. Galilei", Universit\`a di Padova, Via Marzolo 8, 35131 Padova, Italy\label{aff14}
\and
INFN-Padova, Via Marzolo 8, 35131 Padova, Italy\label{aff15}
\and
INAF-Osservatorio Astronomico di Padova, Via dell'Osservatorio 5, 35122 Padova, Italy\label{aff16}
\and
Instituto de Astrof\'{\i}sica de Canarias, V\'{\i}a L\'actea, 38205 La Laguna, Tenerife, Spain\label{aff17}
\and
Universidad de La Laguna, Departamento de Astrof\'{\i}sica, 38206 La Laguna, Tenerife, Spain\label{aff18}
\and
Departamento de F\'isica, FCFM, Universidad de Chile, Blanco Encalada 2008, Santiago, Chile\label{aff19}
\and
Institute Lorentz, Leiden University, Niels Bohrweg 2, 2333 CA Leiden, The Netherlands\label{aff20}
\and
Departamento de F\'isica, Universidad del Pa\'is Vasco UPV-EHU, 48940 Leioa, Spain\label{aff21}
\and
European Space Agency/ESTEC, Keplerlaan 1, 2201 AZ Noordwijk, The Netherlands\label{aff22}
\and
Leiden Observatory, Leiden University, Einsteinweg 55, 2333 CC Leiden, The Netherlands\label{aff23}
\and
Institut d'Astrophysique de Paris, 98bis Boulevard Arago, 75014, Paris, France\label{aff24}
\and
Oskar Klein Centre for Cosmoparticle Physics, Department of Physics, Stockholm University, Stockholm, SE-106 91, Sweden\label{aff25}
\and
Institut d'Astrophysique de Paris, UMR 7095, CNRS, and Sorbonne Universit\'e, 98 bis boulevard Arago, 75014 Paris, France\label{aff26}
\and
Universit\'e Paris-Saclay, CNRS, Institut d'astrophysique spatiale, 91405, Orsay, France\label{aff27}
\and
ESAC/ESA, Camino Bajo del Castillo, s/n., Urb. Villafranca del Castillo, 28692 Villanueva de la Ca\~nada, Madrid, Spain\label{aff28}
\and
School of Mathematics and Physics, University of Surrey, Guildford, Surrey, GU2 7XH, UK\label{aff29}
\and
INAF-Osservatorio Astronomico di Brera, Via Brera 28, 20122 Milano, Italy\label{aff30}
\and
IFPU, Institute for Fundamental Physics of the Universe, via Beirut 2, 34151 Trieste, Italy\label{aff31}
\and
INAF-Osservatorio Astronomico di Trieste, Via G. B. Tiepolo 11, 34143 Trieste, Italy\label{aff32}
\and
INFN, Sezione di Trieste, Via Valerio 2, 34127 Trieste TS, Italy\label{aff33}
\and
SISSA, International School for Advanced Studies, Via Bonomea 265, 34136 Trieste TS, Italy\label{aff34}
\and
Centre National d'Etudes Spatiales -- Centre spatial de Toulouse, 18 avenue Edouard Belin, 31401 Toulouse Cedex 9, France\label{aff35}
\and
Dipartimento di Fisica e Astronomia, Universit\`a di Bologna, Via Gobetti 93/2, 40129 Bologna, Italy\label{aff36}
\and
INFN-Sezione di Bologna, Viale Berti Pichat 6/2, 40127 Bologna, Italy\label{aff37}
\and
Dipartimento di Fisica, Universit\`a di Genova, Via Dodecaneso 33, 16146, Genova, Italy\label{aff38}
\and
INFN-Sezione di Genova, Via Dodecaneso 33, 16146, Genova, Italy\label{aff39}
\and
Department of Physics "E. Pancini", University Federico II, Via Cinthia 6, 80126, Napoli, Italy\label{aff40}
\and
INAF-Osservatorio Astronomico di Capodimonte, Via Moiariello 16, 80131 Napoli, Italy\label{aff41}
\and
Dipartimento di Fisica, Universit\`a degli Studi di Torino, Via P. Giuria 1, 10125 Torino, Italy\label{aff42}
\and
INFN-Sezione di Torino, Via P. Giuria 1, 10125 Torino, Italy\label{aff43}
\and
INAF-Osservatorio Astrofisico di Torino, Via Osservatorio 20, 10025 Pino Torinese (TO), Italy\label{aff44}
\and
INAF-IASF Milano, Via Alfonso Corti 12, 20133 Milano, Italy\label{aff45}
\and
Centro de Investigaciones Energ\'eticas, Medioambientales y Tecnol\'ogicas (CIEMAT), Avenida Complutense 40, 28040 Madrid, Spain\label{aff46}
\and
Port d'Informaci\'{o} Cient\'{i}fica, Campus UAB, C. Albareda s/n, 08193 Bellaterra (Barcelona), Spain\label{aff47}
\and
INAF-Osservatorio Astronomico di Roma, Via Frascati 33, 00078 Monteporzio Catone, Italy\label{aff48}
\and
INFN section of Naples, Via Cinthia 6, 80126, Napoli, Italy\label{aff49}
\and
Institute for Astronomy, University of Hawaii, 2680 Woodlawn Drive, Honolulu, HI 96822, USA\label{aff50}
\and
Dipartimento di Fisica e Astronomia "Augusto Righi" - Alma Mater Studiorum Universit\`a di Bologna, Viale Berti Pichat 6/2, 40127 Bologna, Italy\label{aff51}
\and
Institute for Astronomy, University of Edinburgh, Royal Observatory, Blackford Hill, Edinburgh EH9 3HJ, UK\label{aff52}
\and
Jodrell Bank Centre for Astrophysics, Department of Physics and Astronomy, University of Manchester, Oxford Road, Manchester M13 9PL, UK\label{aff53}
\and
European Space Agency/ESRIN, Largo Galileo Galilei 1, 00044 Frascati, Roma, Italy\label{aff54}
\and
Universit\'e Claude Bernard Lyon 1, CNRS/IN2P3, IP2I Lyon, UMR 5822, Villeurbanne, F-69100, France\label{aff55}
\and
Institut de Ci\`{e}ncies del Cosmos (ICCUB), Universitat de Barcelona (IEEC-UB), Mart\'{i} i Franqu\`{e}s 1, 08028 Barcelona, Spain\label{aff56}
\and
Instituci\'o Catalana de Recerca i Estudis Avan\c{c}ats (ICREA), Passeig de Llu\'{\i}s Companys 23, 08010 Barcelona, Spain\label{aff57}
\and
UCB Lyon 1, CNRS/IN2P3, IUF, IP2I Lyon, 4 rue Enrico Fermi, 69622 Villeurbanne, France\label{aff58}
\and
Mullard Space Science Laboratory, University College London, Holmbury St Mary, Dorking, Surrey RH5 6NT, UK\label{aff59}
\and
Departamento de F\'isica, Faculdade de Ci\^encias, Universidade de Lisboa, Edif\'icio C8, Campo Grande, PT1749-016 Lisboa, Portugal\label{aff60}
\and
Instituto de Astrof\'isica e Ci\^encias do Espa\c{c}o, Faculdade de Ci\^encias, Universidade de Lisboa, Campo Grande, 1749-016 Lisboa, Portugal\label{aff61}
\and
Department of Astronomy, University of Geneva, ch. d'Ecogia 16, 1290 Versoix, Switzerland\label{aff62}
\and
Aix-Marseille Universit\'e, CNRS, CNES, LAM, Marseille, France\label{aff63}
\and
INAF-Istituto di Astrofisica e Planetologia Spaziali, via del Fosso del Cavaliere, 100, 00100 Roma, Italy\label{aff64}
\and
Aix-Marseille Universit\'e, CNRS/IN2P3, CPPM, Marseille, France\label{aff65}
\and
Space Science Data Center, Italian Space Agency, via del Politecnico snc, 00133 Roma, Italy\label{aff66}
\and
Institut d'Estudis Espacials de Catalunya (IEEC),  Edifici RDIT, Campus UPC, 08860 Castelldefels, Barcelona, Spain\label{aff67}
\and
Institute of Space Sciences (ICE, CSIC), Campus UAB, Carrer de Can Magrans, s/n, 08193 Barcelona, Spain\label{aff68}
\and
Universit\"ats-Sternwarte M\"unchen, Fakult\"at f\"ur Physik, Ludwig-Maximilians-Universit\"at M\"unchen, Scheinerstrasse 1, 81679 M\"unchen, Germany\label{aff69}
\and
Max Planck Institute for Extraterrestrial Physics, Giessenbachstr. 1, 85748 Garching, Germany\label{aff70}
\and
Institute of Theoretical Astrophysics, University of Oslo, P.O. Box 1029 Blindern, 0315 Oslo, Norway\label{aff71}
\and
Jet Propulsion Laboratory, California Institute of Technology, 4800 Oak Grove Drive, Pasadena, CA, 91109, USA\label{aff72}
\and
Department of Physics, Lancaster University, Lancaster, LA1 4YB, UK\label{aff73}
\and
Felix Hormuth Engineering, Goethestr. 17, 69181 Leimen, Germany\label{aff74}
\and
Technical University of Denmark, Elektrovej 327, 2800 Kgs. Lyngby, Denmark\label{aff75}
\and
Cosmic Dawn Center (DAWN), Denmark\label{aff76}
\and
Max-Planck-Institut f\"ur Astronomie, K\"onigstuhl 17, 69117 Heidelberg, Germany\label{aff77}
\and
NASA Goddard Space Flight Center, Greenbelt, MD 20771, USA\label{aff78}
\and
Department of Physics and Astronomy, University College London, Gower Street, London WC1E 6BT, UK\label{aff79}
\and
Department of Physics and Helsinki Institute of Physics, Gustaf H\"allstr\"omin katu 2, 00014 University of Helsinki, Finland\label{aff80}
\and
Universit\'e de Gen\`eve, D\'epartement de Physique Th\'eorique and Centre for Astroparticle Physics, 24 quai Ernest-Ansermet, CH-1211 Gen\`eve 4, Switzerland\label{aff81}
\and
Laboratoire d'etude de l'Univers et des phenomenes eXtremes, Observatoire de Paris, Universit\'e PSL, Sorbonne Universit\'e, CNRS, 92190 Meudon, France\label{aff82}
\and
SKA Observatory, Jodrell Bank, Lower Withington, Macclesfield, Cheshire SK11 9FT, UK\label{aff83}
\and
Centre de Calcul de l'IN2P3/CNRS, 21 avenue Pierre de Coubertin 69627 Villeurbanne Cedex, France\label{aff84}
\and
Dipartimento di Fisica "Aldo Pontremoli", Universit\`a degli Studi di Milano, Via Celoria 16, 20133 Milano, Italy\label{aff85}
\and
INFN-Sezione di Milano, Via Celoria 16, 20133 Milano, Italy\label{aff86}
\and
University of Applied Sciences and Arts of Northwestern Switzerland, School of Computer Science, 5210 Windisch, Switzerland\label{aff87}
\and
Universit\"at Bonn, Argelander-Institut f\"ur Astronomie, Auf dem H\"ugel 71, 53121 Bonn, Germany\label{aff88}
\and
INFN-Sezione di Roma, Piazzale Aldo Moro, 2 - c/o Dipartimento di Fisica, Edificio G. Marconi, 00185 Roma, Italy\label{aff89}
\and
Dipartimento di Fisica e Astronomia "Augusto Righi" - Alma Mater Studiorum Universit\`a di Bologna, via Piero Gobetti 93/2, 40129 Bologna, Italy\label{aff90}
\and
Department of Physics, Institute for Computational Cosmology, Durham University, South Road, Durham, DH1 3LE, UK\label{aff91}
\and
Universit\'e Paris Cit\'e, CNRS, Astroparticule et Cosmologie, 75013 Paris, France\label{aff92}
\and
CNRS-UCB International Research Laboratory, Centre Pierre Bin\'etruy, IRL2007, CPB-IN2P3, Berkeley, USA\label{aff93}
\and
Institute of Physics, Laboratory of Astrophysics, Ecole Polytechnique F\'ed\'erale de Lausanne (EPFL), Observatoire de Sauverny, 1290 Versoix, Switzerland\label{aff94}
\and
Telespazio UK S.L. for European Space Agency (ESA), Camino bajo del Castillo, s/n, Urbanizacion Villafranca del Castillo, Villanueva de la Ca\~nada, 28692 Madrid, Spain\label{aff95}
\and
Institut de F\'{i}sica d'Altes Energies (IFAE), The Barcelona Institute of Science and Technology, Campus UAB, 08193 Bellaterra (Barcelona), Spain\label{aff96}
\and
DARK, Niels Bohr Institute, University of Copenhagen, Jagtvej 155, 2200 Copenhagen, Denmark\label{aff97}
\and
Waterloo Centre for Astrophysics, University of Waterloo, Waterloo, Ontario N2L 3G1, Canada\label{aff98}
\and
Department of Physics and Astronomy, University of Waterloo, Waterloo, Ontario N2L 3G1, Canada\label{aff99}
\and
Perimeter Institute for Theoretical Physics, Waterloo, Ontario N2L 2Y5, Canada\label{aff100}
\and
Universit\'e Paris-Saclay, Universit\'e Paris Cit\'e, CEA, CNRS, AIM, 91191, Gif-sur-Yvette, France\label{aff101}
\and
Institute of Space Science, Str. Atomistilor, nr. 409 M\u{a}gurele, Ilfov, 077125, Romania\label{aff102}
\and
Consejo Superior de Investigaciones Cientificas, Calle Serrano 117, 28006 Madrid, Spain\label{aff103}
\and
Universit\"at Innsbruck, Institut f\"ur Astro- und Teilchenphysik, Technikerstr. 25/8, 6020 Innsbruck, Austria\label{aff104}
\and
Satlantis, University Science Park, Sede Bld 48940, Leioa-Bilbao, Spain\label{aff105}
\and
Department of Physics, Royal Holloway, University of London, TW20 0EX, UK\label{aff106}
\and
Instituto de Astrof\'isica e Ci\^encias do Espa\c{c}o, Faculdade de Ci\^encias, Universidade de Lisboa, Tapada da Ajuda, 1349-018 Lisboa, Portugal\label{aff107}
\and
Cosmic Dawn Center (DAWN)\label{aff108}
\and
Niels Bohr Institute, University of Copenhagen, Jagtvej 128, 2200 Copenhagen, Denmark\label{aff109}
\and
Universidad Polit\'ecnica de Cartagena, Departamento de Electr\'onica y Tecnolog\'ia de Computadoras,  Plaza del Hospital 1, 30202 Cartagena, Spain\label{aff110}
\and
Kapteyn Astronomical Institute, University of Groningen, PO Box 800, 9700 AV Groningen, The Netherlands\label{aff111}
\and
Infrared Processing and Analysis Center, California Institute of Technology, Pasadena, CA 91125, USA\label{aff112}
\and
INAF, Istituto di Radioastronomia, Via Piero Gobetti 101, 40129 Bologna, Italy\label{aff113}
\and
Astronomical Observatory of the Autonomous Region of the Aosta Valley (OAVdA), Loc. Lignan 39, I-11020, Nus (Aosta Valley), Italy\label{aff114}
\and
Universit\'e C\^{o}te d'Azur, Observatoire de la C\^{o}te d'Azur, CNRS, Laboratoire Lagrange, Bd de l'Observatoire, CS 34229, 06304 Nice cedex 4, France\label{aff115}
\and
Department of Physics, Oxford University, Keble Road, Oxford OX1 3RH, UK\label{aff116}
\and
Aurora Technology for European Space Agency (ESA), Camino bajo del Castillo, s/n, Urbanizacion Villafranca del Castillo, Villanueva de la Ca\~nada, 28692 Madrid, Spain\label{aff117}
\and
Department of Mathematics and Physics E. De Giorgi, University of Salento, Via per Arnesano, CP-I93, 73100, Lecce, Italy\label{aff118}
\and
INFN, Sezione di Lecce, Via per Arnesano, CP-193, 73100, Lecce, Italy\label{aff119}
\and
INAF-Sezione di Lecce, c/o Dipartimento Matematica e Fisica, Via per Arnesano, 73100, Lecce, Italy\label{aff120}
\and
INAF - Osservatorio Astronomico di Brera, via Emilio Bianchi 46, 23807 Merate, Italy\label{aff121}
\and
INAF-Osservatorio Astronomico di Brera, Via Brera 28, 20122 Milano, Italy, and INFN-Sezione di Genova, Via Dodecaneso 33, 16146, Genova, Italy\label{aff122}
\and
ICL, Junia, Universit\'e Catholique de Lille, LITL, 59000 Lille, France\label{aff123}
\and
ICSC - Centro Nazionale di Ricerca in High Performance Computing, Big Data e Quantum Computing, Via Magnanelli 2, Bologna, Italy\label{aff124}
\and
Technical University of Munich, TUM School of Natural Sciences, Physics Department, James-Franck-Str.~1, 85748 Garching, Germany\label{aff125}
\and
Max-Planck-Institut f\"ur Astrophysik, Karl-Schwarzschild-Str.~1, 85748 Garching, Germany\label{aff126}
\and
Departamento de F{\'\i}sica Fundamental. Universidad de Salamanca. Plaza de la Merced s/n. 37008 Salamanca, Spain\label{aff127}
\and
Instituto de Astrof\'isica de Canarias (IAC); Departamento de Astrof\'isica, Universidad de La Laguna (ULL), 38200, La Laguna, Tenerife, Spain\label{aff128}
\and
Universit\'e de Strasbourg, CNRS, Observatoire astronomique de Strasbourg, UMR 7550, 67000 Strasbourg, France\label{aff129}
\and
Center for Data-Driven Discovery, Kavli IPMU (WPI), UTIAS, The University of Tokyo, Kashiwa, Chiba 277-8583, Japan\label{aff130}
\and
Ludwig-Maximilians-University, Schellingstrasse 4, 80799 Munich, Germany\label{aff131}
\and
Max-Planck-Institut f\"ur Physik, Boltzmannstr. 8, 85748 Garching, Germany\label{aff132}
\and
Dipartimento di Fisica - Sezione di Astronomia, Universit\`a di Trieste, Via Tiepolo 11, 34131 Trieste, Italy\label{aff133}
\and
California Institute of Technology, 1200 E California Blvd, Pasadena, CA 91125, USA\label{aff134}
\and
Department of Physics \& Astronomy, University of California Irvine, Irvine CA 92697, USA\label{aff135}
\and
Departamento F\'isica Aplicada, Universidad Polit\'ecnica de Cartagena, Campus Muralla del Mar, 30202 Cartagena, Murcia, Spain\label{aff136}
\and
Instituto de F\'isica de Cantabria, Edificio Juan Jord\'a, Avenida de los Castros, 39005 Santander, Spain\label{aff137}
\and
Observatorio Nacional, Rua General Jose Cristino, 77-Bairro Imperial de Sao Cristovao, Rio de Janeiro, 20921-400, Brazil\label{aff138}
\and
CEA Saclay, DFR/IRFU, Service d'Astrophysique, Bat. 709, 91191 Gif-sur-Yvette, France\label{aff139}
\and
Institute of Cosmology and Gravitation, University of Portsmouth, Portsmouth PO1 3FX, UK\label{aff140}
\and
Department of Computer Science, Aalto University, PO Box 15400, Espoo, FI-00 076, Finland\label{aff141}
\and
Instituto de Astrof\'\i sica de Canarias, c/ Via Lactea s/n, La Laguna 38200, Spain. Departamento de Astrof\'\i sica de la Universidad de La Laguna, Avda. Francisco Sanchez, La Laguna, 38200, Spain\label{aff142}
\and
Caltech/IPAC, 1200 E. California Blvd., Pasadena, CA 91125, USA\label{aff143}
\and
Ruhr University Bochum, Faculty of Physics and Astronomy, Astronomical Institute (AIRUB), German Centre for Cosmological Lensing (GCCL), 44780 Bochum, Germany\label{aff144}
\and
Department of Physics and Astronomy, Vesilinnantie 5, 20014 University of Turku, Finland\label{aff145}
\and
Serco for European Space Agency (ESA), Camino bajo del Castillo, s/n, Urbanizacion Villafranca del Castillo, Villanueva de la Ca\~nada, 28692 Madrid, Spain\label{aff146}
\and
ARC Centre of Excellence for Dark Matter Particle Physics, Melbourne, Australia\label{aff147}
\and
Centre for Astrophysics \& Supercomputing, Swinburne University of Technology,  Hawthorn, Victoria 3122, Australia\label{aff148}
\and
DAMTP, Centre for Mathematical Sciences, Wilberforce Road, Cambridge CB3 0WA, UK\label{aff149}
\and
Kavli Institute for Cosmology Cambridge, Madingley Road, Cambridge, CB3 0HA, UK\label{aff150}
\and
Department of Astrophysics, University of Zurich, Winterthurerstrasse 190, 8057 Zurich, Switzerland\label{aff151}
\and
Department of Physics, Centre for Extragalactic Astronomy, Durham University, South Road, Durham, DH1 3LE, UK\label{aff152}
\and
IRFU, CEA, Universit\'e Paris-Saclay 91191 Gif-sur-Yvette Cedex, France\label{aff153}
\and
Univ. Grenoble Alpes, CNRS, Grenoble INP, LPSC-IN2P3, 53, Avenue des Martyrs, 38000, Grenoble, France\label{aff154}
\and
INAF-Osservatorio Astrofisico di Arcetri, Largo E. Fermi 5, 50125, Firenze, Italy\label{aff155}
\and
Dipartimento di Fisica, Sapienza Universit\`a di Roma, Piazzale Aldo Moro 2, 00185 Roma, Italy\label{aff156}
\and
Centro de Astrof\'{\i}sica da Universidade do Porto, Rua das Estrelas, 4150-762 Porto, Portugal\label{aff157}
\and
Instituto de Astrof\'isica e Ci\^encias do Espa\c{c}o, Universidade do Porto, CAUP, Rua das Estrelas, PT4150-762 Porto, Portugal\label{aff158}
\and
Dipartimento di Fisica, Universit\`a di Roma Tor Vergata, Via della Ricerca Scientifica 1, Roma, Italy\label{aff159}
\and
INFN, Sezione di Roma 2, Via della Ricerca Scientifica 1, Roma, Italy\label{aff160}
\and
HE Space for European Space Agency (ESA), Camino bajo del Castillo, s/n, Urbanizacion Villafranca del Castillo, Villanueva de la Ca\~nada, 28692 Madrid, Spain\label{aff161}
\and
Theoretical astrophysics, Department of Physics and Astronomy, Uppsala University, Box 516, 751 37 Uppsala, Sweden\label{aff162}
\and
Mathematical Institute, University of Leiden, Einsteinweg 55, 2333 CA Leiden, The Netherlands\label{aff163}
\and
Institute of Astronomy, University of Cambridge, Madingley Road, Cambridge CB3 0HA, UK\label{aff164}
\and
Department of Astrophysical Sciences, Peyton Hall, Princeton University, Princeton, NJ 08544, USA\label{aff165}
\and
Space physics and astronomy research unit, University of Oulu, Pentti Kaiteran katu 1, FI-90014 Oulu, Finland\label{aff166}
\and
Institut de Physique Th\'eorique, CEA, CNRS, Universit\'e Paris-Saclay 91191 Gif-sur-Yvette Cedex, France\label{aff167}
\and
Center for Computational Astrophysics, Flatiron Institute, 162 5th Avenue, 10010, New York, NY, USA\label{aff168}}    

\date{Received \textit{Month} \textit{Day} 2025; accepted \textit{Month} \textit{Day}, 2025} 
 
\abstract 
{The {\em Euclid} mission of the European Space Agency will deliver galaxy and cosmic shear surveys, which will be used to constrain initial conditions and statistics of primordial fluctuations.}
{We present highlights for the {\em Euclid} scientific capability to test initial conditions beyond $\Lambda$CDM
with the 3-dimensional galaxy clustering from the spectroscopic survey, the tomographic approach to $3\times2$pt statistics (information) from photometric galaxy survey, and their combination. We then present how these {\em Euclid} results can be enhanced when combined with current and future measurements of the cosmic microwave background (CMB) anisotropies.}
{We provide Fisher forecasts from the combination of {\em Euclid} spectroscopic and photometric surveys
for spatial curvature, running of the spectral index of the power spectrum of curvature perturbations, isocurvature perturbations, and primordial features. For the parameters of these models we also provide the combination of
{\em Euclid} forecasts (pessimistic and optimistic) with three different CMB specifications, i.e. {\em Planck}, the Simons Observatory (SO), and CMB-S4. We provide Fisher forecasts for how the power spectrum and bispectrum from the {\em Euclid} spectroscopic survey will constrain the local, equilateral, and orthogonal shapes of primordial non-Gaussianity. We also review how Bayesian field-level inference of primordial non-Gaussianity can constrain local primordial non-Gaussianity.}
{We find that the combination of the {\em Euclid} main probes will provide the uncertainty of $\sigma(\Omega_K) = 0.0044~(0.003)$ at the $68 \%$ confidence level (CL) in the pessimistic (optimistic) settings assuming flat spatial sections as fiducial cosmology. We also find that the combination of the {\em Euclid} main probes can detect the running of the scalar spectral index for the fiducial value $\alpha_s = - 0.01$ with approximately $2\sigma$ ($4\sigma$) uncertainty and provide the uncertainty of $\sigma(\alpha_s)  = 0.004~(0.0015)$ for the fiducial value $\alpha_s = - 0.001$ at the $68\%$ CL, always with pessimistic (optimistic) settings. We show how {\em Euclid} will have the capability to provide constraints on isocurvature perturbations with a
blue spectral index that are one order of magnitude tighter than current bounds. The combined power spectrum and bispectrum Fisher forecast for the {\em Euclid} spectroscopic survey leads to $\sigma (f_{\rm NL}^{\rm local}) = 2.2$, $\sigma (f_{\rm NL}^{\rm equil}) = 108$, and $\sigma (\fnl^{\rm ortho}) = 33$ by assuming $k_{\rm max} = 0.15\,h\,{\rm Mpc}^{-1}$ and universality for the halo mass function. We show how the expected uncertainty on the local shape decreases by encompassing summary statistics and using Bayesian field-level inference. For the {\em Euclid} main probes, we find relative errors on the amplitude of primordial oscillations (with a fiducial value of $0.01$), of 21\% (18 \%) for linear frequency and of 22\% (18 \%) for logarithmic frequency at $68.3\%$ CL in the pessimistic (optimistic) case. These uncertainties can be further improved by adding the information from the bispectrum and the non-linear reconstruction.}
{We show how {\em Euclid}, with its unique combination of 3-dimensional galaxy clustering from the spectroscopic survey and $3\times2$pt statistics from the photometric survey, will provide the tightest constraints on low redshift to date. By targeting a markedly different range in redshift and scale, {\em Euclid}'s expected uncertainties are complementary to those obtained by CMB primary anisotropy, returning the tightest combined constraints on the physics of the early Universe.}


\titlerunning{\Euclid preparation. Expected constraints on initial conditions}
\authorrunning{Euclid Collaboration: Finelli et al.}

\maketitle

\section{Introduction}

The initial conditions of primordial fluctuations are the most ancient fossil imprint of the early Universe,
which can be retrieved from the statistical properties of the cosmological data provided by galaxy surveys.
With the large fraction of the sky measured by the current and planned galaxy surveys and their large redshift coverage, not only will they greatly improve our understanding of dark energy and dark matter, but they will also provide the most precise measurement of the initial conditions of primordial fluctuations at low redshift \citep{Amendola:2016saw,Alvarez:2014vva}.

The current observational status is consistent with a spatially flat universe and nearly
Gaussian, adiabatic perturbations whose spectrum is described by a simple power-law function \citep{Planck:2018vyg,Akrami:2018odb,Planck:2019izv,eBOSS:2020yzd}:
\begin{equation}
  {\cal P_R}(k) \equiv \frac{k^3}{2 \pi^2} |{\cal R}(k)|^2 =  A_\mathrm{s} \left(\frac{k}{k_*}\right)^{n_\mathrm{s} - 1} \,,
  \label{Asns}
\end{equation}
where $A_\mathrm{s}$ is the amplitude of the power spectrum of the scalar curvature perturbation ${\cal R}$ at the pivot scale $k_*$ (usually taken to be $k_*=0.05$ Mpc$^{-1}$) and $n_\mathrm{s}$ is the scalar spectral index.

Current cosmological observations are consistent with the key predictions of cosmic inflation \citep{Brout:1977ix, Starobinsky:1980te,Guth:1980zm,Sato:1980yn,Linde:1981mu,Albrecht:1982wi,Hawking:1982ga,Linde:1983gd}, a phase of nearly exponential expansion in the primordial Universe, which diluted any pre-existing inhomogeneity -- this has provided arguably the most elegant solution to the flatness problem and the other puzzles of the standard hot Big Bang cosmology. 
In the simplest models, a single real Klein--Gordon scalar field -- the inflaton -- slowly rolling on a sufficiently flat potential, drives the inflationary stage, during which unavoidable quantum fluctuations in scalar \citep{Mukhanov:1981xt, Hawking:1982cz, Guth:1982ec, Starobinsky:1982ee, 1983PhRvD..28..679B, Mukhanov:1985rz} and tensor \citep{Starobinsky:1979ty} curvature perturbations are stretched to scales relevant for structure formation inheriting a nearly scale-invariant spectrum with statistics that are close to Gaussian. The initial conditions for the primordial perturbations, which evolve into the large-scale structure that we observe today, are therefore connected to the inflaton's Lagrangian, in particular to the inflaton's potential, under minimal assumptions for the simplest models.

The power of cosmological observations to  constrain the initial conditions is currently dominated by measurements of
the anisotropies of the cosmic microwave background (CMB),
which mainly provide a snapshot of initial conditions at photon decoupling, i.e. at $z \sim 1100$, when cosmological fluctuations can be described in the linear regime. However, by the end of the current decade, thanks to the 3-dimensional (3D) information contained in the large-scale structure (LSS), the constraining power on the initial conditions retrieved from statistics of the data provided by the galaxy surveys DESI \citep{2016arXiv161100036D, DESI:2024mwx}, {\em Euclid} \citep{Euclid:2024yrr}, LSST/Vera Rubin Observatory  \citep{2009arXiv0912.0201L}, and SPHEREx \citep{SPHEREx:2014bgr} will increase significantly compared to current LSS measurements. Thanks to the sensitivity to the initial conditions at different redshifts and wavelengths,
and to the access to the higher number of modes contained in the 3D volume, galaxy surveys will help us break the degeneracies between cosmological parameters that are encoded in the CMB constraints. This will lead to combined, 
tighter constraints on initial conditions that are more robust than current bounds on the physics governing the low-redshift cosmological evolution.

{\em Euclid} \citep{Redbook,Euclid:2024yrr,Euclid:2021icp} is a medium-class mission of the European Space Agency, launched on July 1st, 2023, which will map the local Universe to improve our understanding of cosmic expansion history and of the growth of cosmic structures. The telescope will observe approximately 14\,000 deg$^2$ of the sky through two instruments. The Visible Camera \citep{EuclidSkyVIS} will collect optical images of more than one billion galaxies, whose photometric redshift will be determined with supporting  ground-based observations. The Near-Infrared Spectrometer and Photometer \citep{EuclidSkyNISP}, aims to determine emission-line redshifts of between 20 and 30 million galaxies in the range $0.9 < z <1.8$ by using slitless spectroscopy.
The combination of spectroscopy and photometry will allow us to measure galaxy clustering and weak gravitational lensing, aiming at percent accuracy in the corresponding power spectra.

In this paper we study the {\em Euclid} capabilities in testing initial conditions and statistics of primordial fluctuations through their main probes, i.e. 3D clustering from the galaxy spectroscopic sample and the tomographic approach to the joint galaxy clustering - weak gravitational lensing statistics from the photometric galaxy survey. We will also show how these {\em Euclid} forecasts can be enhanced when combined with current and future measurements of the cosmic microwave background (CMB) anisotropies.

This paper is organised as follows. In Sect.~\ref{sec:Euclid}, we summarise the modeling of {\em Euclid} clustering from the spectroscopic survey and of the joint galaxy clustering -- weak lensing analysis -- from the photometric survey.
We present in Sect. \ref{sec:Omegak} the expected
constraints from {\em Euclid}'s main probes on the spatial curvature of the Universe, i.e. on the curvature parameter $\Omega_K$. {\em Euclid} forecasts for the constraints on the spectral index of the spectrum of curvature perturbations and its running are presented in Sect. \ref{sec:powerspectrum}.
We discuss in Sect. \ref{sec:isocurvature} how {\em Euclid} can constrain the presence of additional isocurvature fluctuations. In Sect. \ref{sec:PNG} we discuss the expected
constraints on primordial non-Gaussianity. We first present the Fisher forecasts of the joint power spectrum
and bispectrum analysis of the {\em Euclid} spectroscopic survey for the local, equilateral, and orthogonal main shapes.
In the same section we also summarise what can be achieved for the local shape of primordial non-Gaussianity by Bayesian field-level inference, always for the spectroscopic survey. Section \ref{sec:features} summarises the expected constraints from {\em Euclid} on primordial features, and finally, Sect. \ref{sec:conclusions} presents our conclusions. In Appendix \ref{sec:AppendixA}, we review the methodology adopted to combine the {\em Euclid} forecasts with the information contained in current and future measurements of the CMB anisotropies.

In this paper, unless stated otherwise, all uncertainties (denoted by $\sigma$) are provided at the $68\%$ confidence level (CL).

\section{{\it Euclid}'s main probes}
\label{sec:Euclid}

In this section, we summarise the modelling of the spectroscopic galaxy clustering, the $3\times2$pt analysis from the photometric survey, and their combination,  following~\citet{Euclid:2019clj}, henceforth, EC20 \footnote{Differently from EC20, we use only the direct approach to 3D galaxy clustering used in EC20. We also corrected the inconsistency in the unit conversion from Mpc$^{-1}$ to $h$ Mpc$^{-1}$ in the 3D galaxy clustering that affected forecasts in EC20: thus, the uncertainty in $h$ degrades with respect to EC20 by a factor $\sim 4$ in $\Lambda$CDM, while those of other cosmological parameters are only mildly changed.}. Although the currently foreseen specifications of the \Euclid Wide Survey (EWS) differ from those assumed in EC20, e.g. in terms of the survey area \citep{Euclid:2024yrr}, and the theoretical non-linear modelling within the \Euclid collaboration has made progress since EC20 -- see, e.g., \cite{Euclid:2023bgs,Euclid:2024xfd} for physics beyond $\Lambda$CDM at low redshift -- we use the EC20 methodology here for initial conditions to allow for comparison with similar forecasts in the context of dark energy/modified gravity \citep{Euclid:2023tqw,Euclid:2023rjj}, neutrinos \citep{Euclid:2024imf}, and particle dark matter \citep{Euclid:2024pwi}.

\subsection{Spectroscopic galaxy clustering}\label{sec:spectro_clustering} 

We model the full, anisotropic, non-linear observed galaxy power spectrum as:
\begin{align}
P_\text{obs}(k_{\rm ref},\mu_{\rm ref};\,z) =& 
\frac{1}{q_\perp^2 (z) \, q_\parallel (z)}
\left\{\frac{\left[b (z) \sigma_8(z) + f_{\rm g}(z) \, \sigma_8(z) \, \mu^2\right]^2}{1+k^2\mu^2f_{\rm g}^2(z) \, \sigma_{\rm p}^2(z)}\right\}\nonumber \\ 
&\times \frac{P_\text{dw}(k,\mu;z)}{\sigma^2_8(z)}
F_z(k,\mu;z) +P_\text{s}(z) \,,
\label{eq:GC:pkobs}
\end{align}
where $\mu$ is the cosine of the angle of the wave mode with respect to the line of sight pointing in the $\hat{r}$ direction, and $k$ is the wavenumber of the perturbation.
The subscript ``${\rm ref}$'' refers to the reference (or fiducial) cosmology.

The first term on the right-hand side of Eq. (\ref{eq:GC:pkobs}) is due to the Alcock--Paczynski (AP)  effect \citep{Alcock:1979mp}, which takes into account deviations from the fiducial cosmology by the conversion of redshifts and angles in perpendicular and parallel distances,
\begin{align}\label{eq:AP1}
    q_\perp(z) &= \frac{k_{\perp,{\rm ref}}}{k_\perp} = \frac{D_{\rm A}(z)}{D_{\rm A,\, ref}(z)}\,, \nonumber \\ 
    q_\parallel(z) &= \frac{k_{\parallel,{\rm ref}}}{k_\parallel} = \frac{H_{\rm ref}(z)}{H(z)} \,,
\end{align}
where $D_{\rm A}(z)$ and $H(z)$ are the angular diameter distance and 
the Hubble expansion rate, respectively.

The anisotropic distortion in the density field due to the line-of-sight effects of the peculiar velocity of the observed galaxy redshifts, known as redshift-space distortions (RSD), is modelled in the linear regime by the contribution in the numerator in the curly brackets of Eq. \eqref{eq:GC:pkobs} \citep{Kaiser:1987qv}.
Here, the linear galaxy bias $b$, connecting the dark matter (DM) power spectrum to the one of H$\alpha$ emitter galaxies targeted by NISP, and the growth rate $f_{\rm g}$ are considered scale-independent unless otherwise specified. 
The fingers-of-god (FoG) effect caused by galaxy peculiar velocity is modelled as a Lorentzian factor with the distance dispersion $\sigma_\mathrm{p} (z)$ following \cite{Hamilton:1997zq}.

The dewiggled power spectrum, $P_{\rm dw}(k,\mu;\,z)$, models the smearing of the baryon acoustic oscillations (BAO) signal due to the effects of large-scale bulk flows \citep{Wang:2012bx} as:
\begin{equation} \label{eq:GC:pk_dw}
    P_{\rm dw}(k,\mu;z) = P_{\rm m}^{\rm \,L}(k;z)\,\text{e}^{-g_\mu k^2} + P_{\rm nw}(k;z)\left(1-\text{e}^{-g_\mu k^2}\right) \,,
\end{equation}
where $P_{\rm nw}(k;z) $ is the `no-wiggle' linear power spectrum and $P_{\rm m}^{\rm\, L}(k;z)$ is the linear matter power spectrum. The former is obtained by removing the BAO from the latter \citep{Boyle:2017lzt}, while the linear matter power spectrum is derived by an Einstein-Boltzmann numerical code, e.g. CAMB \citep{CAMB} or CLASS \citep{CLASS}. The amplitude of the nonlinear damping factor is given by \citep{Eisenstein:2006nj} 
\begin{eqnarray}
    g_{\mu}(z, \mu) = \sigma^2_{\rm v}(z)\left[1 - \mu^2 + \mu^2\left[1+f_\mathrm{g}(z)\right]^2\right]\,,\label{eq:gmu}
\label{eq:sigmav}
\end{eqnarray}
where $\sigma_{\rm v}(z)$ is the galaxy velocity dispersion modelled analogously to $\sigma_{\rm p} (z)$ in Eq. (\ref{eq:GC:pkobs}) with four parameters, one for each redshift bin (see EC20 for more details).

The total galaxy power spectrum in Eq.~\eqref{eq:GC:pkobs} includes the errors on redshift through the factor 
\begin{equation} \label{eq:zerrors}
    F_z(k, \mu;z) = \text{e}^{-k^2\mu^2\sigma_\mathrm{r}^2(z)}\,,
\end{equation}
where $\sigma_\mathrm{r}(z) = c(1+z)\sigma_{z,0}/H(z)$ and $\sigma_{z,0} = 0.001$ is the error on the measured 
redshift.

The last term in Eq.~\eqref{eq:GC:pkobs}, $P_{\rm s}(z)$, is the shot noise caused by the Poissonian distribution of measured galaxies on the smallest scales due to the finite total number of observed galaxies. For each bin, we model it with one contribution given by the fiducial inverse number density, plus a second nuisance parameter accounting for the residual shot noise.

The final Fisher matrix for the galaxy clustering observable for one redshift bin $z_i$ is 
\begin{align} \label{eq:fisher-gc}
    F_{\alpha\beta}(z_i) =& \frac{1}{8\pi^2}\int_{-1}^{1}{\rm d}\mu\int_{k_{\rm min}}^{k_{\rm max}} V_{\rm eff}(k,\mu;\,z) \frac{\partial \ln P_{\rm obs}(k,\mu;\,z_i)}{\partial p_\alpha} \nonumber \\
    & \times\frac{\partial \ln P_{\rm obs}(k,\mu;\,z_i)}{\partial p_\beta}  k^2{\rm d}k\,,
\end{align}
where $k_{\rm min} = 0.002 \,h\,{\rm Mpc}^{-1}$. 
The derivatives in Eq. (\ref{eq:fisher-gc}) are evaluated at the parameter values of the fiducial model and $V_{\rm eff}$ is 
the effective volume of the survey given by
\begin{equation}
    V_{\rm eff}(k,\mu;\,z) = V_{\rm s}(z)\left[\frac{n(z)P_{\rm obs}(k,\mu;\,z)}{1+n(z)P_{\rm obs}(k,\mu;\,z)}\right]^2\,,
\end{equation}
with $V_{\rm s}(z)$ the volume of the survey and $n(z)$ the galaxy number density in a redshift bin.

\subsection{Photometric galaxy clustering and weak lensing}

Beyond the spectroscopic survey, {\it Euclid} will also perform a photometric survey, which will enable us to consider at least three main cosmological probes. These are galaxy clustering for galaxies with photometric redshifts, weak lensing, and their cross-correlations, also called galaxy-galaxy lensing. The combination of these three probes is commonly denoted as a $3\times2$pt analysis. As for the spectroscopic case in the previous section, we closely follow the forecasting method and the tools presented in~EC20.
We refer the interested reader to this reference for all the details, but still provide a self-consistent description of the {\it \Euclid} photometric probes below.

We consider as the main photometric observables the angular harmonic power spectra, $C_{ij}^{XY}(\ell)$, between two probes $X$ and $Y$ in two tomographic bins $i$ and $j$. Under the Limber approximation and by setting the $\ell$-dependent prefactor in the flat-sky limit to unity, these spectra can be expressed as
\begin{equation}
    C_{ij}^{XY}(\ell) = c \int_{z_{\rm min}}^{z_{\rm max}}\,\text{d}z\, \frac{W_i^X(z)W_j^Y(z)}{H(z)r^2(z)}P_{\rm m}^{\rm NL}(k_{\ell},z)\,,
    \label{eq:CXYij}
\end{equation}
with $P_{\rm m}^{\rm NL}(k,z)$ the non-linear matter power spectrum evaluated at $k_{\ell}=(\ell+1/2)/r(z)$, where $r(z)$ is the co-moving distance.\footnote{As in EC20, and justified in \cite{Taylor:2018qda}, we use this formalism also for non-flat spatial geometries avoiding the introduction of hyper-spherical Bessel functions and non-Limber approximation by assuming that the changes in the cosmological parameters, the power spectrum, and the reduced window function are enough to capture the effect of a small non-zero spatial curvature.} We further set $z_{\rm min}=0.001$ and $z_{\rm max}=4$, which is enough to cover the entire redshift range of the survey. The kernels for the photometric galaxy clustering and the weak lensing probes are, respectively,
\begin{align}
 W_i^{\rm G}(z) =&\; b_i(z)\,\frac{n_i(z)}{\bar{n}_i}\frac{H(z)}{c}\,, \label{eq:wg_mg}\\  
 W_i^{\rm L}(z) =&\; \frac{3}{2}\Omega_{\rm m} \frac{H_0^2}{c^2}(1+z)\,r(z)
\int_z^{z_{\rm max}}{\de z'\frac{n_i(z')}{\bar{n}_i}\frac{r(z')-r(z)}{r(z')}}\nonumber\\
  &+W^{\rm IA}_i(z)\,, \label{eq:wl_mg}
\end{align}
where $n_i(z)/\bar{n}_i$ stands for the normalised galaxy distribution in the $i$-th bin and $b_i(z)$ denotes the galaxy bias for the same tomographic bin (see Sect. 3 of~EC20 for more details).

Finally, the $W^{\rm IA}_i(z)$ term contains the contribution of the intrinsic alignment of galaxies to weak lensing. As in~EC20, we consider the extended non-linear alignment (eNLA) model with the window function
\begin{equation}\label{eq:IA}
 W^{\rm IA}_i(z)=-\frac{\mathcal{A}_{\rm IA}\,\mathcal{C}_{\rm IA}\,\Omega_{\rm m}\,\mathcal{F}_{\rm IA}(z)}{D(z)}\frac{n_i(z)}{\bar{n}_i(z)}\frac{H(z)}{c}\,,
\end{equation}
where 
\begin{equation}
 \mathcal{F}_{\rm IA}(z)=(1+z)^{\eta_{\rm IA}}\left[\frac{\langle L\rangle(z)}{L_\star(z)}\right]^{\beta_{\rm IA}}\,.
\end{equation}
In the equation above, $\langle L\rangle(z)$ and $L_\star(z)$ are the redshift-dependent mean and the characteristic luminosity of source galaxies as computed from the luminosity function. The growth factor is represented by $D(z)$. 
As in EC20, we keep $\mathcal{C}_{\rm IA}$ fixed to $0.0134$ and we allow to vary the other three parameters of the eNLA model $\{\mathcal{A}_{\rm IA}\,,\eta_{\rm IA}\,,\beta_{\rm IA}\}$ around their fiducial values $=\{1.72\,,-0.41\,,2.17\}$. 

We consider a Gaussian-only covariance whose elements are given by
\begin{align}
    \text{Cov}\left[C_{ij,\ell}^{\rm AB},C_{mn,\ell'}^{\rm CD}\right]=&\frac{\delta_{\ell\ell'}^{\rm K}}{(2\ell+1)f_{\rm sky}\Delta \ell} \notag\\
    &\times\left\{\left[C_{im,\ell}^{\rm AC}+{\cal N}_{im,\ell}^{\rm AC}\right]\left[C_{jn,\ell'}^{\rm BD}+{\cal N}_{jn,\ell'}^{\rm BD}\right]\right. \notag\\
    &+\left.\left[C_{in,\ell}^{\rm AD}+{\cal N}_{in,\ell}^{\rm AD}\right]\left[C_{jm,\ell'}^{\rm BC}+{\cal N}_{jm,\ell'}^{\rm BC}\right]\right\} \,,
\end{align}
where the upper- and lower-case Latin indices run over L and G and all tomographic bins, respectively, $\delta_{\ell\ell'}^{\rm K}$ 
is the Kronecker delta coming from the lack of correlation between different multipoles $(\ell, \ell')$, $f_{\rm sky}$ 
is the survey's sky fraction, and $\Delta \ell$ denotes the width of the logarithmic equi-spaced multipole bins. We 
consider a white noise,
\begin{align}
    {\cal N}_{ij,\ell}^{\rm LL} &= \frac{\delta_{ij}^{\rm K}}{\bar{n}_i}\sigma_\epsilon^2 \,, \quad
    {\cal N}_{ij,\ell}^{\rm GG} = \frac{\delta_{ij}^{\rm K}}{\bar{n}_i} \,, \quad
    {\cal N}_{ij,\ell}^{\rm GL} = 0 \,,
\end{align}
where $\sigma_\epsilon$ is the variance of observed ellipticities.

For evaluating the Fisher matrix $F_{\alpha\beta}^{\rm ph}$ for the observed angular power spectrum, we use
\begin{equation}
    F_{\alpha\beta}^{\rm ph} = \sum_{\ell=\ell_{\rm min}}^{\ell_{\rm max}} \sum_{ij,mn}
    \frac{\partial C_{ij,\ell}^{\rm AB}}{\partial \theta_\alpha}
    \text{Cov}^{-1}\left[C_{ij,\ell}^{\rm AB},C_{mn,\ell}^{\rm CD}\right]
    \frac{\partial C_{mn,\ell}^{\rm CD}}{\partial \theta_\beta} \,.
\end{equation}

\subsection{Probe combination} 

We compute spectroscopic galaxy clustering (\GCsp), photometric galaxy clustering (\GCph), weak lensing (WL), and cross-correlation 
between two photometric probes (XC) in two configurations: a pessimistic setting and an optimistic one, as in EC20. We consider $f_{\rm sky}\simeq 0.36$ for both of \Euclid's main probes \citep{Euclid:2021icp}. For the pessimistic setting, 
we set $k_{\rm max} = 0.25\,h\,{\rm Mpc}^{-1}$ for \GCsp, $\ell_{\rm max} = 750$ for \GCph\ and XC, and $\ell_{\rm max} = 1500$ for WL. In addition, we impose a redshift cut of $z<0.9$ on \GCph\ to neglect any possible cross-correlation between the spectroscopic and photometric data when we combine the 3D \GCsp\ with the $3\times2$pt.
For the optimistic setting, we expand \GCsp\ to $k_{\rm max} = 0.30\,h\,{\rm Mpc}^{-1}$, \GCph\ and XC to $\ell_{\rm max} = 3000$, and WL to $\ell_{\rm max} = 5000$, without imposing any redshift cut on \GCph. For more details, see EC20.

For the cold dark matter isocurvature models in Sect. \ref{sec:isocurvature} that may be more susceptible to the details of their non-linear modeling, we also compute even more conservative results using a pessimistic quasilinear setting. That is $k_{\rm max} = 0.15\,h\,{\rm Mpc}^{-1}$ for \GCsp, $\ell_{\rm max} = 750$ for WL, and $\ell_{\rm max} = 500$ for \GCph\ and XC.

We will consider the results obtained with the pessimistic setting as {\em the baseline} results. Where appropriate, we will also report the results with the EC20 optimistic setting to pair this with the other papers in this series \citep{Euclid:2023tqw,Euclid:2023rjj,2023arXiv230917287B}.

\section{Expected constraints on $\Omega_K$}
\label{sec:Omegak}

In this section, we present the results of the Fisher analysis for the cosmological parameters of interest corresponding to a non-flat $\Lambda$CDM,
\begin{equation}
    \Theta_{\rm final} = \left\{\Omega_{\rm m}, \Omega_{\rm b}, h, n_{\rm s}, \Omega_K, \sigma_8 \right\}\,,
\end{equation}
after marginalisation over spectroscopic and photometric nuisance parameters. 

Spatial flatness is one of the key predictions of the simplest slow-roll models of inflation, which have a stage of accelerated expansion sufficiently long to dilute pre-existing inhomogeneities and spatial curvature to undetectable levels. However, a non-zero spatial curvature larger than the standard deviation of $\sim 2 \times 10^{-5}$ for $\Omega_K$ due to inhomogeneities \citep{Waterhouse:2008vb} would have profound implications in terms of departures from the simplest slow-roll models: a negative curvature would support open inflation \citep{Bucher:1994gb}, whereas a positive curvature would exclude false-vacuum eternal inflation and the simplest slow-roll models \citep{Kleban:2012ph}.

The {\em Planck} measurements of CMB temperature, polarisation, and lensing anisotropies have constrained 
the spatial curvature at percent level: $\Omega_K = -0.011^{+0.013}_{-0.012}$ 
\citep{Planck:2018vyg}. 
This differs from the constraint $\Omega_K = -0.044^{+0.018}_{-0.015}$ \citep{Planck:2018vyg} obtained from the combination of temperature and polarisation data only, as the {\em Planck} CMB lensing breaks the degeneracy 
between the total energy density and $\Omega_{\rm m}$ \citep{Efstathiou:1998xx} and reduces the allowed room for negative values.

The geometric information contained in the LSS breaks this CMB degeneracy between $\Omega_K$ and $\Omega_{\rm m}$ more effectively.
The completed SDSS-IV extended BAO survey alone and in combination with {\em Planck} measurements of CMB temperature and polarisation anisotropies 
gives, respectively, $\Omega_K = 0.078^{+0.086}_{-0.099}$ and $\Omega_K = -0.0001 \pm 0.0018$ \citep{eBOSS:2020yzd}. 
The DESI BAO DR2 has updated these results to $\Omega_K = 0.025 \pm 0.041$ and $\Omega_K = 0.0023 \pm 0.0011$, alone and in combination with the CMB,  respectively \citep{DESI:2025zgx}.
The DES Y3 $3\times2$pt alone and in combination with {\em Planck}, BAO, and supernovae gives, respectively, $\Omega_K > -0.16$ at the $95 \%$ CL and $\Omega_K = 0.001 \pm 0.002$ at the $68 \%$ CL \citep{DES:2022ccp}.

\begin{table*}[htbp]
\caption{Fisher-matrix marginalised uncertainties on the cosmological parameters when the spatial curvature is allowed to vary in the pessimistic and optimistic settings and using \Euclid observations \GCsp,
WL+\GCph+XC, and their combination. 
We quote absolute uncertainties for $\Omega_K$ (for the chosen fiducial value of $\Omega_K=0$) and relative uncertainties for the other parameters (compared to their corresponding non-zero fiducial values).}
\begin{tabularx}{\textwidth}{Xcccccc}
\hline\noalign{\vskip 1.5pt}\hline
\multicolumn{7} {c}{{{\boldsymbol{$\Omega_K = 0$}}}} \\
\hline\noalign{\vskip 1.5pt}\hline

& \multicolumn{1}{c}{$\Omega_{\rm m}$} & \multicolumn{1}{c}{$\Omega_{\rm b}$} & \multicolumn{1}{c}{$h$} & \multicolumn{1}{c}{$n_{\rm s}$} & \multicolumn{1}{c}{$\Omega_K$} & \multicolumn{1}{c}{$\sigma_{8}$} \\
\hline
\multicolumn{7} {c}{Pessimistic setting} \\
\GCsp \, $(k_{\rm max} = 0.25\,h\,{\rm Mpc}^{-1})$ & 1.6\% & 2.9\% & 2.5\% & 2.0\% &  0.015 & 0.99\% \\
WL+\GCph+XC & 0.82\% & 6.1\% & 5.4\%   & 2.9\%   & 0.0072 & 0.44\% \\
\GCsp+WL+\GCph+XC($z<0.9$)                                            &   0.62\%  &   1.8\%  &   1.09\%  &   0.47\%  & 0.0044  & 0.32\%  \\
\GCsp+WL+\GCph+XC($z<0.9$)+$Planck$-like                                            &   0.47\%  &   0.84\%  &   0.35\%  &   0.27\%  & 0.0010  & 0.23\%  \\
\GCsp+WL+\GCph+XC($z<0.9$)+SO-like+$Planck$ low-$\ell$                                           &   0.43\%  &   0.68\%  &   0.32\%  &   0.22\%  & 0.0009  & 0.20\%  \\
\GCsp+WL+\GCph+XC($z<0.9$)+S4-like+$Planck$ low-$\ell$                                             &   0.38\%  &   0.58\%  &   0.28\%  &   0.19\%  & 0.0007  & 0.16\%  \\
\hline
\multicolumn{7} {c}{Optimistic setting} \\

\GCsp\, $(k_{\rm max} = 0.3\,h\,{\rm Mpc}^{-1})$ & 1.6\%    & 2.4\% & 1.92\% & 1.8\% & 0.013 & 0.96\% \\
WL+\GCph+XC                                         & 0.30\% & 4.6\% & 2.0\%    & 0.37\% & 0.0039 & 0.14\% \\
\GCsp+WL+\GCph+XC                                   &   0.27\%  &   1.6\%  &   0.72\%  &   0.19\%  &  0.003 & 0.13\% \\
\GCsp+WL+\GCph+XC+$Planck$-like                                            &   0.24\%  &   0.59\%  &   0.21\%  &   0.15\%  & 0.0007  & 0.11\%  \\
\GCsp+WL+\GCph+XC+SO-like+$Planck$ low-$\ell$                                           &   0.23\%  &   0.44\%  &   0.20\%  &   0.14\%  & 0.0006  & 0.11\%  \\
\GCsp+WL+\GCph+XC+S4-like+$Planck$ low-$\ell$                                           &   0.22\%  &   0.38\%  &   0.18\%  &   0.12\%  & 0.0006  & 0.10\%  \\
\hline\noalign{\vskip 1.5pt}\hline
\end{tabularx}
\label{tab:results_OmegaK}
\end{table*}

\begin{figure*}
\centering
\includegraphics[width=0.49\textwidth]{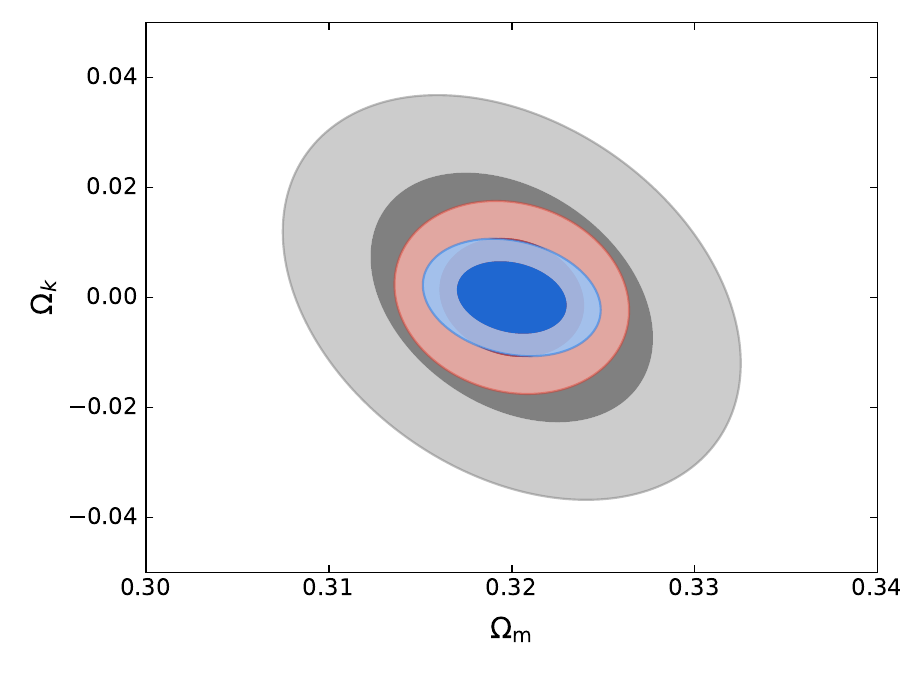}
\includegraphics[width=0.49\textwidth]{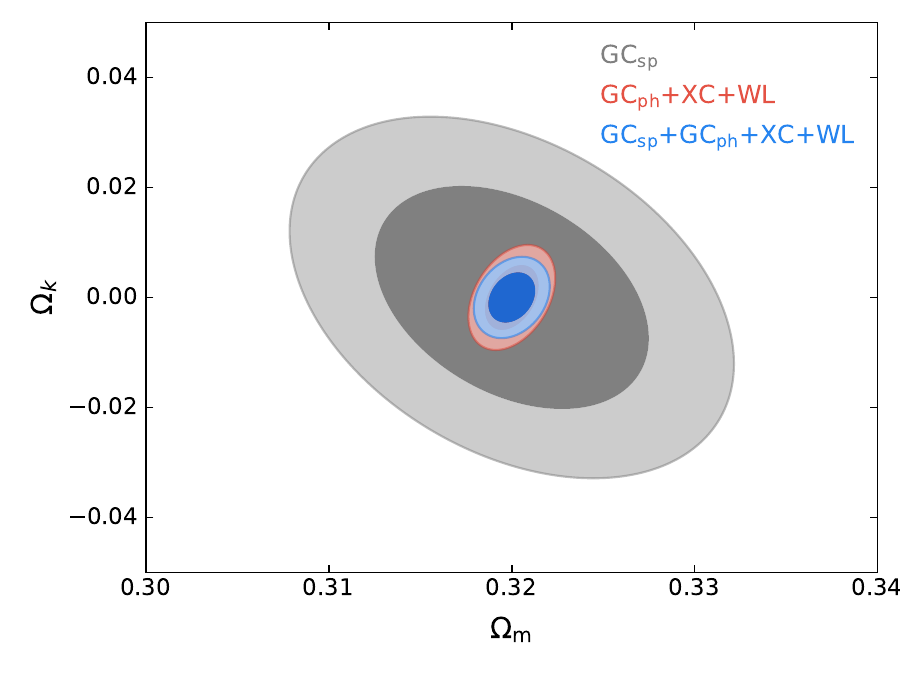}
\caption{{\em Euclid} Fisher-forecast marginalised two-dimensional contours for $\Omega_{\rm m}$ and $\Omega_K$. The left panel corresponds to the pessimistic setting for \GCsp\ (grey), WL+\GCph+XC (red), and their combination (blue). The right panel shows the same constraints, but for the optimistic setting.}
\label{fig:OmegamOmegaK_Euclid}
\end{figure*}

For a fiducial value of $\Omega_K=0$, the forecasted {\em Euclid}  uncertainties obtained with Halofit \citep{Takahashi:2012em} are displayed in detail in Table \ref{tab:results_OmegaK}.  
The forecast uncertainty from \GCsp is $ \sigma (\Omega_K) = 0.015$ ($0.013$) for the pessimistic (optimistic) setting. The photometric survey can provide tighter constraints: $\sigma (\Omega_K) = 0.0072$ ($0.0039$) for the pessimistic (optimistic) setting.
Combining \GCsp\ with the $3\times2$ photometric information, i.e. WL+\GCph+XC, we find $\sigma (\Omega_K) = 0.0044$ ($0.003$) for the pessimistic (optimistic) setting. 
In the left and right panels of Fig. \ref{fig:OmegamOmegaK_Euclid}, we compare 
the expected constraints from \GCsp, WL+\GCph+XC, and their combination for the pessimistic and optimistic settings, respectively.
{\em Euclid}  alone has, therefore, the capability to squeeze the error on $\Omega_K$ an order of magnitude below the current constraint from galaxy surveys \citep{eBOSS:2020yzd}.

\begin{figure*}
\centering
\includegraphics[width=0.49\textwidth]{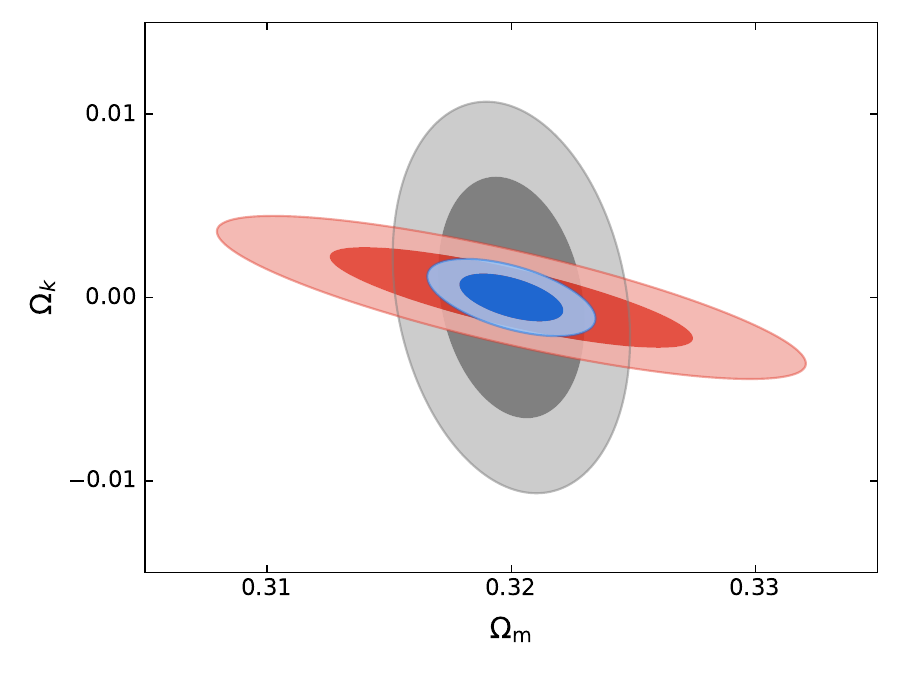}
\includegraphics[width=0.49\textwidth]{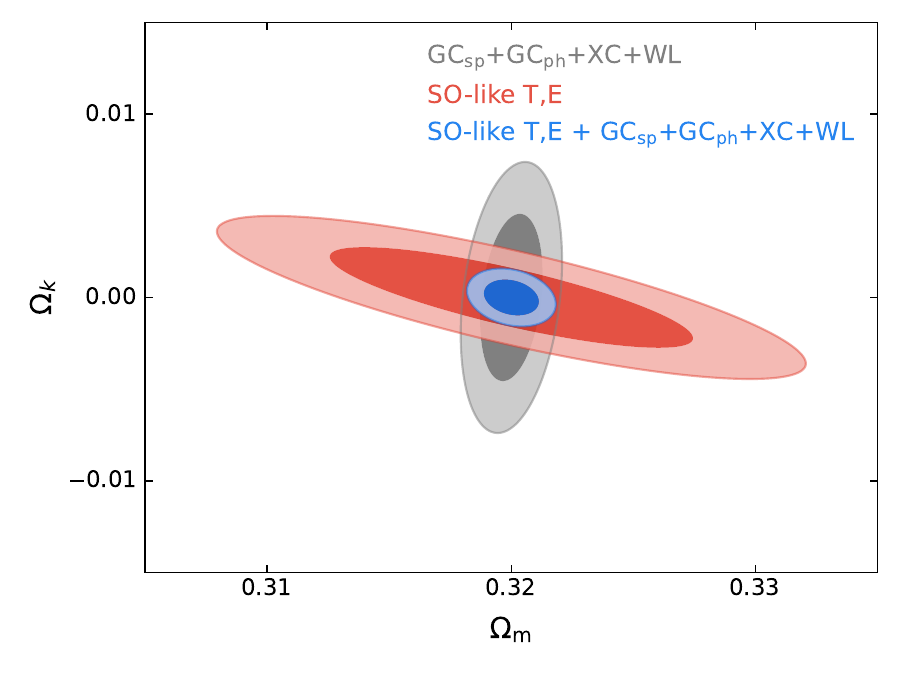}
\caption{Fisher-forecast marginalised two-dimensional contours for $\Omega_{\rm m}$ and $\Omega_K$ from the combination of {\em Euclid} and CMB observations. The left panel corresponds to the pessimistic setting for {\em Euclid}'s \GCsp+WL+\GCph+XC (grey), SO CMB temperature and polarisation (red), and their combination (blue). The right panel shows the same constraints, but for the optimistic setting.}
\label{fig:OmegamOmegaK_Euclid_CMB}
\end{figure*}

In Table \ref{tab:results_OmegaK}, we show the constraints achievable from the combination of {\em Euclid}  with CMB measurements. Following \citet{Euclid:2021qvm} -- see Appendix \ref{sec:AppendixA} for a summary of their methodology -- we consider the constraints expected from the combination of {\em Euclid} with {\em Planck}, SO, and CMB-S4.
In Fig. \ref{fig:OmegamOmegaK_Euclid_CMB}, we show how the expected constraints from {\em Euclid} can break the CMB temperature and polarisation degeneracy in the $(\Omega_\mathrm{m} \,, \Omega_K)$ plane in the case of SO. We forecast $\sigma (\Omega_K) = 0.0009$ ($0.0006$) for the uncertainty on $\Omega_K$ in the pessimistic (optimistic) case from the combination of {\em Euclid}, SO, and {\em Planck}. Table \ref{tab:results_OmegaK} shows that these constraints are improved compared to those obtained from the combination of {\em Euclid} and {\em Planck} only and can be superseded by the combination with CMB-S4 and {\em Planck}.

\section{Expected constraints on the scalar spectral index and its running}
\label{sec:powerspectrum}

In this section, we forecast \Euclid's capability to constrain the scalar power spectrum when the scalar spectral index is assumed to have a non-zero running 
$\alpha_\mathrm{s} \equiv {\rm d} n_\mathrm{s}/{\rm d}\ln k$ \citep{Kosowsky:1995aa}, i.e.
\begin{equation}
  {\cal P_R}(k) =  A_\mathrm{s} \left(\frac{k}{k_*}\right)^{n_\mathrm{s} - 1 + \frac{1}{2}\alpha_\mathrm{s} \ln (k/k_*)} \,.
  \label{running}
\end{equation}
This expression includes the next-to-leading term in the Taylor expansion of $\ln {\cal P_R}(k)$ in $\ln (k/k_*)$ with respect to the simplest power-law form in Eq. (\ref{Asns}). 
In Eq. (\ref{running}), we omit higher-order terms, such as the running of the running $\beta_\mathrm{s} = {\rm d}^2 n_\mathrm{s}/{\rm d}(\ln k)^2$, for simplicity. 
Both $n_\mathrm{s} \ne 1$ and $\alpha_\mathrm{s} \ne 0$ (as well as $\beta_\mathrm{s} \ne 0$) are generic predictions of cosmic inflation and their values help in distinguishing among different models of inflation. However, in the simplest models characterised by an inflaton slowly rolling on a nearly flat potential,
the running $\alpha_\mathrm{s}$ is next-to-leading order with respect to the deviation from scale invariance, i.e. $n_\mathrm{s}-1$, in 
the parameters that characterise the slow-roll dynamics.
\footnote{Within the slow-roll approximation, 
$n_\mathrm{s}-1 \simeq - 2\epsilon_1 - \epsilon_2$, with $\epsilon_1 = - \dot H/H^2$ 
and $\epsilon_{2} = \dot \epsilon_1/(H \epsilon_1) $ as the first two functions of the 
Hubble flow function (HFF) hierarchy. Analogously, always in the slow-roll 
approximation, we can write these deviations in terms of the 
slow-roll parameters $\epsilon = M_\mathrm{pl}^2 V_{ \phi}^2 / (2 V^2)$ and $\eta = M_\mathrm{pl}^2 V_{ \phi 
\phi} / V$ (where $M_\mathrm{pl} = (8 \pi G)^{-1/2}$ is the reduced Planck mass) related to the inflaton potential $V(\phi)$. The running of the spectral index is instead also related to the third HFF $\epsilon_3$, $\alpha_\mathrm{s} \simeq - \epsilon_2 \epsilon_3 - 2 \epsilon_1 \epsilon_3$, or 
equivalently to the third derivative of the potential through $\xi_V = 
M_\mathrm{pl}^4 V_{ \phi \phi \phi} V_{\phi} / V$, i.e. 
$\alpha_\mathrm{s} \simeq - 2 \xi^2_V + 16 \epsilon _V \eta_V- 24 \epsilon^2_V - 2 \xi^2_V$. 
See~\citet{Auclair:2022yxs,Ballardini:2024irx} for third-order slow-roll expressions.} 
As a consequence, this scale dependence is suppressed 
compared to $n_\mathrm{s}-1$ and is expected to be $\lesssim 10^{-3}$ in the simplest single-field slow-roll models. 
On more general grounds, $\alpha_\mathrm{s}$ is the next parameter to measure after the spectral index $n_\mathrm{s}$
of the power spectrum of scalar perturbations and its running, and is of utmost importance for inflation: a detection of 
$|\alpha_\mathrm{s}| \gtrsim$ a few $10^{-3}$ would indicate new physics beyond the simplest models of inflation and a mild violation of 
slow roll even in single-field models.
 
Most of the cosmological observations are consistent with the key predictions of the simplest slow-roll inflationary models: deviation of the scalar spectral index from exact scale invariance
$n_\mathrm{s} - 1 \sim -0.035$ is measured at high statistical significance, while we usually have upper bounds on the absolute value of its running. 
The latest {\em Planck} CMB temperature and polarisation data provide a precise measurement of the spectral index, i.e. $n_\mathrm{s} = 0.9635 \pm 0.0046$, and a tight constraint on its running, i.e. 
$\alpha_{\rm s} = -0.0055 \pm 0.0067$ \citep{Planck:2018vyg}. Including a compilation of BAO measurements 
improves these results only marginally: $n_\mathrm{s} = 0.9658 \pm 0.0041$ and 
$\alpha_{\rm s} = -0.0049 \pm 0.0068$. 
We should, however, note that a non-zero running of $\alpha_{\rm s} = -0.010 \pm 0.004$ may be the physical explanation alternative to unaccounted hidden systematics for the different spectral indices required to fit the Ly$\alpha$ and {\em Planck} data separately \citep{Palanque-Delabrouille:2019iyz}.

\begin{figure}
\centering
\includegraphics[width=0.49\textwidth]{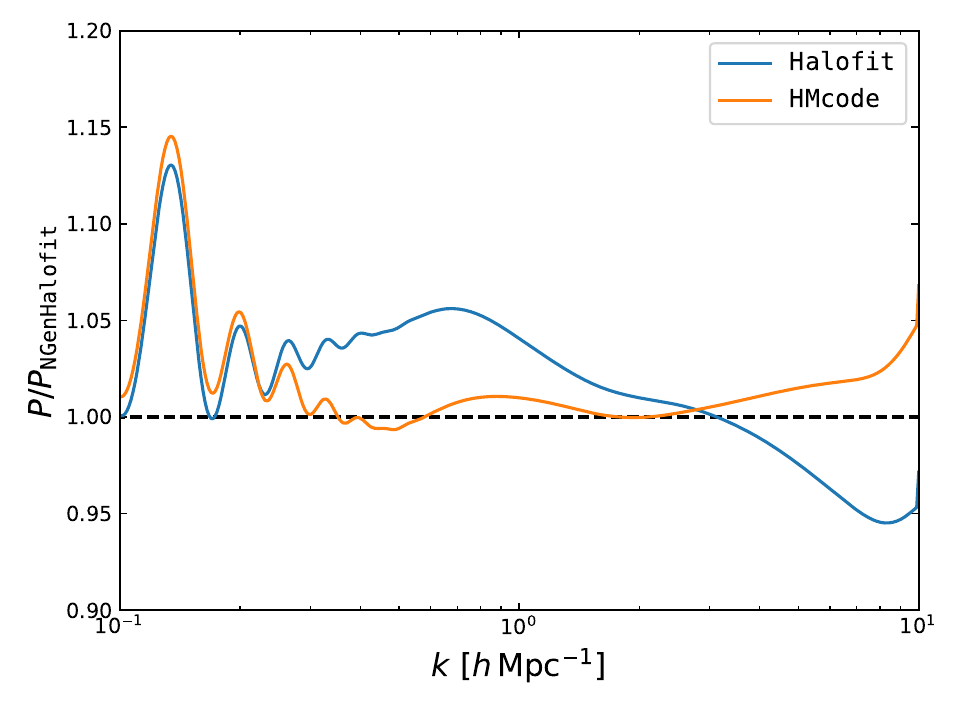}
\caption{Ratio of the non-linear matter power spectrum in Fourier space at redshift $z = 0$ with $\alpha_{\rm s} = - 0.01$ 
calculated using {\tt Halofit} (blue line) and {\tt HMCode} (orange line) to the non-linear matter power spectrum calculated using {\tt NGenHalofit}.}
\label{fig:running_nonlinear}
\end{figure}

In order to model small scales relevant for galaxy surveys, we use a fitting formula designed to capture the enhancement in the power
spectrum compared to the \lcdm\ non-linear power spectrum as a function of the running of the scalar spectral index $\alpha_\mathrm{s}$.
This fitting function has been calibrated using the {\tt NGenHalofit} \citep{Smith:2018zcj}, which extends {\tt Halofit} \citep{Smith:2002dz, Takahashi:2012em} to cosmologies that include a running spectral index,
\begin{align}
    \frac{P^{\rm NL}_\mathrm{m}(\alpha_\mathrm{s},k,z)}{P^{\rm NL}_\mathrm{m}(\alpha_s=0,k,z)} = \left(\frac{k}{k_*}\right)^{\frac{1}{2}\alpha_\mathrm{s} \ln(k/k_*)}
    \frac{1 + \left(1+ c_1\,k\right)\mathrm{e}^{c_2\,k}}{2 + c_1\,k}\,,
\label{ngen}
\end{align}
where $c_1$ and $c_2$ are quadratic polynomials in $\alpha_\mathrm{s}$ and $z$.
Fig.~\ref{fig:running_nonlinear} shows the $P^{\rm NL}_\mathrm{m}$ resulting from the prescription (\ref{ngen}) 
compared to those obtained by {\tt NGenHalofit} \citep{Smith:2002dz} or {\tt HMCode} \citep{Mead:2015yca}.
We note that the differences at large scales are due to perturbation theory corrections to the  
smoothing of the BAO implemented in {\tt NGenHalofit}. This non-linear modelling improves the theoretical treatment adopted in previous forecasts for $\alpha_\mathrm{s}$ from galaxy surveys \citep{Fedeli:2010iw, Huang:2012mr, Basse:2014qqa, Amendola:2016saw, Ballardini:2016hpi, Munoz:2016owz, Bahr-Kalus:2022prj}, in particular at $k \gtrsim 0.2 h/\mathrm{Mpc}$.
 
\begin{table*}[htbp]
\caption{Fisher-matrix marginalised uncertainties on the cosmological parameters relative to their 
corresponding fiducial values in the pessimistic and optimistic settings using \Euclid's observations \GCsp, 
WL+\GCph+XC, and their combination when the running for the scalar spectral index is included.}
\begin{tabularx}{\textwidth}{Xcccccc}
\hline\noalign{\vskip 1.5pt}\hline
\multicolumn{7} {c}{{{\boldsymbol{$\alpha_{\rm s} = -10^{-2}$}}}} \\
\hline\noalign{\vskip 1.5pt}\hline
& \multicolumn{1}{c}{$\Omega_{\rm m}$} & \multicolumn{1}{c}{$\Omega_{\rm b}$} & \multicolumn{1}{c}{$h$} & \multicolumn{1}{c}{$n_{\rm s}$} & \multicolumn{1}{c}{$\alpha_{\rm s}$} & \multicolumn{1}{c}{$\sigma_{8}$} \\
\hline
\multicolumn{7}{c}{{{Pessimistic setting}}} \\ 
\GCsp \, $(k_{\rm max} = 0.25\,h\,{\rm Mpc}^{-1})$ & 1.8\% & 3.1\% & 1.3\% & 1.6\% & 115\% & 1.0\% \\
WL+\GCph+XC & 0.91\% & 6.1\% & 4.6\%   & 2.0\%   & 57\% & 0.45\% \\
\GCsp+WL+\GCph+XC($z<0.9$)                                             &   0.81\%  &   2.3\%  &   0.62\%  &   0.78\%  &   53\% & 0.39\%  \\
\GCsp+WL+\GCph+XC($z<0.9$)+{\em Planck}-like                 &   0.35\%  &   0.32\%  &   0.24\%  &   0.23\%  &   23\% & 0.19\% \\
\GCsp+WL+\GCph+XC($z<0.9$)+SO-like+{\em Planck} low-$\ell$   &   0.32\%  &   0.095\%  &   0.075\%  &   0.21\%  &   14\% & 0.18\% \\
\GCsp+WL+\GCph+XC($z<0.9$)+CMB-S4+{\em Planck} low-$\ell$    &   0.30\%  &   0.084\%  &   0.067\%  &   0.19\%  &   13\% & 0.17\% \\
\hline
\multicolumn{7}{c}{{{Optimistic setting}}}  \\
\GCsp\, $(k_{\rm max} = 0.3\,h\,{\rm Mpc}^{-1})$ & 1.7\%    & 2.9\% & 1.1\% & 1.5\% & 91\% & 0.93\% \\
WL+\GCph+XC                                         & 0.30\% & 5.2\% & 4.1\%    & 1.6\% & 40\% & 0.16\% \\
\GCsp+WL+\GCph+XC                                   &   0.28\%  &   1.8\%  &   0.47\%  &   0.43\%  &   23\% & 0.14\% \\
\GCsp+WL+\GCph+XC+{\em Planck}-like                 &   0.20\%  &   0.28\%  &   0.18\%  &   0.17\%  &   13\% & 0.11\% \\
\GCsp+WL+\GCph+XC+SO-like+{\em Planck} low-$\ell$   &   0.17\%  &   0.084\%  &   0.065\%  &   0.16\%  &   11\% & 0.090\% \\
\GCsp+WL+\GCph+XC+CMB-S4+{\em Planck} low-$\ell$    &   0.17\%  &   0.076\%  &   0.057\%  &   0.15\%  &   10\% & 0.089\% \\
\hline\noalign{\vskip 1.5pt}\hline
\multicolumn{7} {c}{{{\boldsymbol{$\alpha_{\rm s} = -10^{-3}$}}}} \\
\hline\noalign{\vskip 1.5pt}\hline
& \multicolumn{1}{c}{$\Omega_{\rm m}$} & \multicolumn{1}{c}{$\Omega_{\rm b}$} & \multicolumn{1}{c}{$h$} & \multicolumn{1}{c}{$n_{\rm s}$} & \multicolumn{1}{c}{$\alpha_{\rm s}$} & \multicolumn{1}{c}{$\sigma_{8}$} \\
\hline
\multicolumn{7}{c}{{{Pessimistic setting}}} \\ 
\GCsp \, $(k_{\rm max} = 0.25\,h\,{\rm Mpc}^{-1})$ & 1.0\% & 1.7\% & 1.3\% & 1.2\% & 1130\% & 0.75\% \\
WL+\GCph+XC & 0.92\% & 6.0\% & 4.6\%   & 2.0\%   & 569\% & 0.43\% \\
\GCsp+WL+\GCph+XC($z<0.9$)                                   &   0.62\%  &   1.3\%  &   0.58\%  &   0.58\%  &   418\% & 0.28\%  \\
\GCsp+WL+\GCph+XC($z<0.9$)+{\em Planck}-like                 &   0.32\%  &   0.32\%  &   0.24\%  &   0.22\%  &   215\% & 0.17\% \\
\GCsp+WL+\GCph+XC($z<0.9$)+SO-like+{\em Planck} low-$\ell$   &   0.29\%  &   0.094\%  &   0.073\%  &   0.20\%  &   123\% & 0.16\% \\
\GCsp+WL+\GCph+XC($z<0.9$)+CMB-S4+{\em Planck} low-$\ell$    &   0.27\%  &   0.083\%  &   0.065\%  &   0.18\%  &   118\% & 0.16\% \\
\hline
\multicolumn{7}{c}{{{Optimistic setting}}}  \\ 
\GCsp\, $(k_{\rm max} = 0.3\,h\,{\rm Mpc}^{-1})$    & 1.0\%   & 1.6\%    &  1.1\%   & 1.1\%   & 858\% & 0.69\% \\
WL+\GCph+XC                                         & 0.27\%  & 4.7\%    &  3.2\%    & 1.2\%   & 241\% & 0.13\% \\
\GCsp+WL+\GCph+XC                                   & 0.25\%  & 1.2\%    &  0.046\%  & 0.36\%  & 154\% & 0.11\% \\
\GCsp+WL+\GCph+XC+{\em Planck}-like                 & 0.19\%  & 0.29\%   &  0.18\%  & 0.17\%  & 116\% & 0.096\% \\
\GCsp+WL+\GCph+XC+SO-like+{\em Planck} low-$\ell$   & 0.16\%  & 0.083\%  &  0.066\%  & 0.15\%  & 99\%  & 0.082\% \\
\GCsp+WL+\GCph+XC+CMB-S4+{\em Planck} low-$\ell$    & 0.15\%  & 0.075\%  &  0.058\%  & 0.14\%  & 96\%  & 0.080\% \\
\hline\noalign{\vskip 1.5pt}\hline
 \end{tabularx}
\label{tab:results_run}
\end{table*}

Differently from the $\Omega_K$ case presented in the previous section,
here we provide {\em Euclid} forecasts by choosing different non-zero fiducial values for the running of the spectral index consistent with current
constraints. We present the results of the Fisher analysis for the cosmological parameters in the presence of a running spectral index,
\begin{equation}
    \Theta_{\rm final} = \left\{\Omega_{\rm m}, \Omega_{\rm b}, h, n_{\rm s}, \alpha_{\rm s}, \sigma_8 \right\} \,
\end{equation}
after marginalisation over spectroscopic and photometric nuisance parameters.

\begin{figure*}
\centering
\includegraphics[width=0.49\textwidth]{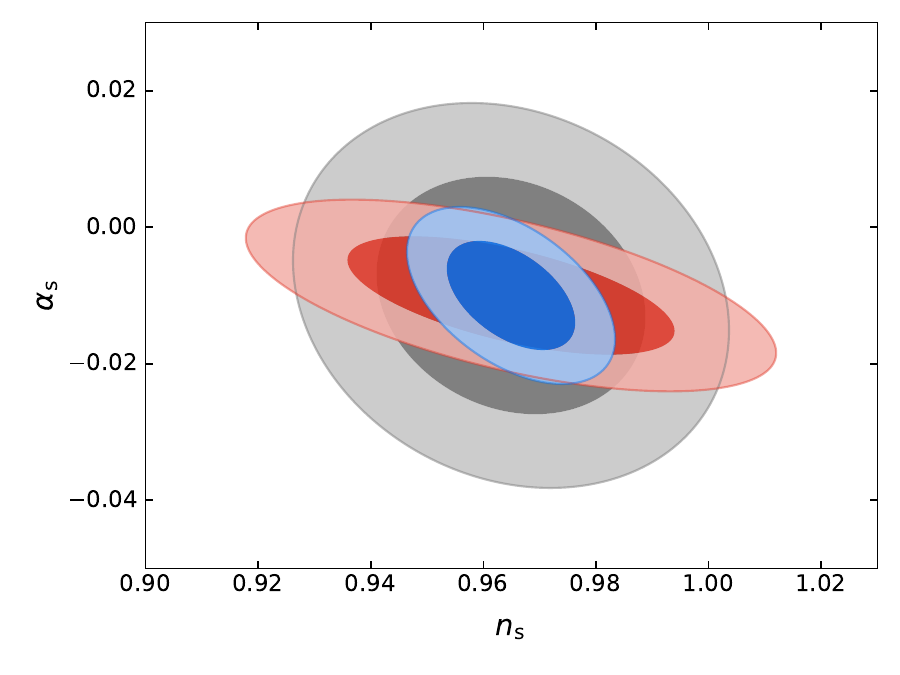}
\includegraphics[width=0.49\textwidth]{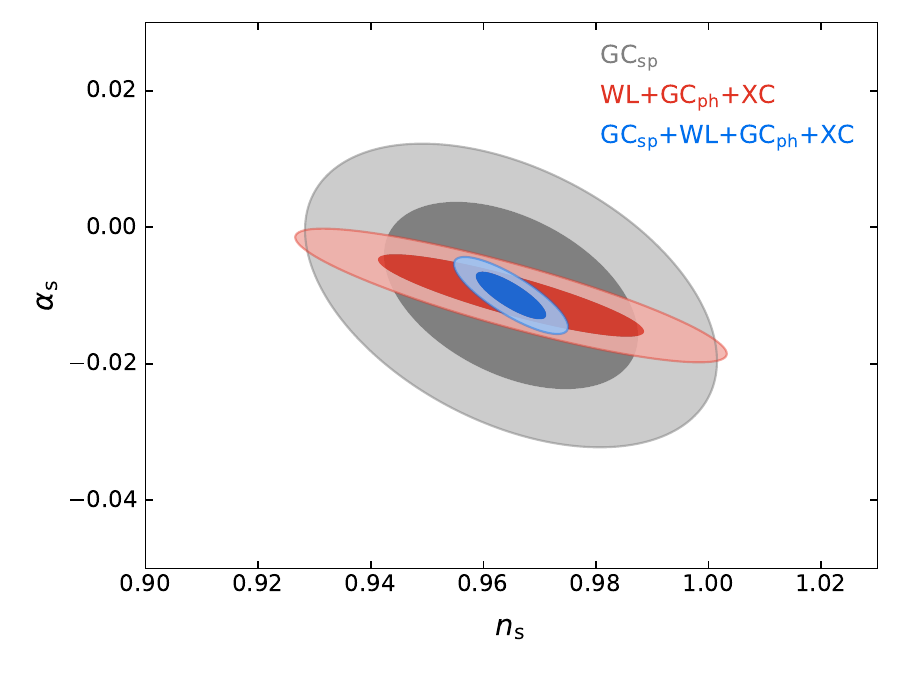}
\caption{{\em Euclid} Fisher-forecast marginalised two-dimensional contours for $n_{\rm s}$ and $\alpha_{\rm s}$. The left panel corresponds to the pessimistic setting for \GCsp\ (grey), WL+\GCph+XC (red), and their combination (blue). The right panel shows the same constraints, but for the optimistic setting.}
\label{fig:nsalphas_Euclid}
\end{figure*}

Our results for the fiducial value $\alpha_{\rm s} = - 0.01$ are presented in Table~\ref{tab:results_run}.
The relative uncertainties forecasted from \GCsp\ are 
1.6\% (1.5\%) and 115\% (91\%) for $n_\mathrm{s}$ and $\alpha_\mathrm{s}$, respectively, with the pessimistic (optimistic) setting. The photometric survey provides tighter uncertainties compared to the spectroscopic survey only for $\alpha_{\rm s}$; we obtain 2.0\% (1.6\%) and 57\% (40\%) for $n_\mathrm{s}$ and $\alpha_\mathrm{s}$, 
respectively, for the pessimistic (optimistic) setting (in qualitative agreement with the LSST forecasts presented in \cite{Ferte:2023wwf}, where it is possible to make a comparison).
Combining \GCsp\ with the $3 \times 2$ photometric information, i.e. WL+\GCph+XC, we find 0.78\% (0.43\%) and 53\% (23\%)
for the pessimistic (optimistic) setting. 
Fig.~\ref{fig:nsalphas_Euclid} shows \Euclid's expected uncertainties in the ($n_\mathrm{s} ,\, \alpha_\mathrm{s}$) two-dimensional parameter space. These results imply that the uncertainty on the running 
expected from \Euclid alone is comparable to or smaller than the current one based on {\em Planck} \citep{Akrami:2018odb}.
Therefore, \Euclid alone will be sensitive to a value of the running of the scalar spectral index that is as large as $\alpha_\mathrm{s} = -0.01$, but still consistent with the {\em Planck} constraints, at $2\sigma$ ($4\sigma$) for the pessimistic (optimistic) setting.

\begin{figure*}
\centering
\includegraphics[width=0.49\textwidth]{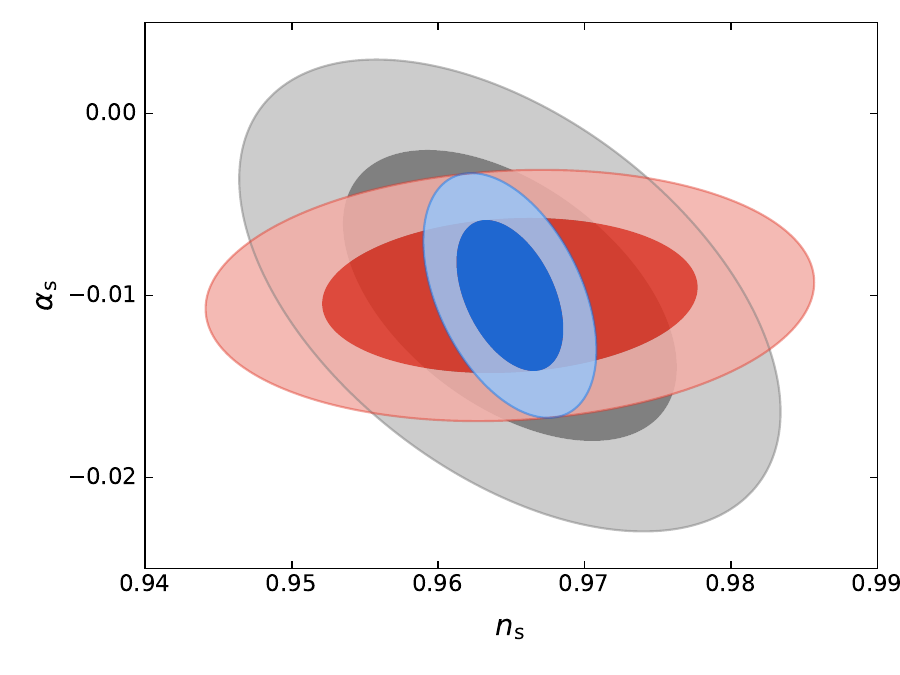}
\includegraphics[width=0.49\textwidth]{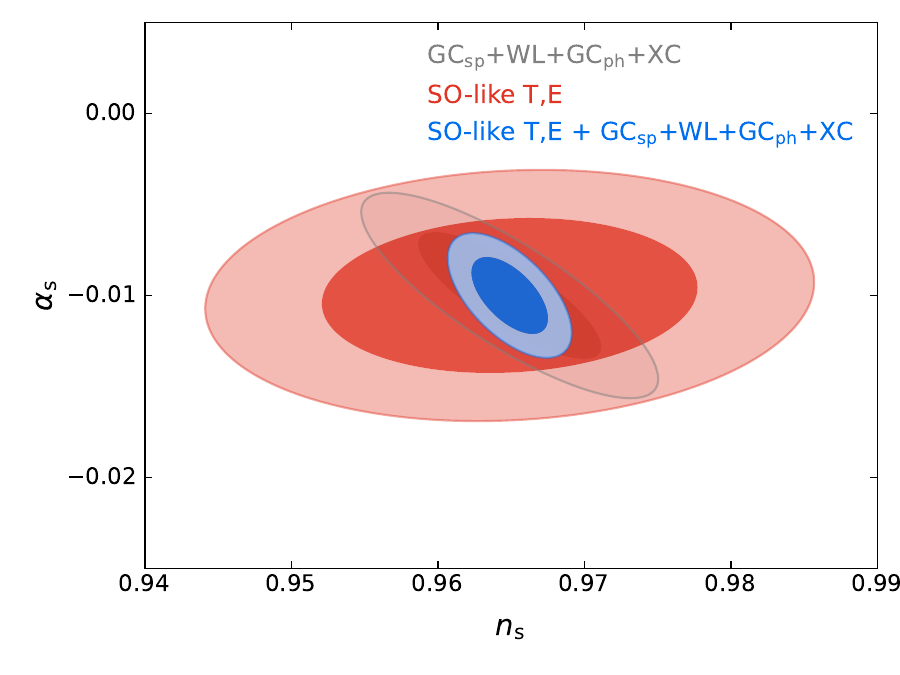}
\caption{Fisher-forecast marginalised two-dimensional contours for $n_{\rm s}$ and $\alpha_{\rm s}$ from the combination of {\em Euclid} and CMB observations. The left panel corresponds to the pessimistic setting for {\em Euclid}'s \GCsp+WL+\GCph+XC (grey), SO CMB temperature and polarisation (red), and their combination (blue). The right panel shows the same constraints, but for the optimistic setting.}
\label{fig:nsalphas_Euclid_CMB}
\end{figure*}

As in the case of spatial curvature, we also show in Table~\ref{tab:results_run} the constraints that will be obtained by {\em Euclid} in combination with current and future CMB measurements. We find that {\em Euclid} with the pessimistic (optimistic) setting can reach a relative uncertainty of 23\%, 14\%, and 13\% 
(13\%, 11\%, and 10\%)
in combination with 
{\em Planck}, SO, and CMB-S4, respectively. Fig. \ref{fig:nsalphas_Euclid_CMB} shows, specifically, our forecasted marginalised two-dimensional contours for $n_\mathrm{s}$ and $\alpha_\mathrm{s}$ from the combination of {\em Euclid} and SO measurements of the CMB temperature and polarisation.

In order to test the possible dependence of our Fisher results on the fiducial value, we repeat our analysis for $\alpha_\mathrm{s} 
= -0.001$, a fiducial value that is smaller by one order of magnitude and closer to the typical 
values predicted by slow-roll inflation on the basis of the current measurements of $n_\mathrm{s}$.\footnote{In a similar way, different 
fiducial values have also been considered in \Euclid forecasts for $f(R)$ cosmology \citep{Euclid:2023tqw} and for scale-independent modifications of gravity \citep{Euclid:2023rjj}.}
The results are presented in Table~\ref{tab:results_run} and 
show that the (absolute) uncertainty on the running is quite stable with respect to the change in the fiducial value by an order of magnitude.
Consequently, it is difficult to detect a running as small as $\alpha_\mathrm{s} = -0.001$ with \Euclid or with any of the studied combinations of \Euclid and CMB observations.

\begin{figure*}
\includegraphics{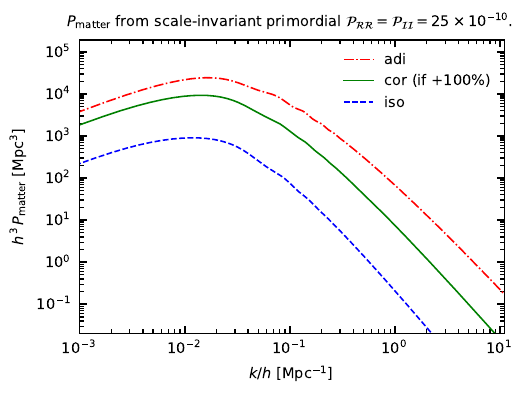}
\includegraphics{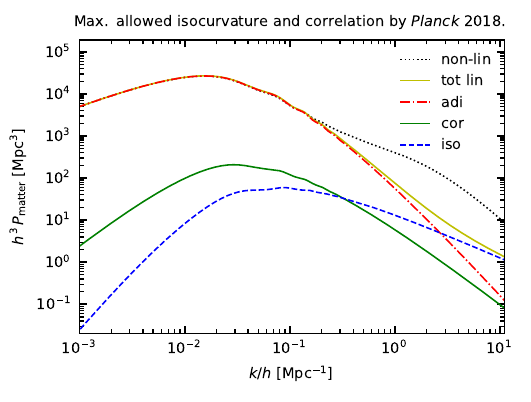}
\caption{\label{fig:isocurvatureexamples}
Matter power spectra at redshift $z=0$.
The left panel shows the linear matter power spectrum today, if the primordial perturbations are scale-invariant pure adiabatic perturbations with amplitude $\mathcal{P}_\mathcal{RR} = 25\times10^{-10}$ (red dotted-dashed), or if the primordial perturbations are scale-invariant pure cold dark matter isocurvature perturbations with amplitude $\mathcal{P}_\mathcal{II} = 25\times10^{-10}$ (blue dashed), and the resulting correlation component to the matter power spectrum if there are these two components and they are $+100\%$ correlated with each other (green solid).
The right panel shows the fiducial model used in the case of ``maximum $2\sigma$ allowed by {\em Planck}.'' The linear total matter power spectrum (tot lin) is the sum of the adiabatic (adi) and blue-tilted isocurvature (iso) contributions, and their $\sim10\%$ correlation (cor). This (tot lin) is then corrected by \texttt{Halofit} to approximate the non-linear spectrum (non-lin) of the fiducial model.}
\end{figure*}

\begin{table*}[htbp]
\caption{\label{tab:isocurvaturefisher}Fisher-matrix $68.3\%$ CL marginalised uncertainties in the pessimistic setting on the cosmological parameters, relative to their 
corresponding fiducial values, when the ``maximum $2\sigma$ allowed by {\em Planck}'', ``scale-invariant uncorrelated isocurvature contribution'', or ``blue-tilted uncorrelated isocurvature contribution'' is respectively included.
%in the pessimistic and optimistic settings, using \Euclid observations \GCsp and
%WL+\GCph+XC.
In Case 2 and Case 3, the fiducial model has a zero isocurvature contribution, and consequently, instead of the percent-level error with respect to the fiducial value, we report the 68.3\% ($1\sigma$) upper limit for $10^{10}\mathcal{P}_{\mathcal{II}1}$. Case 2 and Case 3 have only one isocurvature parameter, $10^{10}\mathcal{P}_{\mathcal{II}1}$. Thus, we indicate in square brackets the ``derived'' value of $10^{10}\mathcal{P}_{\mathcal{II}2} = 10^{10}\mathcal{P}_{\mathcal{II}1}$ in Case 2 and $10^{10}\mathcal{P}_{\mathcal{II}2} =  1690.6\times10^{10}\mathcal{P}_{\mathcal{II}1}$ in Case 3. In the last column, we report the corresponding isocurvature fraction at $k_2=0.1\,$Mpc$^{-1}$ calculated from the other values by $\beta_{\mathrm{iso}2} = \mathcal{P}_{\mathcal{II}2} / (\mathcal{P}_{\mathcal{II}2} + \mathcal{P}_{\mathcal{RR}2,\,\mathrm{fiducial}})$. Here square brackets indicate that this only corresponds to the other values, not a 68.3\% CL marginalised uncertainty for  $\beta_{\mathrm{iso}2}$, which would have a highly non-Gaussian posterior in these cases where $\mathcal{P}_{\mathcal{II}2}$ is of the same order of magnitude as $\mathcal{P}_{\mathcal{RR}2}$.
%\jvc{\bf 26.3.2024 left only "Pessimistic", others can be restored by inputting "FullIsocurvatureTable1" file.}
}%, and their combination.}
\begin{tabularx}{\textwidth}{Xrrrrrrrr}
\hline\noalign{\vskip 1.5pt}\hline
\multicolumn{9} {c}{{\bf Case 1: }{Maximum $2\sigma$ allowed by {\em Planck} isocurvature and correlation contribution at $k_2$}} \\
\hline\noalign{\vskip 1.5pt}\hline
$\phantom{\Big|}$ & \multicolumn{1}{r}{$\Omega_{\rm m}$} & \multicolumn{1}{r}{$\Omega_{\rm b}$} & \multicolumn{1}{r}{$h$} & \multicolumn{1}{r}{$10^{10}\mathcal{P}_{\mathcal{RR}1}$} & \multicolumn{1}{r}{$10^{10}\mathcal{P}_{\mathcal{RR}2}$} & \multicolumn{1}{r}{$10^{10}\mathcal{P}_{\mathcal{II}1}$} &
\multicolumn{1}{r}{$10^{10}\mathcal{P}_{\mathcal{II}2}$} &
\multicolumn{1}{r}{$10^{10}\mathcal{P}_{\mathcal{RI}1}$} \\
\hline
Fiducial value $\phantom{\Big|}$ & 0.315 & 0.049 & 0.674 & 28.990 & 19.449 & 0.017 & 30.916 & 0.076\\
\hline
%\rowcolor{lavender(web)}\multicolumn{9}{l}{{{Pessimistic setting}}}  \\ 
\GCsp\, $(k_{\rm max} = 0.25\,h\,{\rm Mpc}^{-1})$ &
1.0\% & 2.4\% &   1.0\% &   2.1\% &    1.5\% &    39.0\% &   32.5\% &   35.8\% 
\\
WL+\GCph+XC 
& 0.9\% & 7.7\% & 6.2\%   & 3.3\%   & 12.5\% & 338.9\% & 180.5\% & 162.1\% \\
\GCsp+WL+\GCph+XC &
0.3\% &  2.0\% &   0.9\% &   1.1\% &    1.3\% &    31.4\% &   16.0\% &   26.4\%
\\
%%%%%%%%%%%%%%%%%%%%%%%%%%%%%%%%%%%%
\hline\noalign{\vskip 1.5pt}\hline
\multicolumn{9} {c}{{\bf Case 2: }{Scale-invariant ($n_\mathrm{iso}=1.0$) uncorrelated isocurvature contribution; fiducial $\mathcal{P}_{\mathcal{II}1} = 0$}} \\
\hline\noalign{\vskip 1.5pt}\hline
$\phantom{\Big|}$ & \multicolumn{1}{c}{$\Omega_{\rm m}$} & \multicolumn{1}{c}{$\Omega_{\rm b}$} & \multicolumn{1}{c}{$h$} & {$10^{10}\mathcal{P}_{\mathcal{RR}1}$} & \multicolumn{1}{c}{$10^{10}\mathcal{P}_{\mathcal{RR}2}$} & \multicolumn{1}{c}{$10^{10}\mathcal{P}_{\mathcal{II}1}$} &  \multicolumn{1}{c}{[$10^{10}\mathcal{P}_{\mathcal{II}2}$]} &   \multicolumn{1}{c}{[$\beta_\mathrm{iso2}$]}
\\
\hline
Fiducial value $\phantom{\Big|}$ & 0.315 & 0.049 & 0.674 & 23.551 & 20.529 & 0 & [0] & [0]\\
\hline
%\rowcolor{lavender(web)}\multicolumn{9}{l}{{{Pessimistic setting}}} \\ 
\GCsp \, $(k_{\rm max} = 0.25\,h\,{\rm Mpc}^{-1})$ &
1.4\% &	2.2\% &	1.2\% &	5.7\% &	1.7\% &	19.0 & [19.0] & [0.48]
\\
WL+\GCph+XC &
0.9\% & 6.0\% &	4.5\% &	6.7\% &	9.3\% &	24.7 & [24.7] & [0.55]
\\
\GCsp+WL+\GCph+XC &
0.6\% &	1.5\% &	0.8\% &	2.5\% &	1.4\% &	9.4 & [9.4] & [0.31]
\\
\hline\noalign{\vskip 1.5pt}\hline
\multicolumn{9}{c}{{\bf Case 3: }{Blue-tilted ($n_\mathrm{iso}=2.9$) uncorrelated isocurvature contribution; fiducial $\mathcal{P}_{\mathcal{II}1} = 0$}} \\
\hline\noalign{\vskip 1.5pt}\hline
%& \multicolumn{1}{c}{$\Omega_{\rm m,0}$} & \multicolumn{1}{c}{$\Omega_{\rm b,0}$} & \multicolumn{1}{c}{$h$} & $10^{10}\mathcal{P}_{\mathcal{RR}1}$ & \multicolumn{1}{c}{$10^{10}\mathcal{P}_{\mathcal{RR}2}$} & \multicolumn{1}{c}{$10^{10}\mathcal{P}_{\mathcal{II}1}$} &  \multicolumn{1}{c}{[$10^{10}\mathcal{P}_{\mathcal{II}2}$]} &   \multicolumn{1}{c}{[$\beta_\mathrm{iso2}$]}
%\\
%\hline
%Fiducial values $\phantom{\Big|}$ & 0.315 & 0.049 & 0.674 & 23.551 & 20.529 & 0 & [0] & [0]\\
%\hline
%\rowcolor{lavender(web)}\multicolumn{9}{l}{{{Pessimistic setting}}} \\ 
\GCsp \, $(k_{\rm max} = 0.25\,h\,{\rm Mpc}^{-1})$ &
1.5\% &	2.2\% &	1.2\% &	2.7\% &	2.2\% &	0.0200 &	[33.7] & [0.62]
\\
WL+\GCph+XC & 0.8\% & 6.3\% & 4.8\% & 1.7\% & 7.9\% & 0.0011 & [1.9] &  [0.08] \\
\GCsp+WL+\GCph+XC & 
0.6\% &	1.5\%	& 0.8\% &	1.0\% &	1.1\% &	0.0005 & 	[0.9] & [0.04]
\\
\hline\noalign{\vskip 1.5pt}\hline
\multicolumn{9} {c}{{Adiabatic $\Lambda$CDM}} \\
\hline\noalign{\vskip 1.5pt}\hline
%$\phantom{\Big|}$ & \multicolumn{1}{r}{$\Omega_{\rm m,0}$} & \multicolumn{1}{r}{$\Omega_{\rm b,0}$} & \multicolumn{1}{r}{$h$} & \multicolumn{1}{r}{$10^{10}\mathcal{P}_{\mathcal{RR}1}$} & \multicolumn{1}{r}{$10^{10}\mathcal{P}_{\mathcal{RR}2}$} &  \\
%\hline
%Fiducial values $\phantom{\Big|}$ & 0.315 & 0.049 & 0.674 & 23.551 & 20.529  \\
%\hline
%\rowcolor{lavender(web)}\multicolumn{9}{l}{{{Pessimistic setting}}} \\ 
\GCsp \, $(k_{\rm max} = 0.25\,h\,{\rm Mpc}^{-1})$ &
1.4\% &  2.2\% & 1.2\% & 2.5\% &   1.3\%
\\
WL+\GCph+XC & 
0.8\% &   4.9\%  &  2.7\% &   1.5\% &   3.7\%
\\
\GCsp+WL+\GCph+XC & 
0.6\% &   1.4\%  &  0.8\% &   1.0\% &   1.0\%
\\
\hline\noalign{\vskip 1.5pt}\hline
\end{tabularx}
\end{table*}

\section{Expected constraints on isocurvature contribution}
\label{sec:isocurvature}

In the minimal scenario, primordial perturbations are assumed to be purely
adiabatic, i.e. perturbations in the spatial curvature \citep[see, e.g.,][]{1983PhRvD..28..679B}. In this case, the total energy density of the Universe—and thus its
spatial curvature—can vary from point to point on super-Hubble scales, while
the relative abundances of the different species (e.g., the photon-to-cold dark
matter ratio) remain spatially uniform. The simplest single-field inflationary
models naturally generate only this type of perturbation.

However, in more general multi-field inflationary models, there are more degrees of freedom, and orthogonal modes, i.e. isocurvature perturbations, can also be excited \citep{1985PhLB..158..375L,1986MNRAS.218..103E,1996PThPh..95...71S,1996PhRvD..53.5437G,1999ApJ...510..523P}. These isocurvature perturbations generated during inflation can survive after inflation as fluctuations in the composition of the Universe, where the ratios
between species vary spatially, even though the total energy density remains spatially uniform on super-Hubble scales.

This means that the detection of isocurvature modes would be a smoking gun for models beyond the simplest single-field inflation. Moreover, if the trajectory or the kinetic manifold in field space is curved then the curvature and isocurvature perturbations can be correlated \citep{1999PhRvD..59l3512L,2000PhRvD..62d3504L,2000PhRvD..63b3506G,DiMarco:2002eb, 2002PhRvL..88u1302A,2003PhRvL..91m1302V,Byrnes:2006fr}.

\subsection{Motivation and background}

Here we focus on one isocurvature mode out of four theoretically possible ones \citep{2000PhRvD..62h3508B}, the cold dark matter density isocurvature (CDI) perturbation\footnote{The CDI mode is observationally similar to the baryon isocurvature mode. Indeed, the CMB and matter power spectra depend on the total matter isocurvature
$\mathcal{I}_{\mathrm{m}\gamma} = \frac{\Omega_\mathrm{b}}{\Omega_\mathrm{m}} \mathcal{I}_{\mathrm{b}\gamma} +  \frac{\Omega_\mathrm{c}}{\Omega_\mathrm{m}} \mathcal{I}_{\mathrm{c}\gamma}$, where
$\mathcal{I}_{\mathrm{b}\gamma} = \delta_\mathrm{b} - \frac{3}{4}\delta_\gamma$. Thus, our results on the CDI can easily be converted to the constraints on the total matter isocurvature or baryon isocurvature.} 
for which
\begin{equation}
\mathcal{I}_{\mathrm{c}\gamma} = \delta_\mathrm{c} - \frac{3}{4}\delta_\gamma
\end{equation}
is non-zero, i.e. varies spatially (unlike in the pure adiabatic case). Here, $\delta_\gamma = \delta\rho_\gamma / \bar{\rho}_\gamma$ is the density perturbation of photons and $\delta_\mathrm{c}$ is the density perturbation of cold dark matter. From now on, we ignore the subscripts in $\mathcal{I}_{\mathrm{c}\gamma}$.

CMB observations indicate that on scales corresponding to $0.001\,$Mpc$^{-1} \lesssim k \lesssim 0.05\,$Mpc$^{-1}$ the primordial perturbations are predominantly adiabatic and mildly red-tilted, with 
$n_\mathrm{s} \approx 0.964$ \citep{Akrami:2018odb},
which coincides with one of the predictions of single-field slow-roll inflation. However, there is no similar restriction for a CDI isocurvature contribution 
on which we have only upper bounds \citep{Akrami:2018odb}. The isocurvature spectrum, e.g. from multi-field inflation, can be blue-tilted (more perturbation power at small scales than at large scales), i.e. $n_\mathrm{iso} > 1$ \citep{1994PhRvD..50.6123P,1997PhRvD..56..535L} and, on the other hand, the CMB observations are not very sensitive to even a dominant isocurvature contribution at $k \gtrsim 0.05\,$Mpc$^{-1}$. Here the LSS observations can significantly improve the constraints or lead to a detection.

A strong dark matter candidate, the axion \citep{PhysRevLett.40.223,PhysRevLett.40.279,1983PhLB..126..178A}, can acquire even very blue-tilted ($n_\mathrm{iso} \sim 2$--$4$) isocurvature perturbations \citep{2009PhRvD..80b3516K,2016JCAP...08..056K,PhysRevD.98.023525,2023arXiv230917010C}, though at some small scale the spectrum should flatten. For reviews on the axion, see e.g. \citet{2013ARNPS..63...69K,2016PhR...643....1M}. Superheavy dark matter or WIMPzilla is another example of a model that could produce a blue-tilted isocurvature spectrum and for which the detection of the isocurvature mode could be a smoking gun
\citep{2005PhRvD..72b3511C,2017PhRvD..96j3540K}.

In these cases with a blue-tilted isocurvature spectrum, the adiabatic and isocurvature perturbations would be uncorrelated. 
Note that this possibility is very different from the standard curvaton case \citep{PhysRevD.42.313,2002NuPhB.626..395E,2001PhLB..522..215M,2002PhLB..524....5L,2002PhRvD..65l1301B,2003PhRvD..67b3503L,2003PhRvD..67l3513G,2008JCAP...12..004L}. Indeed, a generic feature of almost all curvaton models is that the adiabatic and isocurvature spectral indices are equal, since both perturbations arise from the same source in these models. As the existing data constrain the adiabatic spectral index near the value of $0.964$, the curvaton models do not offer a large isocurvature signal to the CMB or matter power spectra at high $k$, contrary to the blue-tilted axion or WIMPzilla models.

\subsection{CDI models we consider and their observational aspects}

We assume the standard, spatially flat, $\Lambda$CDM cosmology and that the primordial power spectra of the adiabatic perturbation, $\mathcal{R}$, and the CDI perturbation, $\mathcal{I}$,  obey independent power-law forms
\begin{equation}
\mathcal{P}_{\mathcal{RR}}(k) =  \mathcal{P}_{\mathcal{RR}}(k_*) \left(\frac{k}{k_*}\right)^{n_\mathrm{adi}-1},
\qquad
\mathcal{P}_{\mathcal{II}}(k)  =  \mathcal{P}_{\mathcal{II}}(k_*) \left(\frac{k}{k_*}\right)^{n_\mathrm{iso}-1},
\end{equation}
with their correlation power spectrum
\begin{equation}
\mathcal{P}_{\mathcal{RI}}(k) =  \mathcal{P}_{\mathcal{RI}}(k_*) \left(\frac{k}{k_*}\right)^{n_\mathrm{cor}-1},
\end{equation}
where $n_\mathrm{cor}=(n_\mathrm{adi}+n_\mathrm{iso})/2$.
Here we have five parameters: three amplitudes $\mathcal{P}_{\mathcal{RR}}(k_*)$, $\mathcal{P}_{\mathcal{II}}(k_*)$, and  $\mathcal{P}_{\mathcal{RI}}(k_*)$ at the pivot scale $k_*=0.05\,$Mpc$^{-1}$, and two independent constant spectral indices $n_\mathrm{adi}$ and $n_\mathrm{iso}$. 
Each spectrum is a straight line (with a slope $n-1$) in the $[\ln(k/k_*),\,\ln \mathcal{P}]$ space.
However, in this parametrisation the adiabatic model would be recovered with any $n_\mathrm{iso}$ as long as the amplitudes $\mathcal{P}_{\mathcal{II}}(k_*)$ and $\mathcal{P}_{\mathcal{RI}}(k_*)$ are zero. In the absence of detection of a non-zero isocurvature amplitude this parametrisation becomes problematic because the ``adiabatic parameter space volume'' would be arbitrarily large, i.e. the adiabatic model would not be a single point in the space of the extended parameters.
Thus, we parametrise the perturbations mathematically equivalently by five amplitudes given at two scales\footnote{Our choices for the fiducial models assume $h=0.674$, so $k_1 \approx 0.003\,h\,$Mpc$^{-1}$ and $k_2 \approx 0.15\,h\,$Mpc$^{-1}$.} $k_1 = 0.002\,$Mpc$^{-1}$ and $k_2 = 0.1\,$Mpc$^{-1}$:
\begin{equation}
\mathcal{P}_{\mathcal{RR}1},\quad \mathcal{P}_{\mathcal{RR}2},\quad
\mathcal{P}_{\mathcal{II}1},\quad
\mathcal{P}_{\mathcal{II}2},\quad {\rm and}\quad  \mathcal{P}_{\mathcal{RI}1}.
\end{equation}
The first four of these are non-negative, whereas the correlation can be positive, negative, or zero. (In this parametrisation, the adiabatic model is simply recovered by setting all the extra isocurvature parameters  $\mathcal{P}_{\mathcal{II}1}$, $\mathcal{P}_{\mathcal{II}2}$, and $\mathcal{P}_{\mathcal{RI}1}$ to zero.)
In addition, we can calculate derived parameters, such as the primordial isocurvature fraction $\beta_\mathrm{iso}(k) = \mathcal{P}_{\mathcal{II}}(k) / \left[ \mathcal{P}_{\mathcal{RR}}(k) + \mathcal{P}_{\mathcal{II}}(k)\right]$, which can depend on $k$ and is in the range $(0,1)$, and the primordial correlation fraction $\cos\Delta = \mathcal{P}_{\mathcal{RI}} / \left( \mathcal{P}_{\mathcal{RR}} \mathcal{P}_{\mathcal{II}} \right)^{1/2}$, which is constant in $k$ and in the range $(-1, 1)$.
For details, see
\citet{2016A&A...594A..20P} and \citet{Akrami:2018odb}.

The CDI mode is constrained by the {\em Planck} CMB angular power spectra $C_\ell^{TT,TE,EE}$ on large scales (corresponding to $k\sim k_1$) and medium scales (corresponding to $k\sim k_*$) to be less than a few percent of the dominant adiabatic mode. However, if the adiabatic and CDI modes had the same spectral index of primordial perturbations, i.e. $n_\mathrm{iso} = n_\mathrm{adi}$, the observational effects of the CDI mode would be damped compared to the adiabatic one at smaller scales, i.e., at larger multipoles or wavenumbers by $C_{\ell,\mathrm{CDI}} / C_{\ell,\mathrm{ADI}} \propto (\ell / \ell_\mathrm{eq})^{-2} \sim (k / k_\mathrm{eq})^{-2}$, where in the latter we refer to the scale that re-entered the Hubble radius at the radiation--matter equality. Hence, the constraints at the smallest scales ($k\sim k_2$) probed by {\em Planck} remain weaker, allowing for a spectral index as large as $n_\mathrm{iso} \sim 2.9$. At $k=k_2=0.1\, $Mpc$^{-1}$ the {\em Planck} CMB temperature and polarisation data give a 95\% CL upper bound on the ratio of the primordial isocurvature power to the curvature perturbation power $\mathcal{P}_\mathcal{II} / \mathcal{P}_\mathcal{RR} \lesssim 1.4$, i.e., $\beta_\mathrm{iso}(k_2) < 0.58$, which tightens to $\beta_\mathrm{iso}(k_2) < 0.47$ when the {\em Planck} CMB lensing data are included. This is the region of parameter space for the CDI mode in which we expect {\em Euclid} to significantly improve the current CMB constraints.

However, one should keep in mind that the isocurvature contribution is also damped (compared to the primordial ratio $\mathcal{P}_\mathcal{II} / \mathcal{P}_\mathcal{RR}$) in the 
LSS observables. The left panel of Fig.~\ref{fig:isocurvatureexamples} shows the linear matter power spectra $P_\mathrm{m}$ (at redshift $z=0$) resulting from scale-invariant adiabatic and isocurvature perturbations of the same primordial amplitude $\mathcal{P}_\mathcal{RR}(k) = 25\times10^{-10}$ and $\mathcal{P}_\mathcal{II}(k) = 25\times10^{-10}$. In addition, a (possible) 100\% correlation ($\cos\Delta = 1$) is indicated. In this case, the observable CDI to adiabatic ratio in the matter power spectrum would be $P_\mathrm{m}^\mathrm{\,iso} / P_\mathrm{m}^\mathrm{\,adi} \approx 1/13$ at $k/h=0.002\ $Mpc$^{-1}$ and 1/54 at   $k/h=0.1\ $Mpc$^{-1}$, and only 1/244 at $k/h=1\ $Mpc$^{-1}$. The observable effects (among others the modification of the baryon acoustic peak structure) are maximised in the case of a 100\% correlation, which would, in this example, give approximately 1/2, 1/4, and 1/8 contribution, respectively, to the total at the above-mentioned scales. It should be underlined that apart from modifying the overall shape of the matter power spectrum, as is immediately visible in the right panel of Fig.~\ref{fig:isocurvatureexamples}, the CDI and correlation contributions slightly shift the positions (and alter the heights) of BAO peaks in the matter power with respect to the pure adiabatic case \citep[see, e.g., Fig. 15 of][]{2012ApJ...753..151V}. All this is automatically taken into account by our Fisher forecast codes. If the actual Universe had some isocurvature but the pure adiabatic model were fitted to the data, then the interpreted values of the standard cosmological and dark energy parameters might be biased \citep{2012JCAP...07..021M}, but as already \citet{2010JCAP...10..009M} forecasted (and our results confirm it) the error bars on the standard parameters would not be significantly affected by the extra degrees of freedom. The use of the combination of CMB and LSS observations will reduce the risk of biased results for the standard parameters and would simultaneously optimally constrain the isocurvature parameters.

If the spectral index of primordial isocurvature perturbations is $n_\mathrm{iso} \approx n_\mathrm{adi} + 0.5$, then the isocurvature to adiabatic ratio in the observable matter power spectrum stays nearly constant over all wavenumbers. (This can easily be checked by reproducing, e.g. by \texttt{CAMB}, the left panel of Fig.~\ref{fig:isocurvatureexamples}, but with $n_\mathrm{iso} = 1.5$ and plotting the isocurvature to adiabatic ratio.) Therefore, if the data were adiabatic and had the same constraining power at every $k$ in the considered range, then the minimum ``disturbance'' would be achieved with this $n_\mathrm{iso}$. The result is that if the overall isocurvature contribution were to be minimised over the whole data, the posterior probability density function would apparently favour $n_\mathrm{iso} \sim 1.46$. (In the case of the CMB the similar relation is $n_\mathrm{iso} \sim n_\mathrm{adi} + 2$.) On the other hand, this means that {\em Euclid}  has very little constraining power on those CDI models whose spectral index is $n_\mathrm{iso} \lesssim 1.46$. Only for $n_\mathrm{iso} > 1.5$ does {\em Euclid}  have the potential to considerably improve the constraints over the existing CMB constraints.

We present {\em Euclid}  forecasts for three different isocurvature cases:

{\bf Case 1:} The observable effects are the largest for the CDI spectra that are primordially blue-tilted and/or positively correlated with the primordial adiabatic perturbations. Hence, we first derive {\em Euclid}  forecasts for a positively correlated blue-tilted CDI model that lies in terms of the isocurvature contribution close to the maximum 95\% CL allowed region, as well as has the maximum allowed correlation by {\em Planck}. This model has $\beta_\mathrm{iso}(k_1) = 0.00059$,  $\beta_\mathrm{iso}(k_2) = 0.6$, and $\cos\Delta = 0.108$ giving $n_\mathrm{iso} = 2.9$, $n_\mathrm{cor} = 1.9$, and $\beta_{\mathrm{iso}}(k_*) = 0.28$ or $\mathcal{P}_{\mathcal{II}}(k_*)/\mathcal{P}_{\mathcal{RR}}(k_*) = 0.39$.
The values of cosmological parameters\footnote{The values have been taken directly from the {\em Planck} 2018 \texttt{MultiNest} chains, except that all the power spectra amplitudes have been scaled up by a constant in order to obtain $\sigma_8 \approx 0.8111$. (When allowing for CDI, the CMB analysis typically gives $\sigma_8 \sim 0.75$, since the required amplitude for the CMB angular power spectra can be reached by a slightly smaller amplitude of $\mathcal{P}_\mathcal{RR}$ than in the pure adiabatic case due to the CDI power providing the rest at large scales. Then, at small scales, i.e. at high $k$, the matter perturbations originating from the isocurvature component are so much damped that the result is a relatively small $\sigma_8$.) We do this scaling to be able to check the consistency against the other {\em Euclid}  forecasts.}
used to produce this ``maximum 2$\sigma$ allowed by {\em Planck}'' fiducial model are presented in Table~\ref{tab:isocurvaturefisher}. The contributions to the matter power spectrum in this model are presented in the right panel of Fig.~\ref{fig:isocurvatureexamples}. The total linear matter power, which is the sum of the CDI, correlation, and adiabatic components, differs visibly from the pure adiabatic one from $k/h \approx 0.1\,$Mpc$^{-1}$ onward, where $P_\mathrm{m}^\mathrm{\,tot\ lin} / P_\mathrm{m}^\mathrm{\,adi} \approx 1.033$. At $k/h = 0.3\,$Mpc$^{-1}$, the ratio is 1.089. The effects of this magnitude should be detectable by {\em Euclid}. We also show the total linear spectrum run through {\tt Halofit} (setting \texttt{halofit\_version = 4} in \texttt{CAMB}) to estimate the resulting total non-linear matter power spectrum. 
This is used in the WL+\GCph+XC analysis and has been shown to be inaccurate in the case of extreme isocurvature spectral indices, $n_\mathrm{iso}\ge3$, studied in detail by \citet{2023PhRvD.108j3542C}. However, our fiducial $n_\mathrm{iso}$ is more modest,
and when doing the Fisher forecasts we are dealing with models in the vicinity of the fiducial model, and hence even if the non-linearities were modelled slightly inaccurately, these should not considerably affect the results. We also report results with an extra conservative quasilinear setting, leaving the full non-linear modelling of admixtures of nearly scale-invariant adiabatic contribution and a highly blue-tilted isocurvature contribution for future N-body simulations. It is important to note that on sub-Hubble scales there is no isocurvature left, which means that the simulations themselves could be the standard ones as long as adiabatic initial conditions with a linear spectrum similar to the yellow curve ("tot lin") in the right panel of Fig.~\ref{fig:isocurvatureexamples} were specified at a redshift 50--100 instead of the ones similar to the standard red curve (``adi'').

{\bf Case 2:} In our second isocurvature case we go to the opposite extreme. Our fiducial model is now adiabatic $\Lambda$CDM and we forecast a model that has only one isocurvature parameter, a scale-invariant amplitude $\mathcal{P}_\mathcal{II}$, i.e. we have $n_\mathrm{iso} = 1$ and no correlation between the CDI and the adiabatic components. As explained above, due to the damping of the primordial CDI component, {\em Euclid}  is expected to be relatively insensitive to this type of CDI contribution. In addition, here we can safely use the standard adiabatic non-linear modelling and have high confidence in the results even in the optimistic case that picks contributions from high $k$ (because the isocurvature contribution decreases rapidly towards high $k$).

\begin{figure*}
\includegraphics{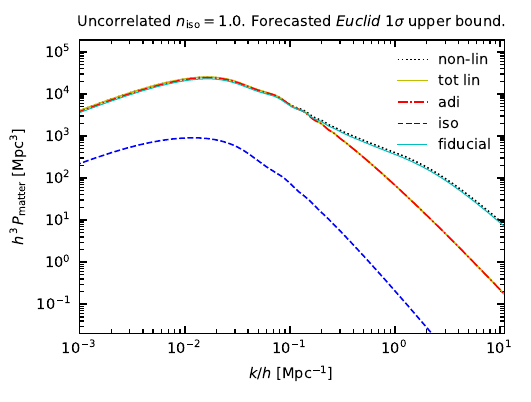}
\includegraphics{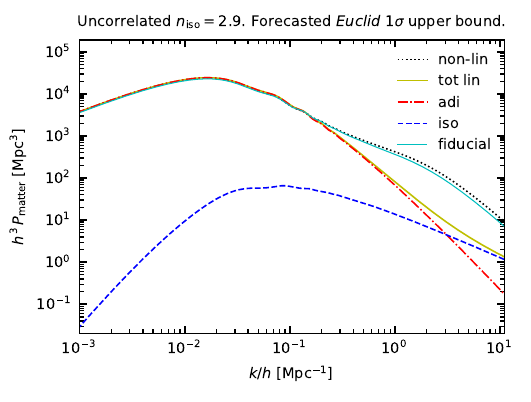}
\caption{\label{fig:1paramisocurvaturecases} Matter power spectra at redshift $z=0$ for Case 2 (left panel) and Case 3 (right panel). The left panel shows the different contributions for our forecasted pessimistic \GCsp{} $1\sigma$ upper bound on our Case 2 uncorrelated isocurvature model with $n_\mathrm{iso}=1.0$ when the underlying true model (fiducial, cyan colour) is purely adiabatic. The upper bound on the isocurvature contribution is so small that the linear pure adiabatic (adi) and total linear (tot lin) matter power spectra are indistinguishable by eye in this figure. The right panel shows the same quantities for the forecasted \GCsp{} $1\sigma$ pessimistic upper bound on our Case 3 one-isocurvature-parameter uncorrelated isocurvature model with $n_\mathrm{iso}=2.9$ when the underlying true model (fiducial) is purely adiabatic. Note how, for Case 3, the isocurvature contribution visibly modifies the spectrum for $k/h \gtrsim 0.1\,\mathrm{Mpc}^{-1}$.}
\end{figure*}

\begin{figure}
\includegraphics{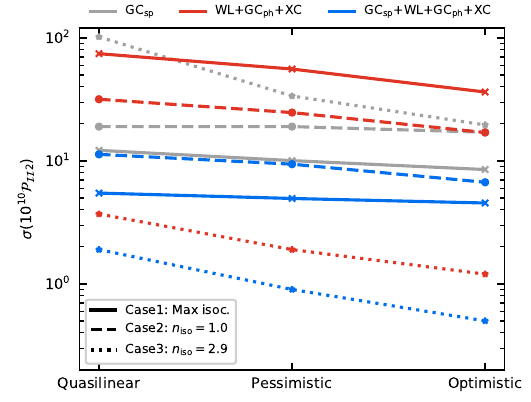}
\caption{Fisher-matrix marginalised uncertainty for $10^{10}$ times the primordial isocurvature perturbation amplitude $\mathcal{P}_\mathcal{II}$ at $k_2 = 0.1\,$Mpc$^{-1}$ for Case 1 ``maximum $2\sigma$ allowed by {\em Planck}'' (solid lines), Case 2 ``scale-invariant uncorrelated $n_\mathrm{iso}=1.0$'' (dashed lines), and Case 3 ``blue-tilted uncorrelated $n_\mathrm{iso}=2.9$'' (dotted lines).
Our baseline results of Table~\ref{tab:isocurvaturefisher} are for the pessimistic setting. To test the effect of the choice of $k_{\rm max}$ and $\ell_{\rm max}$ on the results we repeat the analysis with an even more conservative quasilinear setting ($k_{\rm max} = 0.15\,h\,{\rm Mpc}^{-1}$ for \GCsp, $\ell_{\rm max} = 750$ for WL, and $\ell_{\rm max} = 500$ for \GCph\ and XC) and with the optimistic setting ($k_{\rm max} = 0.30\,h\,{\rm Mpc}^{-1}$ for \GCsp, $\ell_{\rm max} = 5000$ for WL, and $\ell_{\rm max} = 3000$ for \GCph\ and XC).}
\label{fig:isoc_quasi_pess_optimistic}
\end{figure}

{\bf Case 3:} In our third isocurvature case we keep the adiabatic $\Lambda$CDM fiducial model and forecast the expected constraint for an isocurvature fraction with a fixed spectral index $n_\mathrm{iso} = 2.9$. Unlike in Case 1, we consider zero correlation. This case most directly answers the question of how much {\em Euclid}  can improve over the {\em Planck} isocurvature constraints, if the true underlying Universe has adiabatic primordial perturbations. This case is most sensitive to the non-linear scales and, therefore, to non-linear modelling, as the fiducial model is the standard adiabatic model, but then we add to this a blue-tilted component whose contribution grows rapidly towards small scales. For \GCsp{} we impose three different $k$ cuts and see that, as expected, the constraints improve as a function of $k$ cut. For WL+\GCph+XC the situation is more complicated as some contribution from very high $k$ enters to the constraints no matter how aggressive an $\ell$ cut one imposes because $P_\mathrm{m}(k)$ is not rapidly decreasing as in the pure adiabatic case or in the $n_\mathrm{iso}=1$ case (see Fig.~\ref{fig:1paramisocurvaturecases}). Thus, we deem, in particular, forecasts with the optimistic setting for WL+\GCph+XC with $n_\mathrm{iso}=2.9$ in Fig.~\ref{fig:isoc_quasi_pess_optimistic} less reliable than the other results.

\begin{figure}
\includegraphics[width=0.45\textwidth]{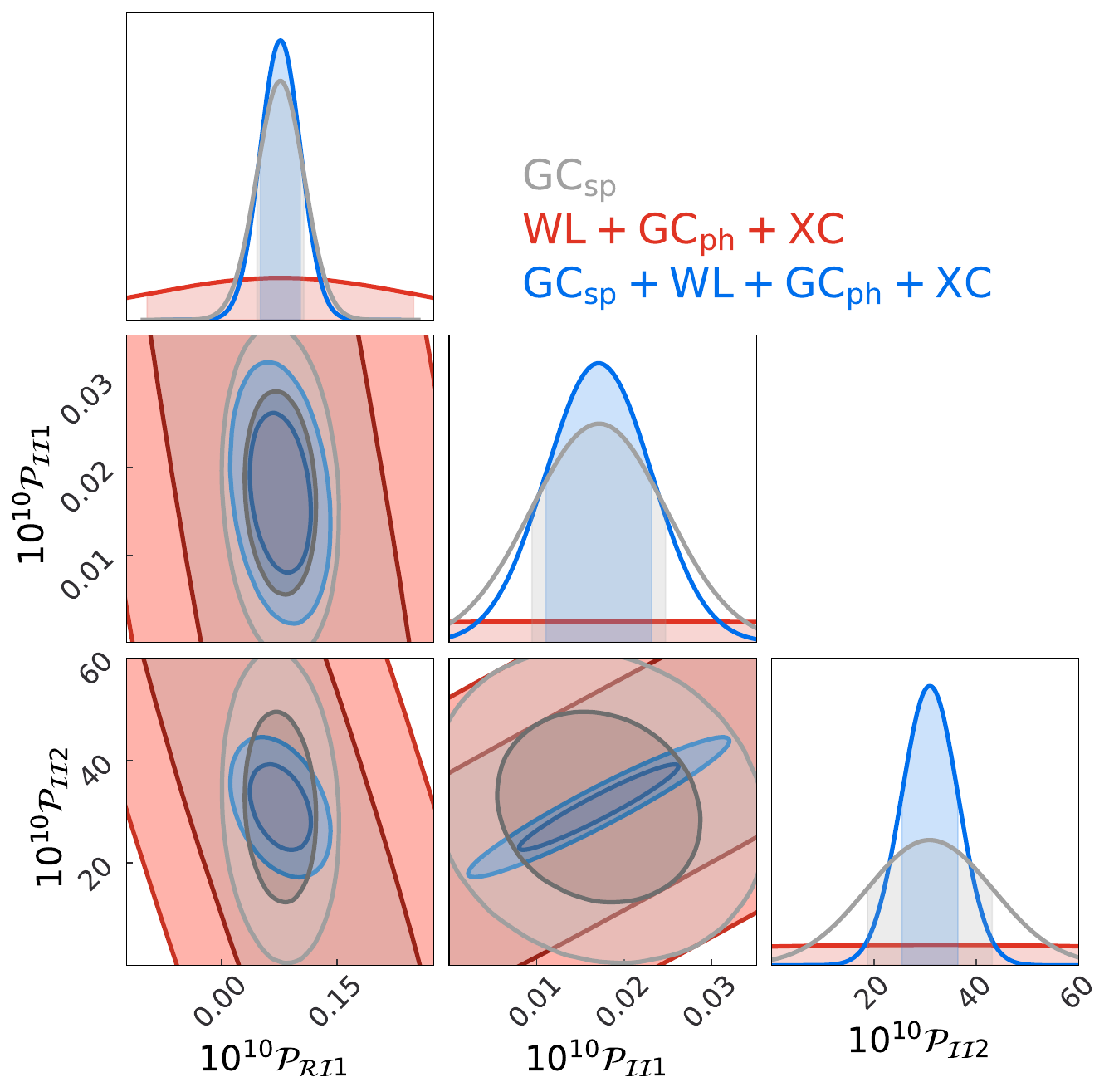}
\caption{\label{fig:IsocCase1Contours}{\em Euclid} marginalised  posterior probability density functions for Case 1 ``maximum 2$\sigma$ allowed by {\em Planck}'' isocurvature model using the pessimistic {\em Euclid}  setting.}
\end{figure}

\subsection{{\em Euclid} forecasts for the CDI mode}

We report the fiducial parameter values and our forecasts with the pessimistic setting for the above three isocurvature cases in Table~\ref{tab:isocurvaturefisher}. For comparison, we show the standard adiabatic results which indicate that allowing for an isocurvature contribution does not significantly affect the $1\sigma$ uncertainty of the background density parameters $\Omega_\mathrm{m}$, $\Omega_\mathrm{b}$, nor the current expansion rate $h$ with \GCsp. However, when using WL+\GCph+XC, the parameters $\Omega_\mathrm{b}$ and, in particular, $h$ are somewhat degenerate with the isocurvature parameters, and hence their uncertainty increases with respect to the adiabatic case. The same is true for the amplitude of adiabatic perturbations ($\mathcal{P}_{\mathcal{RR}1}$ and $\mathcal{P}_{\mathcal{RR}2}$). A smaller primordial adiabatic amplitude can fit the data since part of the perturbation power comes from the isocurvature perturbations.

The results of Table~\ref{tab:isocurvaturefisher} for $\mathcal{P}_{\mathcal{II}2}$ with the pessimistic setting are displayed in Fig.~\ref{fig:isoc_quasi_pess_optimistic}. For comparison, this figure shows the results for more conservative quasilinear settings and for the optimistic setting. Apart from Case 3 ($n_\mathrm{niso} = 2.9$), the dependence on $k_\mathrm{max}$ or $\ell_\mathrm{max}$ is moderate. This fact was expected for Case 2 since the isocurvature contribution with $n_\mathrm{iso} = 1.0$ rapidly decreases towards high $k$ and hence adding information at higher $k$ only mildly improves the constraints. 
For Case 1, the fiducial model has a large blue-tilted isocurvature contribution and including higher $k$ slightly helps to improve the constraints on the isocurvature contribution.
Case 3 ($n_\mathrm{iso} = 2.9$) is very different: as the data do not have any isocurvature but the isocurvature contribution in the theoretical model rapidly increases toward high $k$, we get only an upper bound for the isocurvature amplitude. Adding data up to higher $k$ improves this upper bound significantly. Moreover, in Case 1 and Case 2, \GCsp{} gives tighter constraints compared to WL+\GCph+XC, but for Case 3 \GCsp{} leads to more than one order of magnitude weaker constraints compared to WL+\GCph+XC, which picks up the isocurvature contribution from high $k$.

For Case 1, our main result is that if the actual Universe had a CDI contribution that is of the order of the maximum 2$\sigma$ allowed by {\em Planck} CMB data, then {\em Euclid}  \GCsp{} would be able to detect a non-zero $\mathcal{P}_{\mathcal{II}2}$ at the 3$\sigma$ level (1$\sigma$ is 32.5\% of the fiducial value according to Table~\ref{tab:isocurvaturefisher}). Combining all the {\em Euclid}  data studied in this paper, \GCsp+WL+\GCph+XC, and using the pessimistic setting, would lead to a 6.3$\sigma$ detection of a non-zero isocurvature amplitude at $k_2=0.1\,$Mpc$^{-1}$ (1$\sigma$ is 16.0\% of the fiducial value according to Table~\ref{tab:isocurvaturefisher}). This 6.3$\sigma$ detection is only mildly weakened (improved) to $5.6\sigma$ (6.8$\sigma$) if quasilinear (optimistic) settings are adopted. This great detection potential of {\em Euclid}  is illustrated in Fig.~\ref{fig:IsocCase1Contours}, which also shows that for this model the constraints mainly come from \GCsp{} while the role of WL+\GCph+XC is to confine the isocurvature spectral index close to the fiducial value without giving tight constraints on the overall amplitude, even (0,0) being inside the 68\% CL region in the $(\mathcal{P}_{\mathcal{II}1},\, \mathcal{P}_{\mathcal{II}2})$ plane with WL+\GCph+XC.

Case 2 confirms our expectations that a primordial isocurvature amplitude of the same order as the adiabatic amplitude around $k_2$ is required in order for {\em Euclid}  to be sensitive to it. (Recall that the equal primordial amplitude means about 2\% CDI contribution to the observable matter power spectrum at $k_2$.) The left panel of Fig.~\ref{fig:1paramisocurvaturecases} shows various contributions to the matter power spectrum in a model with the forecasted maximum (1$\sigma$, \GCsp) scale-invariant ($n_\mathrm{iso}=1.0$) CDI contribution. Naturally, in this case, {\em Euclid}  does not improve the constraints over {\em Planck}. 

In Case 3, our $2\sigma$ upper bound $\mathcal{P}_{\mathcal{II}}(k_*) / \mathcal{P}_{\mathcal{RR}}(k_*) < 0.86$ for the strongly blue-tilted uncorrelated CDI ($n_\mathrm{iso} = 2.9$, pessimistic \GCsp, $k_\mathrm{max} = 0.25\,h\,{\rm Mpc}^{-1}$) fully matches the recent findings of \citet{2023PhRvD.108j3542C} who used galaxy correlation and bispectrum with a more rigorous non-linear model, but did not specify $k_\mathrm{max}$. The forecasted {\em Euclid}  \GCsp{} upper bound at $k_2=0.1$\,Mpc$^{-1}$ is 20\% tighter than the {\em Planck} 2018 95\% CL upper bound for the uncorrelated CDI with non-scale-invariant spectrum, ``axion II'' of \citet{Akrami:2018odb}.

We conclude that, in the case of a fiducial model with adiabatic initial conditions, {\em Euclid}  \GCsp{} alone will constrain the blue-tilted CDI at medium and high $k$ at the same level as {\em Planck} CMB observations. Adding WL+\GCph+XC can improve the containing power of {\em Euclid}  by an order of magnitude (see the second last column in Table~\ref{tab:isocurvaturefisher} and the dotted lines in Fig.~\ref{fig:isoc_quasi_pess_optimistic}), should the modelling of non-linearities used here provide a suitable approximation of the photometric and weak lensing observables in the presence of blue-tilted CDI.
This high gain from WL+\GCph+XC merits a detailed future N-body study of blue-tilted CDI models.

\section{Expected constraints on primordial non-Gaussianity}
\label{sec:PNG}

One of the most common predictions of inflation is the presence of a small amount of non-Gaussianity in the distribution of the seed perturbations, generated during the early stages of the Universe. In the simplest models, these primordial fluctuations are nearly perfectly Gaussian, while more complex scenarios predict a deviation from Gaussian distribution \citep[see, e.g.,][for a review]{Achucarro_2022}. Measuring the amount of this primordial non-Gaussianity (PNG), parametrised in terms of a dimensionless amplitude parameter $\fnl$, has emerged as one of the most powerful probes of the early Universe \citep{Bartolo2004,Meerburg:2019qqi}.

Currently, the tightest constraints on $\fnl$ come from the CMB anisotropies \citep{Planck:2019izv}, where the constraints on the so-called local-type PNG are $\fnll=-0.9\pm5.1$ at the $68\%$ CL. Most of the information on PNG within the CMB has been extracted. Next-generation LSS surveys, due to their inherently 3-dimensional structure and large cosmic volumes, are catching up fast \citep{Leistedt:2014zqa,castorina_2019,Mueller:2021jvn,DAmico:2022gki,Cabass_2022,Rezaie_2023,Cagliari:2023mkq} and could soon supersede the CMB measurements \citep[e.g.][]{Karagiannis:2018jdt,Karagiannis:2019jjx}.

In this section, we present forecasted uncertainties on the amplitude parameter of three PNG shapes by utilising the spectroscopic galaxy clustering \Euclid sample (\GCsp). Our main results follow from the Fisher-matrix methodology described in EC20.
Additionally, we summarise, for comparison, the findings on the amplitude of local PNG presented in \citet{andrews_euclid_2024}, which are based on an advanced field-level inference technique. The constraints on $\fnl$ presented in this section follow the CMB convention as in \citet{Planck:2019izv}. 

\subsection{Fisher analysis} \label{sec:fisher_PNG}
\label{sect:fisher_forecasts}
\subsubsection{Matter bispectrum}

In most inflationary scenarios, the generated amount of PNG leads to a non-zero bispectrum\footnote{The presence of PNG leads to all high-order correlation functions of the primordial perturbations being non-zero. However, in most inflationary models, the three-point function and its Fourier transformation, the bispectrum, are the largest.} of the primordial curvature fluctuation field. The resulting bispectrum is characterised by an amplitude parameter $\fnl$, which defines the magnitude of the PNG signal, and by the shape, which describes the functional dependence on the various Fourier space triangles. Both of them depend strongly on the inflationary model at hand. Thus, constraining the amplitude of the primordial bispectrum can rule out different classes of inflationary models and shed light on the early phases of the Universe.

Violating different conditions of the standard inflation generates a non-zero primordial bispectrum, whose signal peaks at distinct triangle configurations. In this section, we consider the most studied PNG shapes:
 \begin{itemize}
 \item {\em local} \citep{Salopek1990,Gangui1993,Verde1999,Komatsu2001}, peaking at squeezed triangles ($k_3\ll k_2\simeq k_1$); 
 \item {\em equilateral} \citep{Creminelli2005}, peaking at equilateral configurations ($k_3\simeq k_2\simeq k_1$); 
 \item {\em orthogonal} \citep{Senatore2009}, peaking at both equilateral and folded triangles ($k_1\simeq k_2 \simeq k_3/2$). Here we consider both approximations of the exact shape derived in \citet{Senatore2009}: \emph{orthogonal-CMB}, more used in CMB analysis because of its 2-dimensional projection accuracy, and \emph{orthogonal-LSS}, which is important for LSS studies.
 \end{itemize}
 The corresponding bispectra of the Bardeen gauge-invariant primordial gravitational potential $\Phi$ are defined respectively as
  \begin{align}
B_{\Phi}^{\text{loc}}(k_1,k_2,k_3)~=&~2 \fnll\Big[P_{\Phi}(k_1)P_{\Phi}(k_2)+\text{2 perms} \Big] \, , \label{eq:fnl_local}\\
B_{\Phi}^{\text{equil}}(k_1,k_2,k_3)~=&~6 \fnle\bigg(-\Big[P_{\Phi}(k_1)P_{\Phi}(k_2)+\text{2 perms} \Big] \nonumber \\
&-2\Big[P_{\Phi}(k_1)P_{\Phi}(k_2)P_{\Phi}(k_3)\Big]^{2/3} \nonumber \\
&+\Big[P_{\Phi}^{1/3}(k_1)P_{\Phi}^{2/3}(k_2)P_{\Phi}(k_3)+\text{5 perms}\Big] \bigg) \,, \\ 
 B_\Phi^\text{ortho-CMB}(k_1,k_2,k_3)~=&~6\fnloC\bigg( -3 \Big[P_{\Phi}(k_1)P_{\Phi}(k_2)+\text{2 perms} \Big] \nonumber \\
 &+3\Big[P_\Phi^{1/3}(k_1)P_\Phi^{2/3}(k_2)P_\Phi(k_3) +5\text{ perms}\Big] \nonumber \\
 &-8 \Big[P_\Phi(k_1)P_\Phi(k_2)P_\Phi(k_3)\Big]^{2/3}
\bigg) \,, \\
 B_\Phi^\text{ortho-LSS}(k_1,k_2,k_3)~=&~6\fnloL\Big[P_\Phi(k_1)P_\Phi(k_2)P_\Phi(k_3)\Big]^{2/3}\nonumber \\
 &\bigg( (1+p)\frac{\Delta}{k_1k_2k_3}-p\frac{\Gamma^3}{k_1^2k_2^2k_3^2}
\bigg) \,,
\end{align}
where 
\begin{align}
p &= \frac{27}{-21 + \dfrac{743}{7 \left(20\pi^2 - 193\right)}}\,,\,\Delta=(k_T-2k_1)(k_T-2k_2)(k_T-2k_3) \nonumber \\
\Gamma&=\frac{2}{3}(k_1k_2+k_2k_3+k_3k_1)-\frac{1}{3}(k_1^2+k_2^2+k_3^2),
\end{align}
and $k_T=k_1+k_2+k_3$. The power spectrum $P_\Phi(k)$ is directly related to the power spectrum of the primordial curvature perturbations generated during inflation. The primordial potential perturbations $\Phi$ are associated with the linear dark matter field via the Poisson equation $\delta_{\rm m}^{\rm L}(\bk,z)=M(k,z)\Phi(\bk)$, where $M(k,z)= 2c^2D(z)\,T(k)\, k^2/(3\Omega_{\rm m} H_0^2)$, with $D(z)$ the linear growth factor and $T(k)$ the matter transfer function normalised to unity at large scales $k \rightarrow 0$. Thus, $P_\Phi(k)=P_{\rm m}(k,z)/M^2(k,z)$. The leading order PNG contribution to the matter density bispectrum is
 \begin{equation} \label{eq:bisng}
B_I(k_1,k_2,k_3;z)=M(k_1,z)M(k_2,z)M(k_3,z)B_\Phi(k_1,k_2,k_3) \,.
 \end{equation}
Due to the non-linear nature of gravity the bispectrum has additional terms at leading order besides the primordial contribution shown above -- for LSS it is an intrinsically non-zero quantity, independently of the initial conditions.  

\subsubsection{Galaxy power spectrum and bispectrum}\label{sec:galaxy_PS_BS}

In a biased tracer of the dark matter field, like the \GCsp{} \Euclid sample considered here, there are two main sources of PNG signal: the primordial part of the galaxy bispectrum (Eq.~\ref{eq:bisng}) and the scale-dependent bias signature \citep{Dalal2008,Slosar2008,Matarrese2008,Verde2009,Afshordi2008,Desjacques2010} induced at linear order in the galaxy power spectrum and bispectrum. Therefore, we use both summary statistics in order to extract the PNG information within the \GCsp{} sample. In redshift space, the tree-level model for the observed galaxy power spectrum and bispectrum, in the presence of PNG, is given by
     \begin{align}
     &P_{\rm obs}(\bk^{\rm ref};z)=\frac{1}{q_\perp^2 (z) \, q_\parallel (z)}\bigg[D^{\rm P}(\bk;z)Z_1(\bk)^2P_{\rm m}^{\rm\, L}(k;z)+P_\mathrm{s}(z)\bigg] \label{eq:Pgs}\,,\\ 
   &B_{\rm obs}(\bk_1^{\rm ref},\bk_2^{\rm ref},\bk_3^{\rm ref};z)= \frac{1}{q_\perp^4 (z) \, q_\parallel^2 (z)}\Bigg[ \nonumber \\
   & D^{\rm B}(\bk_1,\bk_2,\bk_3;z)\;\bigg[\;Z_1(\bk_1)Z_1(\bk_2)Z_1(\bk_3)B_{I}(k_1,k_2,k_3;z) \nonumber \\ 
   &+\Big(2Z_1(\bk_1)Z_1(\bk_2)Z_2(\bk_1,\bk_2)P_{\rm m}^{\rm\, L}(k_1;z)P_{\rm m}^{\rm\, L}(k_2;z)+2~ \text{perm}\Big)\bigg] \nonumber \\
   &+2P_{\veps\veps_{\delta}}(z)\Big(Z_1(\bk_1)P_{\rm m}^{\rm\, L}(k_1;z)+2~ \text{perm}\Big)+B_{\veps}(z)\Bigg]\,, \label{eq:Bgs} 
  \end{align}
where the redshift dependence of the kernels has been depreciated for simplicity. The redshift kernels up to second order for a general type of PNG are
  \begin{align}
   &Z_1(\bk_i,z)=b_1(z)+f_\mathrm{g}(z)\mu_i^2+\fnl b_{\Psi}(z)\frac{k_i^{\alpha}}{M(k_i,z)}\,, \label{eq:Z1}\\
   &Z_2(\bk_i,\bk_j,z)=b_1(z)F_2(\bk_i,\bk_j)+f_\mathrm{g}(z)\mu_{ij}^2G_2(\bk_i,\bk_j)+\frac{b_2(z)}{2} \nonumber \\ &+\frac{b_{s^2}(z)}{2}S_2(\bk_i,\bk_j) +\frac{f_\mathrm{g}(z)\mu_{ij}k_{ij}}{2}\left[\frac{\mu_i}{k_i}Z_1(\bk_j,z)+\frac{\mu_j}{k_j}Z_1(\bk_i,z)\right]
  \nonumber \\ &
   +\fnl b_{\Psi}(z)\frac{1}{2}\,\left[\frac{k_i^{\alpha}}{k_j^2\,M(k_i)}+\frac{k_j^{\alpha}}{k_i^2\,M(k_j)}\right]\,\bk_{i}\cdot\bk_{j} \nonumber \\
   &+\fnl b_{\Psi\delta}(z)\,\frac{1}{2}\,\left[\frac{k_i^{\alpha}}{M(k_i)}+\frac{k_j^{\alpha}}{M(k_j)}\right]\,, \label{eq:Z2}
  \end{align}
where $\mu_i=\hat\bk_i\cdot\hat{\bm{z}}$ with $\hat{\bm z}$ the line-of-sight vector, $\mu_{ij}=(\mu_ik_i+\mu_jk_j)/k_{ij}$, and $k_{ij}^2=(\bk_i+\bk_j)^2$.  The kernels $F_2(\bk_i,\bk_j)$ and $G_2(\bk_i,\bk_j)$ are the second-order symmetric standard perturbation theory  kernels \citep{Bernardeau:2001qr}, while $S_2(\bk_1,\bk_2) = (\bk_1\cdot\bk_2)^2-1/3$ is the  tidal kernel \citep{McDonald:2009dh}. In addition to the triangle shape, defined by the three sides $k_1$, $k_2$, $k_3$, the redshift space bispectrum has a dependence on the orientation of the triangle with respect to the line-of-sight direction. Two angles characterise this orientation, which are chosen to be the polar angle $\theta$ of $\bk_1$ [$\cos(\theta)=\mu_1$] and the azimuthal angle $\phi$ around $\bk_1$, which is involved in the definition of $\mu_2$ \citep{Scoccimarro:1999ed}.  

The factors $D^{\rm P}(\bk,z)$ and $D^{\rm B}(\bk_1,\bk_2,\bk_3,z)$ incorporate the redshift errors and FoG effect (see Sect. \ref{sec:spectro_clustering}) for the power spectrum and bispectrum, respectively, while they are given by $D^{\rm P}(\bk,z)=\exp\{-k\,\mu\,[f_\mathrm{g}^2(z)\,\sigma_{\rm p}^2(z)+\sigma_{\rm r}^2(z)]\}$ and 
$D^{\rm B}(\bk_1,\bk_2,\bk_3,z) = \exp\{-(k_1^2\mu_1^2+k_2^2\mu_2^2+k_3^2\mu_3^2)[f_\mathrm{g}^2(z)\,\sigma_{\rm p}^2(z)+\sigma_{\rm r}^2(z)]/2\}$. The fiducial values of the 
stochastic terms in Eqs.~\eqref{eq:Pgs} and \eqref{eq:Bgs} are taken to be those predicted by Poisson statistics 
\citep{desjacques_large-scale_2018}, i.e. $P_\mathrm{s} = 1/\bar{n}_{\rm g}$, 
$P_{\varepsilon\varepsilon_{\delta}}=b_1/(2\bar{n}_{\rm g})$, and $B_{\varepsilon} = 1/\bar{n}_{\rm g}^2$.

The fiducial values of the linear and quadratic galaxy bias coefficients, needed in Eqs. \eqref{eq:Z1} and \eqref{eq:Z2}, are derived by using the peak-background split argument applied on the best-fit halo mass function of \cite{Tinker2010}, together with the halo occupation distribution (HOD) model of \citet{Yankelevich_2018}. Moreover, the second-order tidal field bias term, following the convention of \citep{Baldauf:2012hs}, is given by $b_{s^2}(z)=-4[b_1(z)-1]/7$.  In the presence of PNG, the peak-background split argument is used to derive the non-Gaussian halo bias coefficients for the shapes considered here \citep{Schmidt2010,Giannantonio2010,Scoccimarro2011,Desjacques2011b,Schmidt2013,Desjacques2011a,Schmidt2013,desjacques_large-scale_2018}:
 \begin{equation}\label{eq:bpsi}
     b_\Psi(M,z)=A\left[b_\Phi+4\left(\frac{\dd\ln\sigma_{R,-\alpha}^2}{\dd\ln\sigma_{R}^2}-1\right)\right]\frac{\sigma_{R,-\alpha}^2}{\sigma_{R}^2}
  \end{equation}
  and 
   \begin{align}\label{eq:bpsidE}
   b_{\Psi\delta}(M,z)&=A\Bigg[\delta_c\left(b_2-\frac{8}{21}(b_1-1)\right) \nonumber \\
   &+(b_1-1)\left(2\frac{\dd\ln\sigma_{R,-\alpha}^2}{\dd\ln\sigma_{R}^2}-3\right)\Bigg]\frac{\sigma_{R,-\alpha}^2}{\sigma_{R}^2}+b_\Psi\,,
  \end{align}
where $\delta_c=1.686$, $ \sigma_{R,n}^2=(2\pi)^{-3}\int \dd^3\bk\, k^n W_R(k)^2P_{\rm m}^{\rm \, L}(k;z) $, with $W_R$ being the top-hat filter. For local PNG, $\alpha=0$ and $A=1$, the above expressions reduce to $b_\Psi\equiv b_\Phi$ and $b_{\Psi\delta}\rightarrow b_{\Phi\delta}= \delta_c \, \left [b_2 - 8/21 \, (b_1 - 1) \right ] - b_1 + 1 + b_\Phi$ \citep{desjacques_large-scale_2018, Barreira:2021ueb,Cabass_2022}. Moreover, for equilateral and orthogonal-CMB PNG they are $\alpha=2,A=3$, and $\alpha=1,A=-3$, respectively \citep{Schmidt2010,Giannantonio2012}. The distinct squeezed-limit behaviour of the orthogonal-LSS template, compared to orthogonal-CMB, has important implications for the scale dependence of PNG bias terms. Like the equilateral case, this template does not lead to a scale-dependent bias correction (i.e. $\alpha = 2$) \citep{Schmidt2010}. These results are used together with the HOD model to derive the fiducial values for the PNG galaxy bias coefficients at redshift $z_i$. 

The exact expression for $b_\Phi$ is still unresolved and under investigation within the community \citep[see, e.g.,][for a discussion]{Barreira:2022sey}. The conventional derivation relies on the universality relation of the halo mass function, i.e. $b_\Phi$ depends only on the halo mass and redshift. Recent work has shown that it depends on the halo formation history \citep{Barreira:2020kvh,Barreira:2021ueb,Lazeyras:2022koc,2023arXiv231110088F}, i.e. on properties beyond total mass. Here, we adopt the modified universality relation $b_\Phi=2\delta_c(b_1-p)$ with a fixed $p=0.55$ \citep{Barreira_2020b,Barreira:2022sey,Cabass_2022}. 
 
Note that in the case of equilateral and orthogonal-LSS PNG templates ($\alpha=2$), the last term in Eq. \eqref{eq:bpsi} and the last two terms in Eq. \eqref{eq:bpsidE} become scale-independent on large scales (i.e. $b_\Psi\propto 1$). This indicates that these terms do not have constraining power on equilateral and orthogonal-LSS PNG \cite[see, e.g.,][for a discussion]{Schmidt2010,Scoccimarro2011,Schmidt2012,Assassi_2015}. Hence, we exclude these terms from the linear and quadratic redshift kernels. This means that the sole contribution in the equilateral and orthogonal-LSS PNG forecasts is the primordial bispectrum (i.e. $B_I$ in Eq. \ref{eq:Bgs}).  

\subsubsection{Fisher forecasts}
\label{sec:fisher_results}

The Fisher matrix for the bispectrum in one redshift bin $z_i$ is 
   \begin{align}\label{eq:fisherBs}
  F_{\alpha\beta}^{B}(z_i)=&\frac{1}{4\pi}\!\sum_T\!\int_{-1}^1\!\! {\rm d}\mu_1 \!\! \int_0^{2\pi}\!\! {\rm d} \phi \frac{\partial B_{\rm obs}(\bk_1,\bk_2,\bk_3,z_i)}{\partial p_{\alpha}}\;[\text{Cov}^{\rm G}]^{-1} \nonumber \\
  &\times\frac{\partial B_{\rm obs}(\bk_1,\bk_2,\bk_3,z_i)}{\partial p_{\beta}}\,,
  \end{align}
where $\Sigma_T$ indicates the sum over all triangles formed by the Fourier modes $k_1$, $k_2$ and $k_3$. Here we only consider the Gaussian part of the bispectrum covariance matrix. In the thin shell limit ($k\gg \Delta k$), it is given by
\begin{align}\label{eq:Cov_G}
    \text{Cov}_{ij}^{\rm G} =\,\frac{(2\pi)^6}{V_s\,V_{123}}\,s_{123}\,\delta_{ij}\,P_{\rm obs}(\bk_1)\,P_{\rm obs}(\bk_2)\,P_{\rm obs}(\bk_3)\, ,
  \end{align}
 where $s_{123}=6,2,1$ for equilateral, isosceles, and scalene triangles, respectively. In this limit, the volume of the fundamental triangle bins is $V_{123}= 8\,\pi^2\,k_1\,k_2\,k_3\,\Delta k ^3$ \citep{Sefusatti2006}, where the bin size $\Delta k$ is taken to be the fundamental frequency of the survey, i.e. $\Delta k=k_{\rm f}=2\pi/V_{\rm s}^{1/3}$. The expression for $V_{123}$ requires some corrections for the extreme cases of flattened ($k_1=k_2+k_3$) and open ($k_1\ne k_2+k_3$) triangles, which are presented in \citet{Biagetti:2021tua} and considered here as well.

\begin{table}[t]
\caption{Forecasted 1$\sigma$ uncertainties on the three PNG shapes from the \GCsp{} sample of \Euclid from the power spectrum, bispectrum, and their joint signal. Inside the parenthesis we present the forecasts assuming the PNG bias coefficients to be free in the presence of priors (see Sect. \ref{sec:fisher_results} for a discussion). All the results consider {\em Planck} priors for the cosmological parameters.}
\centerline{
\begin{tabular} { l c c c }
\noalign{\vskip 3pt}\hline\noalign{\vskip 1.5pt}\hline\noalign{\vskip 6pt}
  & \bf P & \bf B & \bf P+B \\
\noalign{\vskip 3pt}\hline\noalign{\vskip 1.5pt}\hline
\noalign{\vskip 6pt}
\multicolumn{4}{c}{  Pessimistic quasilinear setting} \\ 
\multicolumn{4}{c}{\GCsp \, $(k_{\rm max} = 0.15\,h\,{\rm Mpc}^{-1})$}  \\
\noalign{\vskip 3pt}\hline\noalign{\vskip 6pt}
$\fnll$ &  2.8 (-) & 3.3 (12.7) & 2.2 (12.5) \\  
$\fnle$ &  - & 126 (126) & 108 (108)  \\  
$\fnloC$ &  214 (-) & 38 (40) & 33 (34) \\ 
$\fnloL$ &  - & 37 (37) & 33 (33) \\
\noalign{\vskip 3pt}\hline\noalign{\vskip 1.5pt}\hline\noalign{\vskip 6pt}
\multicolumn{4}{c}{  Pessimistic setting} \\ 
\multicolumn{4}{c}{\GCsp \, $(k_{\rm max} = 0.25\,h\,{\rm Mpc}^{-1})$}  \\
\noalign{\vskip 3pt}\hline\noalign{\vskip 6pt}
$\fnll$ &  2.7 (-) & 2.1 (9.1) & 1.7 (8.7)  \\  
$\fnle$ &  - & 74 (74) & 63 (63) \\  
$\fnloC$ & 177 (-) & 30 (31) & 24 (25)  \\ 
$\fnloL$ &  - & 23 (23) & 21 (21) \\
\noalign{\vskip 3pt}\hline\noalign{\vskip 1.5pt}\hline\noalign{\vskip 6pt}
\multicolumn{4}{c}{  Optimistic setting} \\ 
\multicolumn{4}{c}{\GCsp \, $(k_{\rm max} = 0.3\,h\,{\rm Mpc}^{-1})$}  \\
\noalign{\vskip 3pt}\hline\noalign{\vskip 6pt}
$\fnll$ &  2.7 (-) & 1.9 (8.5) & 1.6 (8.1)  \\  
$\fnle$ &  - & 64 (64) & 53 (53)  \\  
$\fnloC$ &  130 (-) & 27 (29) & 22 (24)  \\ 
$\fnloL$ &  - & 20 (20) & 19 (19) \\
\noalign{\vskip 3pt}\hline\noalign{\vskip 1.5pt}\hline\noalign{\vskip 5pt}
\end{tabular}
}\label{table:PNGresults}
\end{table}
 
The Gaussian covariance is assumed to be accurate enough up to linear and mildly non-linear scales, for high-density samples \citep{Howlett:2017vwp,Barreira:2017kxd,Li:2018scc,Blot:2018oxk,Chan2017}. However, higher-order corrections to the bispectrum variance still have an effect even at the mildly non-linear scales at low redshifts. Hence, we consider these corrections in the bispectrum Fisher analysis by following the prescription of \citet{Chan2017}. 

The stochastic contributions, the amplitude of the FoG effect, and the bias coefficients are treated as nuisance parameters and are marginalised over in the analysis. We use the Fisher analysis to generate  power spectrum and bispectrum 
forecasts on the parameters of interest for $\Lambda$CDM with non-Gaussian initial conditions,
 \begin{equation}
    \Theta_{\rm final} = \left\{\Omega_{\rm m}, \Omega_{\rm b}, h, n_{\rm s}, \sigma_8,  f_{\rm NL}\right\} \,.
\end{equation}  
The forecasted 1$\sigma$ uncertainties on the PNG amplitude of the three types considered here, coming from the \GCsp{} sample, are presented in Table \ref{table:PNGresults}. The forecasts from the galaxy power spectrum, bispectrum, and their combined signal are shown, where the latter is derived by summing the Fisher matrices of the first two without considering their cross-covariance, since it would have a minimal impact on our results \citep{Chan2017,Yankelevich_2018}.  These findings highlight the potential of the \Euclid mission to tightly constrain the primordial Universe.
  
The forecasts are presented for different small-scale cutoffs (i.e. $k_{\rm max}$), with an increasing value, venturing further into the non-linear regime. In the case of local PNG, the power spectrum dominates over the bispectrum as the main source of signal for the small $k_{\rm max}$ value, while the reverse is observed as we allow smaller scales to enter the analysis. For the equilateral and orthogonal PNG types, bispectrum drives the forecasts for all the settings considered. This indicates the importance of the bispectrum in the analysis of the \GCsp{} sample in order to achieve tight constraints on the PNG amplitude.

In addition to the main forecasts presented in Table \ref{table:PNGresults}, for which the PNG bias coefficients (i.e. $b_\Psi$ and $b_\Psi\delta$) are considered known, we present the results when these parameters are marginalised over (values inside the parenthesis). Following recent developments in the literature in effectively constraining the PNG bias coefficients \citep{2023arXiv231110088F}, we use Gaussian priors for $b_\Psi$ and $b_\Psi\delta$ centred around their fiducial values (see the previous section for a discussion) with a $40\%$ relative standard deviation \citep[following][]{andrews_euclid_2024}. Note that in this case the power spectrum only provides constraints on the $\fnl\, b_{\Psi}$ product. The impact of marginalising over the PNG bias parameters, given the chosen priors, is significant for forecasts on local PNG, whereas it remains minimal for the other two types.

\subsection{Bayesian analysis (\borg)}
\label{sect:borg_analysis}

In this section, we highlight the results of \citet{andrews_euclid_2024} regarding the inference of {\em local} primordial non-Gaussianity using the Bayesian origin reconstruction from galaxies (\borg) algorithm \citep{jasche_bayesian_2013,2015JCAP...01..036J,2016MNRAS.455.3169L,2019A&A...625A..64J,Lavaux2019}.

The \borg algorithm is a Bayesian hierarchical field-level inference framework for analysing cosmic structure in cosmological surveys \citep{jasche_bayesian_2013,2015JCAP...01..036J,2016MNRAS.455.3169L,2019A&A...625A..64J,Lavaux2019}. It employs physical forward modeling of 3-dimensional galaxy fields to link the observed data to the 3-dimensional primordial matter fluctuation field. The algorithm proposes a set of initial conditions $\epsilon$, simulates the gravitational evolution of the matter field, and populates the evolved density field through a galaxy bias model \citep{Assassi_2015,desjacques_large-scale_2018,Barreira_2020b}. This process generates model predictions compared to observed galaxy redshift data and includes the primordial perturbation with $\fnll$ and scale-dependent biases under the assumption of the universal mass approximation \citep{barreira_local_2021,Lucie_Smith_2023,2023arXiv231110088F,2023arXiv231212405A}. The target posterior distribution explored, which is generated by the data model, is formally expressed as
\begin{align}
&\mathcal{P}_{\rm{post}}\left(\epsilon, \fnl, \left\{b_i^\textrm{g} \right\}|N_{\textrm{g}}^\textrm{o}\right) \propto \nonumber \\ & \quad \quad \quad \quad \quad \mathcal{P}_{f}\left(\fnl\right) \, \mathcal{P}_{\epsilon}\left(\epsilon\right)\, \mathcal{P}_{b}\left\{b_i^\textrm{g} \right\} \, \mathcal{P}_{\rm{like}}\left(N_{\textrm{g}}^\textrm{o}|\epsilon, \fnl, \left\{b_i^\textrm{g} \right\}\right) \, , 
\label{eq:full_posterior}
\end{align}
where $N_{\textrm{g}}^\textrm{o}$ is the observed galaxy count data, $\mathcal{P}_{\epsilon}\left(\epsilon\right)$ is the Gaussian white-noise prior with zero mean and unit standard variance, $\mathcal{P}_{\rm{like}}\left(N_{\textrm{g}}^\textrm{o}|\epsilon, \fnl, \left\{b_i^\textrm{g} \right\}\right)$ is a likelihood distribution, and $\{b_i \}$ are the bias parameters \citep{andrews_euclid_2024}. Using Hamiltonian Monte Carlo and slice sampling, \borg explores the joint posterior of data, and thus infers the primordial initial conditions, the $\fnl$ parameter, and marginalises over nuisance parameters \citep{jasche_bayesian_2013,andrews_bayesian_2023}. This approach consistently accounts for observational and survey systematic effects and the incorporation of multiple PNG probes into the analysis \citep{jasche_bayesian_2013,andrews_bayesian_2023}.

For mock data generation based on \Euclid forecast specifications, the forward model is applied to a white-noise field and incorporates structure formation effects, redshift selection functions, sky masks, and other realistic observational aspects. The Markov chain starts at a random point for the initial conditions and converges with the target posterior distribution during the burn-in phase. Then, \borg{}, operating at resolutions down to $62.5 \,\Mpch$, infers the initial conditions and $\fnll$ that align closely with the expected confidence intervals of fiducial mock data. We analyse multiple mock data sets, each addressing various factors, such as different resolutions, different ground truth $\fnll$ values, and sampling of additional bias parameters. The overall scope of runs contributes to a comprehensive understanding of inference processes.

We give an overview of the results in Table \ref{tab:fnl_table}. The study demonstrates successful sampling of $\fnll$ using \Euclid data, achieving 1$\sigma$ constraints across all runs. Notably, at a $62.5 \, \Mpch$ resolution, we attain a constraining power of $\sigma (\fnll) = 2.3$, showcasing the method's ability to be competitive with other methods. These findings include handling \Euclid-like specifications in the analysis framework, high-precision $\fnl$ inference at multiple resolutions, and the impact of different parameter settings on the constraints. These results demonstrate the opportunity to use field-level inference as a complementary and independent approach to constrain the parameter space of plausible models for the early Universe.

\begin{table}[]
\caption{A table displaying $\fnll$ inferences for each run in the \borg{} analysis, highlighting how changes in the data model affect constraints on $\fnll$ for the given mock data. The last column provides a brief description of the analysis, with examples including resolution study or a change in the bias parameters. All $\fnll$ estimates fall within $1\sigma$ of fiducial values, indicating consistent and reliable inference. Runs \#1--\#6 assume the universal mass function while Run \#7 samples scale-dependent bias parameters. $k_{\rm max}$ values are in $h\,{\rm Mpc}^{-1}$.
}
\begin{tabular}{llllll}
\hline\noalign{\vskip 1.5pt}\hline
\vspace{0cm}Run \# & $f_{\mathrm{NL}}^{\mathrm{fid.}}$ & $\langle \fnl \rangle$ & $\sigma_{\fnl}$ & ${k_{\rm max}}^{*}$ & Description \\ \hline\noalign{\vskip 1.5pt}\hline
1         & 0 & $-2.8$  & 6.0 & 0.025 &  Low resolution \\
2         & 0 & $\phantom{-}2.6$  & 4.0 & 0.05\phantom{0} &  Mid resolution \\
3         & 0 & $-1.3$  & 2.3 & 0.1\phantom{0}\phantom{0} &  High resolution \\
4         & 5 & $\phantom{-}6.0$  & 2.5 & 0.1\phantom{0}\phantom{0}  &  Non-zero fiducial $\fnll$ \\
5         & 0 & $\phantom{-}1.7$  & 2.6 & 0.05\phantom{0} &  Fixed bias \\
6         & 0 & $\phantom{-}0.9$  & 6.3 & 0.05\phantom{0} &  $p=1$ \\
7         & 0 & $\phantom{-}0.4$  & 5.0 & 0.05\phantom{0} &  Sampling of $b_\phi$, $\bpd$ \\ \hline\noalign{\vskip 1.5pt}\hline
\end{tabular}
\\{\tiny * in units of $h\,{\rm Mpc}^{-1}$}
\label{tab:fnl_table}
\end{table}

\begin{figure}[!]
	\centering
    \includegraphics[width=0.9\columnwidth]{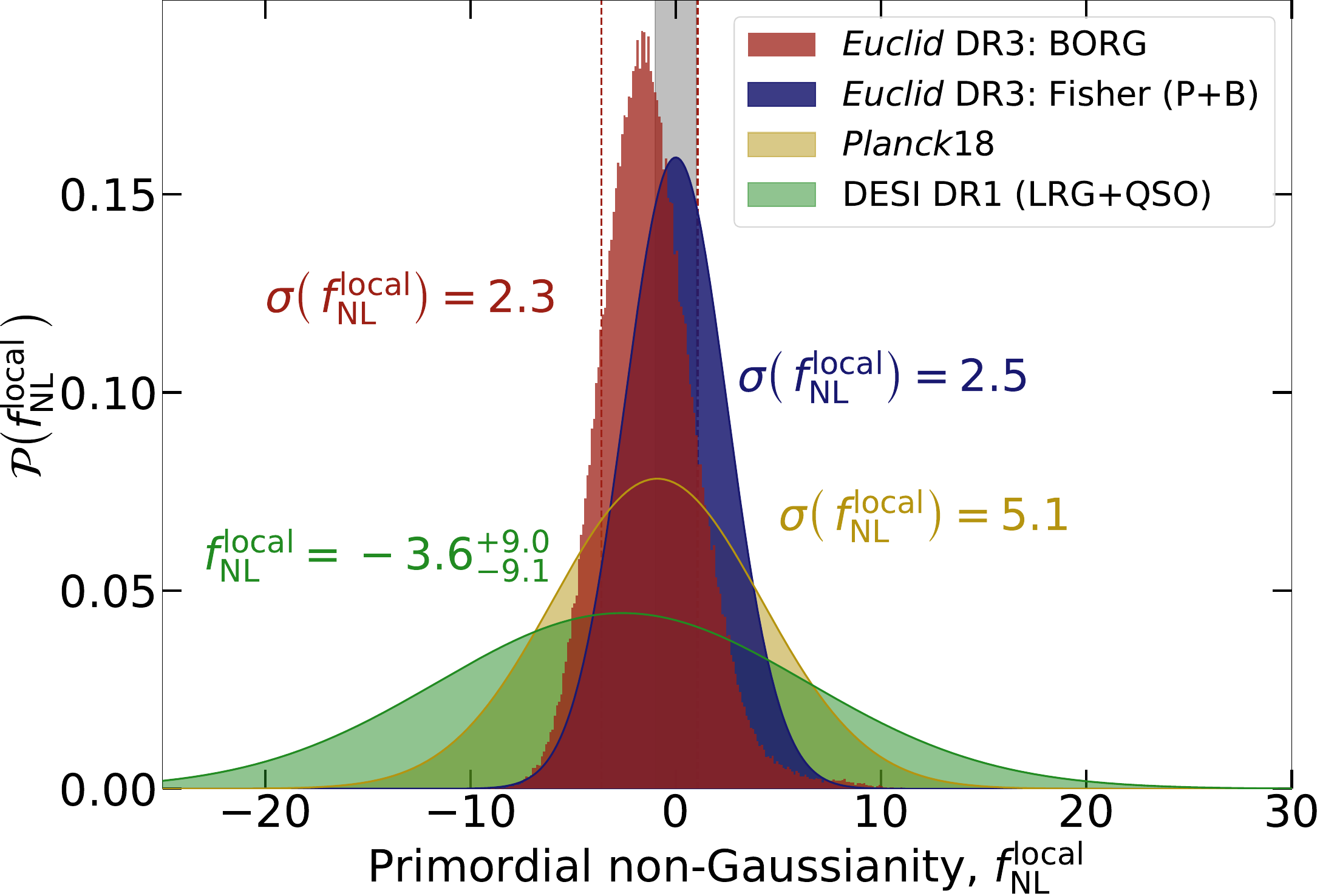}
	\caption{Forecasted constraints on $\fnll$ in \textit{Euclid}-like data for the Fisher matrix formalism (Sect. \ref{sect:fisher_forecasts}) and the \borg{} method (Sect. \ref{sect:borg_analysis}). These results correspond to a fixed $k_{\rm max}=0.1 h\,{\rm Mpc}^{-1}$ (i.e. Run \# 3 of the \borg{} method). For comparison, we also include the current state-of-the-art constraints on $\fnll$ from {\it Planck} \citep{Planck:2019izv}.
 }
\label{fig:local_png_pdfs}
\end{figure}

In Fig.~\ref{fig:local_png_pdfs}, we provide the main results of this section, a forecast of the achievable constraints for the \Euclid-like mock with the method. Moreover, complementary to the main \borg{} results, we show the forecasted constraints from the Fisher matrix approach, presented in the previous section. These results correspond to the summed power spectrum and bispectrum signal after fixing the cosmology and marginalising over the nuisance parameters, as well as keeping a fixed $k_{\rm max}=0.1$ $h\,{\rm Mpc}^{-1}$ over all redshifts, in order to match the specifications of the main \borg{} results. The 1$\sigma$ forecasted errors on $\fnll$ generated by the two methods are in good agreement.

\section{Beyond slow roll: searches for primordial features}
\label{sec:features}

In this section, we summarise the findings of the analysis and forecasts performed in \citet{2023arXiv230917287B}. The search for primordial features in the primordial power spectrum and bispectrum will benefit greatly from future 
LSS measurements expected from \Euclid 
\citep{Wang:1998gb,Zhan:2005rz,Huang:2012mr,Chen:2016vvw,Chen:2016zuu,Ballardini:2016hpi,Xu:2016kwz,Fard:2017oex,Palma:2017wxu,Ballardini:2017qwq,Ballardini:2018noo,Beutler:2019ojk,Ballardini:2019tuc,Debono:2020emh,Li:2021jvz,Ballardini:2022vzh,Ballardini:2022wzu,Mergulhao:2023ukp,2023arXiv230917287B}.

In general, and depending on their origin, features in the primordial power spectrum of 
density perturbations are the sign of conditions and dynamics at the very early Universe connected to new 
physical scales \cite[see][for details and references]{Achucarro_2022}. Modifications to the slow-roll 
dynamics such as an abrupt change in the background functions $\epsilon_1$ \citep{Starobinsky:1992ts,Adams:2001vc} 
and $c_{\rm s}$ \citep{Achucarro:2010da,Chen:2011zf} at a particular moment during inflation, the periodicity 
in the background functions \citep{Chen:2008wn,Flauger:2009ab,Flauger:2010ja,Chen:2010bka}, and deviations from the standard Bunch-Davies 
vacuum of perturbations, (as well as combination of them) all result in oscillatory primordial features. 

Following the Fisher-matrix-based methodology described in 
EC20 and here, we calculate
%\citet{Euclid:2019clj} and here, we calculate 
the forecasted uncertainties from \Euclid's primary probes, i.e. spectroscopic galaxy clustering and the 
combination of photometric galaxy clustering and photometric weak lensing, for parametrised 
primordial features in the primordial power spectrum of density perturbations.
We study two models characterised by a correction to the primordial power spectrum and to the matter 
power spectrum, $1 + {\cal A}_{\rm X} \sin\left(\omega_{\rm X}\Xi_{\rm X} + 2\pi\phi_{\rm X}\right)$, 
where ${\cal A}_{\rm X}$ is the amplitude of the primordial feature, $\omega_{\rm X}$ is the dimensionless frequency, 
and $\phi_{\rm X}$ is the normalised phase. In these models, we have a superimposed undamped oscillatory signal with 
$\Xi_{\rm lin} \equiv k/k_*$ for linearly spaced oscillations (LIN) and with $\Xi_{\rm log} \equiv \ln(k/k_*)$ for logarithmic ones (LOG).

In addition to the \Euclid's primary probes, we consider further information available from \Euclid 
measurements including a numerical reconstruction of non-linear spectroscopic galaxy clustering \citep[][PS rec]{Beutler:2019ojk,Li:2021jvz}, the galaxy clustering bispectrum \citep{Karagiannis:2018jdt},\footnote{We note that primordial features 
generated in single-field models are expected to show correlated signals between the primordial bispectrum and 
the primordial power spectrum \citep{Chen:2006xjb,Chen:2008wn,2014PhRvD..89f3540G,2015JCAP...10..062M} and can be studied through a combined analysis 
\citep{Fergusson:2014tza,Meerburg:2015owa,Akrami:2018odb,Karagiannis:2018jdt}.} and the expected additional 
information (without including the cross-correlation) from CMB experiments \citep{Euclid:2021qvm} such as {\em Planck} 
\citep{Planck:2018vyg}, SO \citep{SimonsObservatory:2018koc}, and CMB-S4 \citep{Abazajian:2019eic}.

We use perturbation theory (PT) predictions to model the galaxy clustering non-linear matter power spectrum 
\citep{Blas:2016sfa,Vasudevan:2019ewf,Beutler:2019ojk,Chen:2020ckc,2024arXiv241102261B} plus a modified version of {\tt HMCODE} 
\citep{Mead:2016zqy} for the photometric probes. In order to assess the accuracy of the PT predictions at leading order 
and next-to-leading order, we generate a set of N-body simulations based on the COmoving Lagrangian Approximation (COLA) 
method \citep{Tassev:2013pn,Tassev:2015mia,Winther:2017jof,Wright:2017dkw} with primordial oscillations injected to their 
initial conditions. We also use simulations produced in \citet{Ballardini:2019tuc} with the N-body code {\tt GADGET-3}, 
a modified version of the publicly available code {\tt GADGET-2} \citep{Springel:2000qu,Springel:2005mi}, to validate 
the COLA-based simulations and we use the N-body code {\tt ramses} \citep{teyssier2002} to generate snapshopts of dark 
matter to reconstruct the density fields as done in \citet{Li:2021jvz}.

\begin{figure}
\centering
\includegraphics[width=0.49\textwidth]{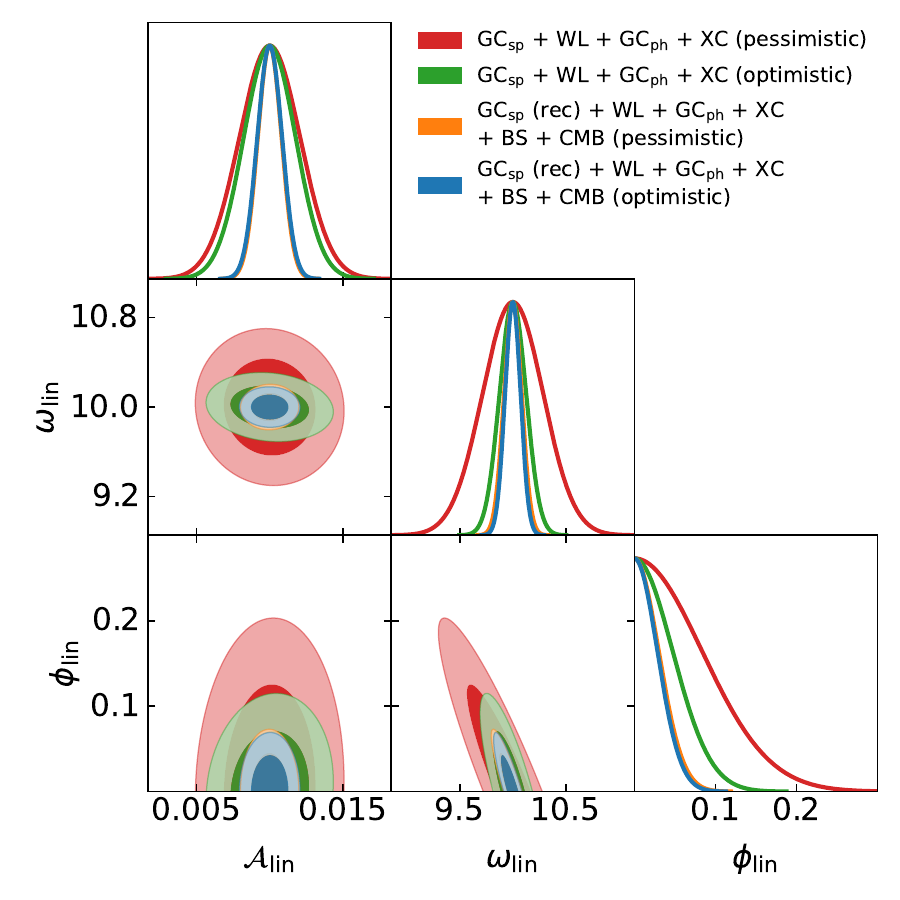}
\includegraphics[width=0.49\textwidth]{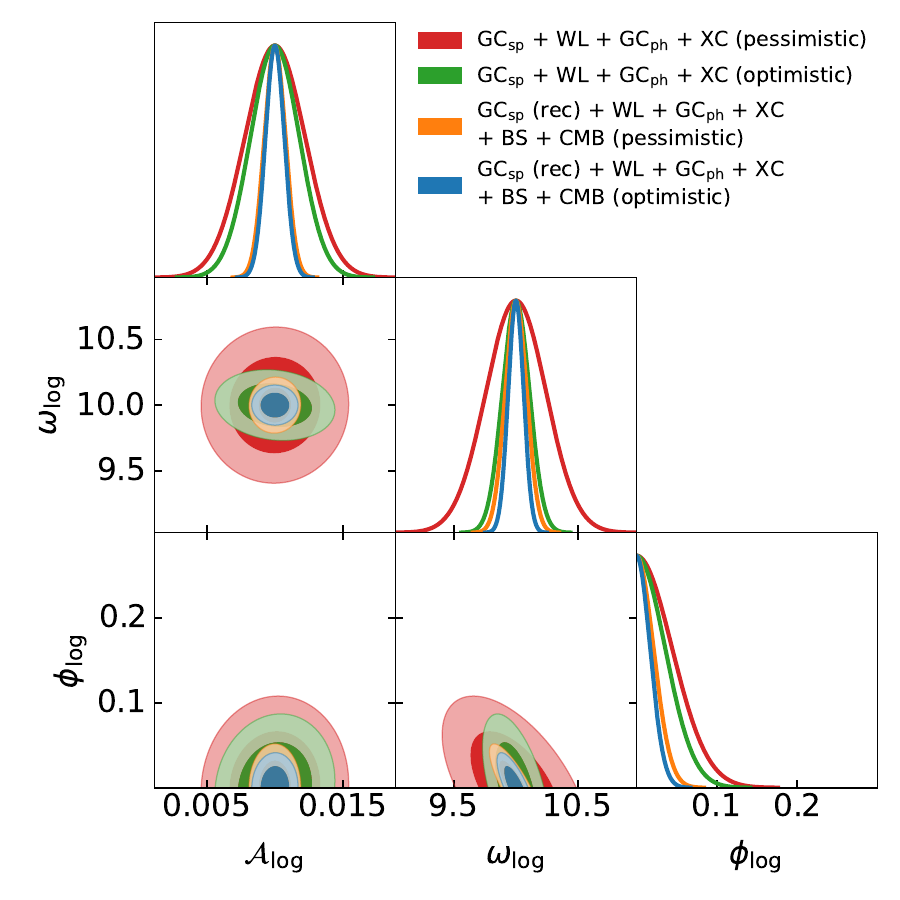}
\caption{Fisher-forecast marginalised two-dimensional contours and one-dimensional probability distribution functions 
from \Euclid on the primordial feature parameters for the LIN model with $\omega_{\rm lin} = 10$ (top panel) and the 
LOG model with $\omega_{\rm log} = 10$ (bottom panel); in both cases we have ${\cal A}_{\rm X} = 0.01$ and $\phi_{\rm X} = 0$. 
We show \GCsp+WL+\GCph+XC with the pessimistic (optimistic) setting in red (green) and \GCsp(rec)+WL+\GCph+XC in combination 
with \Euclid\ \GCsp\ bispectrum and SO-like CMB measurements with the pessimistic (optimistic) setting in orange (blue).}
\label{fig:triangle_features}
\end{figure}

\begin{table}[h]
\centering
\caption{Fisher-matrix-based forecasted marginalised uncertainties for the feature model parameters, relative to 
their corresponding fiducial values (${\cal A}_{\rm X} = 0.01$, $\omega_{\rm X} = 10$, $\phi_{\rm X} = 0$), for the LIN and 
LOG models in the pessimistic and optimistic settings, using \Euclid's primary probes (\GCsp+WL+\GCph+XC) alone and in combination 
with non-linear reconstruction of \GCsp, galaxy clustering bispectrum, and SO-like CMB.}
\begin{tabular} {lcc}
\hline
\hline
 &  \Euclid  &  \GCsp(rec) + WL + \GCph + XC  \\ 
 &                     & + BS + SO-like                  \\ 
 &                     & + {\em Planck} low-$\ell$   \\
\hline
\multicolumn{3}{l}{Pessimistic} \\
\multicolumn{3}{l}{setting} \\
\hline
$A_{\rm lin}$      &  21\%  & 8.5\%   \\  
$\omega_{\rm lin}$ &  2.9\% & 0.83\%  \\  
$\phi_{\rm lin}$   &  0.083 & 0.030  \\  
\hline
$A_{\rm log}$      &  22\%  & 7.3\%  \\  
$\omega_{\rm log}$ &  2.4\% & 0.81\%  \\  
$\phi_{\rm log}$   &  0.044 & 0.019  \\ 
\hline
\hline
\multicolumn{3}{l}{Optimistic} \\
\multicolumn{3}{l}{setting} \\
\hline
$A_{\rm lin}$      &  18\%  & 8.1\% \\  
$\omega_{\rm lin}$ &  1.2\% & 0.84\% \\  
$\phi_{\rm lin}$   &  0.047 & 0.028  \\  
\hline
$A_{\rm log}$      &  18\%  & 6.6\% \\  
$\omega_{\rm log}$ &  1.1\% & 0.60\% \\  
$\phi_{\rm log}$   &  0.035 & 0.016\\ 
\hline
\hline
\end{tabular}
\label{tab:results_PF}
\end{table}
While the uncertainties on the primordial feature amplitude depend strongly on the value of the dimensionless frequency 
$\omega_{\rm X}$ considering single \Euclid probes, the combination of the spectroscopic and photometric probes (as well 
as the further addition of non-primary probes) allows us to reach robust uncertainties on the entire range of frequencies 
$(10^{0.2},\,10^{2.1})$ we consider here.
Our uncertainties, combining all the sources of information expected from \Euclid in combination with future CMB experiments, 
e.g. SO complemented with {\em Planck} at low $\ell$, correspond to ${\cal A}_{\rm lin} = 0.0100 \pm 0.0008$ and ${\cal A}_{\rm log} = 0.0100 \pm 0.0008$ for \GCsp\ (PS rec + BS)+WL+\GCph+XC+SO-like for both the pessimistic and 
optimistic settings. 
In Table~\ref{tab:results_PF}, we summarise the marginalised uncertainties of the primordial feature parameters, 
in percentage relative to the corresponding fiducial values, for the full combination of \Euclid's primary probes, i.e. 
\GCsp+WL+\GCph+XC, and in combination with the reconstruction of non-linear spectroscopic galaxy clustering, galaxy clustering bispectrum, and CMB information; see Fig.~\ref{fig:triangle_features}.

\section{Conclusions}
\label{sec:conclusions}

In this paper, we have presented highlights for the {\em Euclid} capability to test the cosmic initial conditions. 
In most of the cases, we have exploited the uniqueness of {\em Euclid}, i.e. the combination of the 3-dimensional galaxy clustering from the spectroscopic survey and the $3 \times 2$pt from the photometric survey. This combination guarantees the removal of many degeneracies among the cosmological parameters that affect each probe separately
and ensures constraints that are tighter than those from any other galaxy surveys in operation.

We have presented Fisher forecasts following EC20 for the spatial curvature, the running of the spectral index of the power spectrum of curvature perturbations,
isocurvature perturbations, features in the primordial power spectrum from the main probes, and forecasts for primordial non-Gaussianity from the spectroscopic survey.

We have found that the combination of the {\em Euclid}'s  main probes will provide the uncertainty of $\sigma(\Omega_K) = 0.0044$ $ (0.003)$ in the pessimistic (optimistic) setting assuming flat spatial sections as fiducial cosmology. These expected constraints can powerfully break
the parameter degeneracy in what is achievable through the measurements of CMB temperature and polarisation anisotropies and return tighter limits such as
$\sigma (\Omega_K) = 0.0009$ ($0.0006$) for the {\em Euclid}'s  pessimistic (optimistic) setting in combination with the Simons Observatory.

For the fiducial value
$\alpha_s = - 0.01$, we have found that the combination of the {\em Euclid}'s main probes can reach the uncertainties $\sigma (n_{\rm s}) \sim 0.008$ $(0.0044)$ for the scalar spectral index
 and $\sigma (\alpha_{\rm s}) \sim 0.005$ $ (0.002)$ for its running in the pessimistic (optimistic) setting.
We have also found the uncertainty of $\sigma(\alpha_s) = 0.004$ $(0.0015)$ for a smaller fiducial value $\alpha_s = - 0.001$, with
$\sigma (n_{\rm s}) \sim 0.006$ $(0.0036)$, always in the pessimistic (optimistic) setting.
These uncertainties are comparable with current constraints from {\em Planck} \citep{Akrami:2018odb}.
In addition, we have shown how {\em Euclid}, in
combination with future CMB experiments, will probe $\alpha_{\rm s} \lesssim {\cal O} (10^{-3})$, therefore nearing the running expected in the currently preferred simplest slow-roll inflationary models.

By leveraging the smaller scales that {\em Euclid}  has access to compared to the CMB measurements, we have explored what can be achieved for testing the adiabaticity of the initial conditions.
Even though {\em Euclid}  cannot achieve the CMB sensitivity to nearly scale-invariant CDM isocurvature perturbations,
it has the capability to provide constraints on isocurvature perturbations with a
blue spectral index that are at least one order of magnitude better than the current ones based on {\em Planck}. In order to achieve these tight limits, the use
of the $3\times2$ pt is essential, and therefore, more work is needed to model non-linear scales in the presence of isocurvature perturbations at $k$ beyond the EFT
treatment suitable for \GCsp\ adopted in \citet{2023PhRvD.108j3542C}.

We have also forecasted the constraints on
primordial non-Gaussianity from the spectroscopic survey. The combined power spectrum and bispectrum Fisher forecast leads to
$\sigma (f_{\rm NL}^{\rm local}) = 2.2$, $\sigma (f_{\rm NL}^{\rm equil}) = 108$, and
$\sigma (f_{\rm NL}^{\rm ortho}) = 33$ by assuming $k_{\rm max} = 0.15\,h\,{\rm Mpc}^{-1}$
and universality for the halo mass function. We have shown how the expected uncertainty on the
local shape can be improved up to 20\% by encompassing summary statistics and using Bayesian field-level inference, following \citet{andrews_bayesian_2023}. 

We have analysed {\em Euclid}'s sensitivity to features in the primordial power spectrum.
For {\em Euclid}'s  main probes, we have found the relative errors 21\% (22\%) in the pessimistic setting and 18\% (18\%) in the optimistic setting for the amplitude of
the linear (logarithmic) superimposed oscillations, with a fiducial value of $0.01$. These uncertainties can be further
improved by adding the information from the bispectrum and the non-linear reconstruction.

There are at least two important aspects that we have not studied in this paper.

The first aspect is the inclusion of {\em Euclid}-CMB cross-correlation \citep{Euclid:2021qvm} for the physics of the early Universe. There are certain parameters, such as primordial local non-Gaussianity, which can be constrained by the CMB cross-correlation -- see, e.g., \cite{Bermejo-Climent:2021jxf}.

The second aspect is the interplay between {\em Euclid}'s capability and the CMB $B$-mode polarisation in addition to the temperature and $E$-mode polarisation. The CMB $B$-mode polarisation is the imprint of inflationary
gravitational waves, whose amplitude is proportional to the energy scale at which inflation occurred, and is one of the main targets of next-generation CMB experiments dedicated to polarisation measurement \citep{SimonsObservatory:2018koc,Abazajian:2019eic,LiteBIRD:2022cnt}.
The search for the $B$-mode polarization will lead either to a detection or to an upper bound on $r$, {\em Euclid}'s constraints on $n_{\rm s}$ and $\alpha_{\rm s}$ will be highly complementary to the CMB measurements in constraining inflation.

To conclude, {\em Euclid}'s measurements will have profound implications for the physics of inflation.
The precision in the determination of $n_{\rm s}$ will pin down not only the slope of the inflationary potential, but also the uncertainties in the reheating stage after inflation, which affect the theoretical predictions of any inflationary model. The expected sensitivity of $\sigma (\alpha_{\rm s}) \sim {\cal O} (10^{-3})$ 
will take us unprecedentedly closer to a generic prediction of inflation, such as the small
but non-zero running of the scalar tilt, down to the expected values predicted by the simplest slow-roll inflationary models.
As discussed in the paper, {\em Euclid}'s \GCsp\ has the capability of measuring $\sigma (f_{\rm NL}^{\rm local})$
below the {\em Planck}'s constraint when assuming universality for the halo mass function. Expected  {\em Euclid}'s galaxy clustering measurements of
the local, as well as the other types of, primordial non-Gaussianity will be able to shed light on inflaton interactions and on the presence 
of massive fields during inflation, making the idea of testing inflation as a cosmological collider \citep{Arkani-Hamed:2015bza} operational.

\begin{acknowledgements}
FF acknowledges partial financial support from 
the contract ASI/ INAF for the {\em Euclid}  mission n.2018-23-HH.0, INFN InDark initiative and from the COSMOS network ({\tt www.cosmosnet.it}) through the ASI (Italian Space Agency) rants 2016-24-H.0 and 2016-24-H.1-2018, as well as 2020-9-HH.0 (participation in LiteBIRD phase A).
YA acknowledges support by the Spanish Research Agency (Agencia Estatal de Investigaci\'on)'s grant RYC2020-030193-I/AEI/10.13039/501100011033, by the European Social Fund (Fondo Social Europeo) through the  Ram\'{o}n y Cajal program within the State Plan for Scientific and Technical Research and Innovation (Plan Estatal de Investigaci\'on Cient\'ifica y T\'ecnica y de Innovaci\'on) 2017-2020, by the Spanish Research Agency through the grant IFT Centro de Excelencia Severo Ochoa No CEX2020-001007-S funded by MCIN/AEI/10.13039/501100011033, and by the Spanish National Research Council (CSIC) through the Talent Attraction grant 20225AT025.
DK acknowledge support by the MUR PRIN2022 Project “BROWSEPOL: Beyond standaRd mOdel With coSmic microwavE background POLarization”-2022EJNZ53 financed by the European Union - Next Generation EU. MB acknowledges financial support from the INFN InDark initiative and from the COSMOS network ({\tt www.cosmosnet.it}) through the ASI (Italian Space Agency) Grants 2016-24-H.0, 2016-24-H.1-2018, 2020-9-HH.0 (participation in LiteBIRD phase A). 
ZS acknowledge ITP institute funding from DFG project 456622116 and IRAP lab support.
DS acknowledges financial support from the Fondecyt Regular project number 1200171.
GCH acknowledges support through the ESA research fellowship programme.
JJ acknowledges support by the Swedish Research Council (VR) under the project 2020-05143 -- `Deciphering the Dynamics of Cosmic Structure'.
GL acknowledges support by the ANR BIG4 project, grant ANR-16-CE23-0002 of the French Agence Nationale de la Recherche.
AA, JJ, and GL acknowledge the support of the Simons Collaboration on ``Learning the Universe''. 
DP acknowledges financial support by agreement n. 2020-9-HH.0 ASI-UniRM2.
JV was supported by Ruth och Nils-Erik Stenb{\"a}cks Stiftelse.\\

\AckEC
\end{acknowledgements}

\bibliographystyle{aa_url}
%\nolinenumbers
\bibliography{Biblio,MarioBib,AdamBib}
%\linenumbers

\appendix

%\begin{appendix}

\section{Combination with current and future CMB measurements}
\label{sec:AppendixA}

In this appendix, we summarise how we compute and add the cosmological information from current and future 
measurements of CMB temperature, $E$-mode polarisation, and lensing.
Comparing and folding in the information contained in CMB anisotropies is important for the forecasts studied here, and more generally,
for the physics of the early Universe. The comparison shows how well {\em Euclid}  can probe initial conditions in the 
low-redshift and non-linear regime compared to the high-redshift and linear regime of the CMB, which is currently the 
most precise probe of the early Universe. The combination
of CMB measurements and {\em Euclid}  will show the complementarity of the two probes and will break many of the degeneracies present in each independent probe, leading to tighter
constraints.

We follow the recipes described in more detailed in \citet{Euclid:2021qvm}, where the interested reader can find the joint study of
{\em Euclid}  and CMB probes.
Note, however, that in this paper we do not include
the cross-correlation between {\em Euclid} and CMB measurements as in \citet{Euclid:2021qvm}, but we limit ourselves to probe combination, i.e. to the sum of the CMB and {\em Euclid}  Fisher matrices:
\begin{equation}
\label{eq:fishersum}
F_{\alpha \beta}^\mathrm{tot} = F_{\alpha \beta}^\mathrm{spec} + F_{\alpha \beta}^\mathrm{ph} + F_{\alpha \beta}^\mathrm{CMB}\,.
\end{equation}

The Fisher matrix $F_{\alpha \beta}^\mathrm{CMB}$ is defined as
\begin{eqnarray}
\label{eq:fishertrace2}
F_{\alpha \beta}^\mathrm{CMB} &=& 
\left\langle \frac{\partial^2 {\cal L}^\mathrm{CMB}}{\partial \theta_\alpha \partial \theta_\beta}
\right\rangle = \frac{1}{2} {\rm Tr} \left[  \frac{\partial {\cal C}}{\partial \theta_\alpha} {\cal C}^{-1}
\frac{\partial {\cal C}}{\partial \theta_\beta}  {\cal C}^{-1} \right] \nonumber \\
&=& \sum_{\ell_{\rm min}}^{\ell_{\rm max}} \sum_{abcd} \frac{2 \ell+1}{2}
f_{\rm sky}^\mathrm{CMB} \frac{\partial {C_\ell^{ab}}}{\partial \theta_\alpha} ({\cal C}^{-1})^{bc}
\frac{\partial {C_\ell^{cd}}}{\partial \theta_\alpha} ({\cal C}^{-1})^{da}\,,
\end{eqnarray}
where $abcd$ $\in$ $\{T,E,\phi \}$, $f_{\rm sky}^\mathrm{CMB}$ is the effective sky fraction, $C_\ell^{cd}$ is the $cd$ angular power spectrum,
and the theoretical covariance matrix ${\cal C}$ is defined as
\begin{equation}
\label{eq:cov}
{\cal C} =
\begin{bmatrix}
   \bar{C}_{\ell}^{TT} & C_{\ell}^{TE} & C_{\ell}^{T\phi} \\
    C_{\ell}^{TE} &    \bar{C}_{\ell}^{EE} & C_{\ell}^{E\phi} \\
        C_{\ell}^{T\phi} & C_{\ell}^{E\phi} &    \bar{C}_{\ell}^{\phi\phi}
\end{bmatrix} \,.
\end{equation}
Here $\bar{C}_{\ell}^{aa}$ include the isotropic noise deconvolved with the instrument beam \citep{Knox:1995dq},
\begin{equation}
{\cal N}_{\ell}^{aa} = w_{aa}^{-1} b_{\ell}^{-2}, \qquad b_{\ell} = \mathrm{e}^{-\ell(\ell+1)\theta_{\rm FWHM}^2/16 \ln 2}\,,
\label{isotropic}
\end{equation}
where $a \in \{T , E\}$, $\theta_{\rm FWHM}$ is the full width half maximum (FWHM)
of the beam in radians, and $w_{TT}$ and $w_{EE}$ are the inverse square of the detector noise level
for temperature and polarisation in arcmin$^{-1}$ $\mu$K$^{-1}$. For CMB lensing,
we use the resulting ${\cal N}_{\ell}^{TT}$ and ${\cal N}_{\ell}^{EE}$ to reconstruct
the minimum variance estimator for the noise ${\cal N}_{\ell}^{\phi\phi}$
\citep{Okamoto:2003zw} and use the publicly available
code \texttt{quicklens}.\footnote{\href{https://github.com/dhanson/quicklens}{https://github.com/dhanson/quicklens}}

We consider three different specifications for the CMB experiments with which {\em Euclid}  could be complemented.
We first consider a specification that reproduces the {\em Planck} 2018
uncertainties for the $\Lambda$CDM parameters \citep{Planck:2018vyg} with a simplified
setting as in Eq. (\ref{isotropic}): we use noise specifications corresponding to in-flight performances of the high-frequency
instrument (HFI) $143\,{\rm GHz}$ channel \citep{Planck:2018nkj} with the sky fraction $f_{\rm sky} = 0.7$
and the multipole range $\ell_{\rm min}=2$ to $\ell_{\rm max}=1500$ for temperature and
polarisation. The $E$-mode polarisation noise
is further inflated by a factor of 8
for $2 \le \ell \le 30$ to reproduce the uncertainty of the optical depth parameter at reionization $\tau$; see \citet{Bermejo-Climent:2021jxf}. Finally, CMB
lensing is obtained by combining the $143\,{\rm GHz}$ and $217\,{\rm GHz}$ HFI channels assuming a conservative multipole
range of 8--400. For the CMB, we also consider the optical depth at reionisation $\tau$ and then we marginalise over it before combining
the CMB Fisher matrix with the \Euclid ones.

For the SO-like experiment, we use noise curves provided by the SO collaboration in \citet{SimonsObservatory:2018koc}
taking into account residuals and noise from component separation.\footnote{We use version 3.1.0 available at \\ \url{https://github.com/simonsobs/so\_noise\_models}.}
We use spectra from $\ell_{\rm min}=40$ to $\ell_{\rm max}=3000$ for the temperature and temperature-polarisation
cross-correlation, and $\ell_{\rm max}=5000$ for the $E$-mode polarisation. The CMB lensing and temperature-lensing
cross-correlation spectra cover the multipole range of 2--3000. We consider the sky fraction $f_{\rm sky} = 0.4$.
We complement the SO-like information with the {\em Planck}-like large-scale data in the multipole range of 2--40
in both temperature and polarisation.

For CMB-S4, we use noise sensitivities of $1\,\mu{\rm K}\,{\rm arcmin}$ in temperature and
$\sqrt{2}\,\mu{\rm K}\,{\rm arcmin}$ in polarisation, with resolution of $\theta_{\rm FWHM} = 1\,{\rm arcmin}$.
We assume data over the same multipole ranges of SO and with the same sky coverage.
In all these three cases, we do not keep $\tau$ fixed for the CMB and then we marginalise over it before combining
the CMB Fisher matrix with the \Euclid ones.

\end{document}